\documentclass[usenatbib]{mnras}

\usepackage{etoolbox}
\newtoggle{postreferee}
\togglefalse{postreferee} 

\newtoggle{postrefereeB}

\newtoggle{removesomefits}

\newtoggle{CompareCentres}

\usepackage{newtxtext,newtxmath}

\usepackage[T1]{fontenc}
\usepackage{ae,aecompl}

\newcommand{\projectversion}{97bd6dc}
\newcommand{\projectzenodoid}{zenodo.4699702}
\newcommand{\projectzenodohref}{\href{https://zenodo.org/record/4699702}{zenodo.4699702}}
\newcommand{\projectzenodofilesbase}{https://zenodo.org/record/4699702/files}
\newcommand{\projectgitrepository}{\url{https://codeberg.org/boud/elaphrocentre}}
\newcommand{\projectgitrepositoryarchived}{\href{https://archive.softwareheritage.org/swh:1:snp:54f00113661ea30c800b406eee55ea7a7ea35279}{swh:1:dir:54f00113661ea30c800b406eee55ea7a7ea35279}}

\newcommand{\mpgraficname}{{\sc mpgrafic}}
\newcommand{\mpgraficversion}{0.3.19-4b78328}
\newcommand{\ramsesscalavname}{{\sc ramses-scalav}}
\newcommand{\ramsesscalavversion}{0.0-482f90f}
\newcommand{\rockstarname}{{\sc rockstar}}
\newcommand{\rockstarversion}{0.99.9-RC3+-6d16969}
\newcommand{\ctreesname}{{\sc ctrees}}
\newcommand{\ctreesversion}{1.01-e49cbf0}
\newcommand{\sagename}{{\sc sage}}

\newcommand{\convertctreesname}{{\sc convertctrees}}
\newcommand{\convertctreesversion}{0.0-522dac5}
\newcommand{\revolvername}{{\sc revolver}}

\newcommand{\cosmdistname}{{\sc cosmdist}}
\newcommand{\cosmdistversion}{0.3.8.2}

\newcommand{\OmegaMvalue}{0.3}

\newcommand{\OmegaLvalue}{0.7}

\newcommand{\Hubblevalue}{70.0}

\newcommand{\Ncrootvalue}{128}

\newcommand{\Lboxvalue}{80}

\newcommand{\TimeStartvalue}{10}

\newcommand{\TimeStepvalue}{100}

\newcommand{\ramsesparamLevelmaxname}{{\tt levelmax}}
\newcommand{\ramsesparamLevelmaxvalue}{12}

\newcommand{\rockstarparamFOFLINKINGLENGTHvalue}{0.28}

\newcommand{\rockstarparamMINHALOOUTPUTSIZEvalue}{5}

\newcommand{\rockstarparamMASSDEFINITIONvalue}{"200c"}

\newcommand{\sageparamBaryonFracvalue}{0.20}

\newcommand{\sageparamPartMassvalue}{2.03}

\newcommand{\revolverparamiRadFracElaphrovalue}{2.0}

\newcommand{\NHaloesFinalTimevalue}{5329}

\newcommand{\PlotsSizeiRAvgNonvoidsvalue}{4.21}

\newcommand{\PlotsSizeiErrorRAvgNonvoidsvalue}{0.07}

\newcommand{\PlotsSizeiRAvgVoidsvalue}{3.77}

\newcommand{\PlotsSizeiErrorRAvgVoidsvalue}{0.08}

\newcommand{\PlotsSizeiRAvgNonvoidsTotalvalue}{4.19}

\newcommand{\PlotsSizeiErrorRAvgNonvoidsTotalvalue}{0.09}

\newcommand{\PlotsSizeiRAvgVoidsTotalvalue}{3.37}

\newcommand{\PlotsSizeiErrorRAvgVoidsTotalvalue}{0.08}

\newcommand{\SpiniSpinAvgNonvoidsvalue}{0.0413}

\newcommand{\SpiniErrorSpinAvgNonvoidsvalue}{0.0007}

\newcommand{\SpiniSpinAvgVoidsvalue}{0.0421}

\newcommand{\SpiniErrorSpinAvgVoidsvalue}{0.0008}

\newcommand{\SpiniSpinAvgNonvoidsTotalvalue}{0.0414}

\newcommand{\SpiniErrorSpinAvgNonvoidsTotalvalue}{0.0008}

\newcommand{\SpiniSpinAvgVoidsTotalvalue}{0.0420}

\newcommand{\SpiniErrorSpinAvgVoidsTotalvalue}{0.0009}

\newcommand{\RviriRvirAvgNonvoidsvalue}{150.5}

\newcommand{\RviriErrorRvirAvgNonvoidsvalue}{1.4}

\newcommand{\RviriRvirAvgVoidsvalue}{134.8}

\newcommand{\RviriErrorRvirAvgVoidsvalue}{1.3}

\newcommand{\RviriRvirAvgNonvoidsTotalvalue}{145.2}

\newcommand{\RviriErrorRvirAvgNonvoidsTotalvalue}{2.2}

\newcommand{\RviriRvirAvgVoidsTotalvalue}{120.5}

\newcommand{\RviriErrorRvirAvgVoidsTotalvalue}{1.4}

\newcommand{\PlotsSizeiNGalsTotalvalue}{4817}

\newcommand{\PlotsSizeiNGalsInIntervalvalue}{3848}

\newcommand{\PlotsSizeiNVoidGalsTotalvalue}{1998}

\newcommand{\PlotsSizeiNVoidGalsvalue}{1588}

\newcommand{\ElaphroAcciAccRadMedvalue}{0.09}

\newcommand{\ElaphroAcciAccRadStderrMedvalue}{0.01}

\newcommand{\ElaphroAcciAccTanMedvalue}{1.90}

\newcommand{\ElaphroAcciAccTanStderrMedvalue}{0.04}

\newcommand{\ElaphroAcciCountedVoidsPotvalue}{1570}

\newcommand{\nvoidsiNvoidsvalue}{2198}

\newcommand{\InfallMedianValueAmplitudeVoidsvalue}{0.31}

\newcommand{\InfallMedianValueAmplitudeNonVoidsvalue}{0.29}

\newcommand{\InfallRobustErrorAmplitudeVoidsvalue}{0.02}

\newcommand{\InfallRobustErrorAmplitudeNonVoidsvalue}{0.02}

\newcommand{\InfallMedianValueDecayVoidsvalue}{3.67}

\newcommand{\InfallMedianValueDecayNonVoidsvalue}{3.35}

\newcommand{\InfallRobustErrorDecayVoidsvalue}{25.40}

\newcommand{\InfallRobustErrorDecayNonVoidsvalue}{5.18}

\newcommand{\VoidFitiGoodParametervalue}{1263}

\newcommand{\VoidFitiBadParameterThresholdvalue}{324}

\newcommand{\VoidFitiBadParameterFailedFitvalue}{1}

\newcommand{\VoidFitiBadParameterTotalvalue}{325}

\newcommand{\VoidFitiParameterThresholdInputvalue}{1000}

\newcommand{\NonVoidFitiGoodParametervalue}{1877}

\newcommand{\NonVoidFitiBadParameterThresholdvalue}{524}

\newcommand{\NonVoidFitiBadParameterFailedFitvalue}{1}

\newcommand{\NonVoidFitiBadParameterTotalvalue}{525}

\newcommand{\GalRvirFracSlopevalue}{-100}

\newcommand{\GalRvirFracSigSlopevalue}{9}

\newcommand{\GalRvirFracZerovalue}{201}

\newcommand{\GalRvirFracSigZerovalue}{6}

\newcommand{\GalRvirReldistSlopevalue}{-15.8}

\newcommand{\GalRvirReldistSigSlopevalue}{5.9}

\newcommand{\FracRreffSlopevalue}{-0.041}

\newcommand{\FracRreffSigSlopevalue}{0.011}

\newcommand{\FracRreffZerovalue}{0.667}

\newcommand{\FracRreffSigZerovalue}{0.015}

\newcommand{\GalspinFracSlopevalue}{0.021}

\newcommand{\GalspinFracSigSlopevalue}{0.004}

\newcommand{\GalspinFracZerovalue}{0.028}

\newcommand{\GalspinFracSigZerovalue}{0.003}

\newcommand{\GalspinReldistSlopevalue}{0.007}

\newcommand{\GalspinReldistSigSlopevalue}{0.002}

\newcommand{\InfallAmpFracSlopevalue}{1.24}

\newcommand{\InfallAmpFracSigSlopevalue}{0.51}

\newcommand{\InfallAmpFracZerovalue}{1.25}

\newcommand{\InfallAmpFracSigZerovalue}{0.30}

\newcommand{\InfallTauFracSlopevalue}{-2.94}

\newcommand{\InfallTauFracSigSlopevalue}{1.63}

\newcommand{\InfallTauFracZerovalue}{8.91}

\newcommand{\InfallTauFracSigZerovalue}{1.05}

\newcommand{\GalsizeFracSlopevalue}{-1.44}

\newcommand{\GalsizeFracSigSlopevalue}{0.46}

\newcommand{\GalsizeFracZerovalue}{4.72}

\newcommand{\GalsizeFracSigZerovalue}{0.33}

\newcommand{\InfallAmpReldistSlopevalue}{0.54}

\newcommand{\InfallAmpReldistSigSlopevalue}{0.15}

\newcommand{\InfallAmpReldistZerovalue}{1.47}

\newcommand{\InfallAmpReldistSigZerovalue}{0.21}

\newcommand{\InfallTauReldistSlopevalue}{-0.82}

\newcommand{\InfallTauReldistSigSlopevalue}{1.14}

\newcommand{\InfallTauReldistZerovalue}{7.86}

\newcommand{\InfallTauReldistSigZerovalue}{1.33}

\newcommand{\GalsizeReldistSlopevalue}{0.11}

\newcommand{\GalsizeReldistSigSlopevalue}{0.22}

\newcommand{\GalsizeReldistZerovalue}{3.65}

\newcommand{\GalsizeReldistSigZerovalue}{0.26}

\newcommand{\ElaphcenGalsizeReldistSlopevalue}{0.11}

\newcommand{\ElaphcenGalsizeReldistSigSlopevalue}{0.22}

\newcommand{\ElaphcenGalsizeReldistZerovalue}{3.65}

\newcommand{\ElaphcenGalsizeReldistSigZerovalue}{0.26}

\newcommand{\ElaphcenGalRvirReldistZerovalue}{152}

\newcommand{\ElaphcenGalRvirReldistSigZerovalue}{7}

\newcommand{\ElaphcenGalRvirReldistSlopevalue}{-15.8}

\newcommand{\ElaphcenGalRvirReldistSigSlopevalue}{5.9}

\newcommand{\ElaphcenGalspinReldistSlopevalue}{0.007}

\newcommand{\ElaphcenGalspinReldistSigSlopevalue}{0.002}

\newcommand{\ElaphcenGalspinReldistZerovalue}{0.034}

\newcommand{\ElaphcenGalspinReldistSigZerovalue}{0.003}

\newcommand{\CirccenGalsizeReldistSlopevalue}{0.12}

\newcommand{\CirccenGalsizeReldistSigSlopevalue}{0.28}

\newcommand{\CirccenGalsizeReldistZerovalue}{3.65}

\newcommand{\CirccenGalsizeReldistSigZerovalue}{0.31}

\newcommand{\CirccenGalRvirReldistZerovalue}{152}

\newcommand{\CirccenGalRvirReldistSigZerovalue}{6}

\newcommand{\CirccenGalRvirReldistSlopevalue}{-15.6}

\newcommand{\CirccenGalRvirReldistSigSlopevalue}{5.7}

\newcommand{\CirccenGalspinReldistSlopevalue}{0.007}

\newcommand{\CirccenGalspinReldistSigSlopevalue}{0.002}

\newcommand{\CirccenGalspinReldistZerovalue}{0.034}

\newcommand{\CirccenGalspinReldistSigZerovalue}{0.003}

\newcommand{\GeomcenGalsizeReldistSlopevalue}{0.11}

\newcommand{\GeomcenGalsizeReldistSigSlopevalue}{0.27}

\newcommand{\GeomcenGalsizeReldistZerovalue}{3.66}

\newcommand{\GeomcenGalsizeReldistSigZerovalue}{0.30}

\newcommand{\GeomcenGalRvirReldistZerovalue}{152}

\newcommand{\GeomcenGalRvirReldistSigZerovalue}{6}

\newcommand{\GeomcenGalRvirReldistSlopevalue}{-15.6}

\newcommand{\GeomcenGalRvirReldistSigSlopevalue}{6.1}

\newcommand{\GeomcenGalspinReldistSlopevalue}{0.007}

\newcommand{\GeomcenGalspinReldistSigSlopevalue}{0.002}

\newcommand{\GeomcenGalspinReldistZerovalue}{0.034}

\newcommand{\GeomcenGalspinReldistSigZerovalue}{0.003}

\newcommand{\AccRadSlopevalue}{-0.029}

\newcommand{\AccRadSigSlopevalue}{0.010}

\newcommand{\AccRadZerovalue}{0.191}

\newcommand{\AccRadSigZerovalue}{0.040}

\newcommand{\AccTanSlopevalue}{0.012}

\newcommand{\AccTanSigSlopevalue}{0.031}

\newcommand{\AccTanZerovalue}{1.841}

\newcommand{\AccTanSigZerovalue}{0.122}

\newcommand{\HaloMassVoidsMedvalue}{4.1}

\newcommand{\HaloMassVoidsStderrMedvalue}{0.4}

\newcommand{\HaloMassNonvoidsMedvalue}{5.7}

\newcommand{\HaloMassNonvoidsStderrMedvalue}{2.6}

\newcommand{\FormationTimeVoidsMedvalue}{4.1}

\newcommand{\FormationTimeVoidsStderrMedvalue}{0.1}

\newcommand{\FormationTimeNonvoidsMedvalue}{3.3}

\newcommand{\MassgapLowerLimitvalue}{ 10^{11}}

\newcommand{\MassgapUpperLimitvalue}{ 10^{13}}

\newcommand{\ElaphrocentreHaloOriginalRadiusMpcvalue}{1.20}

\newcommand{\HaloIsInVoidMinFractionvalue}{0.50}

\newcommand{\CentreAdjectivevalue}{elaphrocentric}
 \newcommand{\PlotsSizeiRAvgVoidsTotalvalueCosmStdDev}{0.05}
\newcommand{\RviriRvirAvgVoidsTotalvalueCosmStdDev}{1.73}
\newcommand{\ElaphroAcciAccTanMedvalueCosmStdDev}{0.29}
\newcommand{\InfallAmpFracSlopevalueCosmStdDev}{0.40}
\newcommand{\InfallTauFracSlopevalueCosmStdDev}{1.15}
\newcommand{\InfallAmpReldistSlopevalueCosmStdDev}{0.30}
\newcommand{\GalsizeFracSlopevalueCosmStdDev}{0.42}
\newcommand{\GalspinFracSlopevalueCosmStdDev}{0.005}
\newcommand{\ElaphcenGalspinReldistSlopevalueCosmStdDev}{0.002}

\newcommand{\ElaphcenGalRvirReldistSlopevalueCosmStdDev}{6.7}

\newcommand{\AccRadSlopevalueCosmStdDev}{0.006}
\newcommand{\AccTanSlopevalueCosmStdDev}{0.036}
\newcommand{\FracRreffSlopevalueCosmStdDev}{0.012}
\newcommand{\CosmicVarCommitID}{{\tt e93569c}}
\newcommand{\NVerifySigmas}{5.0}

 \newcommand{\machinearchitecture}{x86\_64}
\newcommand{\machinebyteorder}{Little Endian}

 \togglefalse{postrefereeB}   \togglefalse{removesomefits}   \toggletrue{CompareCentres}

\usepackage{amsmath}

\usepackage{csquotes}

\iftoggle{postreferee}{
  \newcommand\postrefereechanges[1]{{\bf \color{myred} \large #1}}

  \newcommand\postrefereechangestitle[1]{\bf \textcolor{myred}{#1}}
  \usepackage{color}  \definecolor{myred}{rgb}{0.7,0.0,0.2}
}{
  \newcommand\postrefereechanges[1]{#1}
  \newcommand\postrefereechangestitle[1]{#1}
    
}

\iftoggle{postrefereeB}{
  \newcommand\postrefereechangesB[1]{{\bf \color{myred} \large #1}}

  \usepackage{color}  \definecolor{myred}{rgb}{0.7,0.0,0.2}
}{
  \newcommand\postrefereechangesB[1]{#1}

}

\usepackage{graphicx}

\usepackage[font={footnotesize}]{caption}

\usepackage{xcolor}

\usepackage[many]{tcolorbox}

\usepackage{tipa}

\newcommand{\mktab}[1]{\textcolor{black!30!white}{\_\_\_TAB\_\_\_}}

\usepackage{tikz}
\usetikzlibrary{graphs}
\usetikzlibrary{external}
\usetikzlibrary{positioning}
\tikzsetexternalprefix{tikz/}

\usepackage{pgfplots}
\pgfplotsset{compat=newest}
\usepgfplotslibrary{groupplots}
\pgfplotsset{
  axis line style={thick},
  tick style={semithick},
  tick label style = {font=\footnotesize},
  every axis label = {font=\footnotesize},
  legend style = {font=\footnotesize},
  label style = {font=\footnotesize}
  }

\tikzset{node-terminal/.style={
  rectangle,
  very thick,
  draw=blue!50,
  text centered,
  top color=white,
  minimum size=6mm,
  text width=2.1cm,
  rounded corners=3mm,
  bottom color=blue!20,
  font=\ttfamily}}

\tikzset{node-nonterminal/.style={
  rectangle,
  very thick,
  text centered,
  top color=white,
  text width=2.1cm,
  minimum size=6mm,
  draw=green!50!black!50,
  bottom color=green!80!black!50,
  font=\ttfamily}}

\tikzset{node-nonterminal-thin/.style={
  rectangle,
  thick,
  text centered,
  top color=white,
  text width=2cm,
  minimum size=2mm,
  draw=green!50!black!50,
  bottom color=green!80!black!50,
  font=\ttfamily\scriptsize}}

\tikzset{node-makefile/.style={
  thick,
  rectangle,
  anchor=south,
  minimum width=2.6cm,
  minimum height=5cm,
  draw=green!50!black!50,
  fill=black!10!green!12!white,
}}

\tikzset{node-point/.style={
  circle,
  black!50,
  inner sep=0pt,
  minimum size=0pt,
  fill=white}}

\tikzset{ bbox/.style={
  rectangle,
  minimum width=2.5cm,
  rounded corners=2mm,
  very thick,draw=blue!50,
  top color=white,
  bottom color=blue!20 } }

\tikzset{ rbox/.style={
    rectangle,
    dotted,
    minimum width=2.5cm,
    rounded corners=2mm,
    very thick,draw=red!50!black!50,
    top color=white,
    bottom color=red!50!black!20 } }

\tikzset{ gbox/.style={
    rectangle,
    minimum width=2.5cm,
    very thick,
    draw=green!50!black!50,
    top color=white,
    bottom color=green!50!black!20 } }

\tikzset{ dirbox/.style={
    thick,
    rectangle,
    anchor=north,
    text centered,
    font=\ttfamily,
    minimum width=15cm,
    minimum height=7.5cm,
    draw=brown!50!black!50,
    fill=brown!10!white }}

\providecommand\pasa{PASA}
\providecommand\pasa{Proc.~Astr.~Soc.~Austr.}

\providecommand\jcap{JCAP}

\providecommand\PRL{Physical Review Letters}
\providecommand\prd{Physical Review D}

\providecommand\APPS{Acta Physica Polonica Supplement}
\providecommand\ijmpd{International Journal of Modern Physics D}

\providecommand\aap{A\&A}
\providecommand\aj{AJ}
\providecommand\apj{ApJ}
\providecommand\apjs{ApJS}
\providecommand\mnras{MNRAS}
\providecommand\apjl{ApJL}
\providecommand\apjl{Astrophys.J.Lett.}

\providecommand\newjphys{New Journal of Physics}
\providecommand\CiSE{Comp.~in~Sci.~Eng.}

\providecommand\SSS{Sect.~}
\providecommand\diffd{\mathrm{d}}

\newcommand\gtapprox{\,\lower.6ex\hbox{$\buildrel >\over \sim$}\,}
\newcommand\ltapprox{\,\lower.6ex\hbox{$\buildrel <\over \sim$}\,}

\newcommand\hMpc{\mbox{Mpc/$h^{-1}$}}

\newcommand\Ommzero{\Omega_{\mathrm{m0}}}   \newcommand\OmLamzero{\Omega_{\Lambda0}}

\newcommand\haloinvoidfrac{f_{\,\cal{H}\cap \cal{V}}}
\newcommand\haloinvoidminfrac{f_{\,\cal{H}\cap \cal{V}}^{\min}}
\newcommand\relaphro{r}

\newcommand\mvoidgood{m_{\mathrm{v}}}
\newcommand\mvoidbadtot{{m}_{\mathrm{v}}^{-}}
\newcommand\mvoidbadfit{{m}_{\mathrm{vf}}^{-}}
\newcommand\mvoidbadthresh{{m}_{\mathrm{vu}}^{-}}

\newcommand\mnonvoidgood{m_{\mathrm{v}}}
\newcommand\mnonvoidbadtot{{m}_{\mathrm{v}}^{-}}
\newcommand\mnonvoidbadfit{{m}_{\mathrm{vf}}^{-}}
\newcommand\mnonvoidbadthresh{{m}_{\mathrm{vu}}^{-}}

\newcommand\rdiskscale{r_{\mathrm{disk}}}
\newcommand{\InfallAmplitude}{A}
\newcommand{\InfallDecayRate}{\tau}

\newcommand\testAccel{-3.05}
\newcommand\rtest{r^{\mathrm{test}}}
\newcommand\tagTest{^{\mathrm{test}}}

\newcommand\sigmarandom{\sigma_{\mathrm{ran}}}
\newcommand\pmcosvar{\pm_{\mathrm{cv}}}
\newcommand\sigmacosvar{\sigma_{\mathrm{cv}}}

\title[Elaphrocentre and void galaxy formation]{The role of the elaphrocentre in \protect\postrefereechangestitle{void} galaxy formation}

\author[Peper \& Roukema]{Marius Peper,$^{1}$
         Boudewijn F. Roukema$^{1,2}$
         \\
         $^1$ Institute of Astronomy, Faculty of Physics,
           Astronomy and Informatics, Nicolaus Copernicus
           University, Grudziadzka 5, 87-100 Toru\'n, Poland
         \\ $^2$ Univ Lyon, Ens de Lyon, Univ Lyon1, CNRS, Centre de
           Recherche Astrophysique de Lyon UMR5574, F--69007, Lyon,
           France}

\date{Accepted \ldots Received \ldots; in original form \ldots}
\pubyear{2020}

\begin{document}

\maketitle

\begin{abstract}
{\postrefereechanges{Voids may affect galaxy formation via weakening \postrefereechangesB{mass} infall or increasing disk sizes, which could potentially play a role in the formation of giant low surface brightness galaxies (LSBGs).}}
{If a dark matter halo forms at the potential hill corresponding to a void of the cosmic web, which we denote the \enquote*{elaphrocentre} in contrast to a barycentre, then the elaphrocentre should weaken the infall rate to the halo when compared to infall rates towards barycentres.
    We investigate this hypothesis numerically.}
{We present a complete software pipeline to simulate galaxy formation, starting from a power spectrum of initial perturbations and an $N$-body simulation through to merger-history-tree based mass infall histories.
    The pipeline is built from well-established, free-licensed cosmological software packages, and aims at highly portable \postrefereechanges{long-term} reproducibility.}
{We find that the elaphrocentric accelerations tending to oppose mass infall are modest.
    We do not find evidence of location in a void or elaphrocentric position weakening mass infall towards a galaxy.
    However, we find indirect evidence of voids influencing \postrefereechanges{galaxy} formation: while void galaxies are of lower mass \postrefereechangesB{compared to galaxies in high density environments}, their spin parameters are typically higher.
    For a fixed mass, the implied disk scale length would be greater.
    Tangential accelerations in voids are found to be high and might significantly contribute to the higher spin parameters.
    We find significantly later formation epochs for void galaxies; this should give lower matter densities and may imply lower surface densities of disk galaxies.}
{Thus, void galaxies have higher spin parameters and later formation epochs; both are factors that may increase the probability of \postrefereechangesB{forming LSBGs} in voids.}
\end{abstract}

\begin{keywords}
  methods: numerical, galaxies: evolution, cosmology: dark matter
\end{keywords}

\section{Introduction}

\postrefereechangesB{The role of the void environment in galaxy formation is worth exploring.
  A particular case of interest is that of giant low surface brightness galaxies (LSBGs).
  The formation mechanisms of LSBGs} (\citealt*{Sandage84,Bothun87}) with high masses and low star formation rates remain unclear.
\postrefereechanges{\citet*{HoffmanSilkWyse92} presented a peaks-in-peaks structure-formation calculation arguing that voids are likely to play a major role in the formation of giant low surface brightness galaxies.}
Here, we numerically investigate \postrefereechanges{the role of voids in galaxy formation, with a particular emphasis of how voids may provide some of the characteristics leading to LSBG formation,} by developing a reproducible pipeline that combines and builds on existing tools, \postrefereechanges{which have developed considerably over the intervening three decades}.

\postrefereechanges{It has long been known} that a high volume fraction of the Universe consists of large underdense regions \citep*[e.g.][]{GregoryThompson78,deLappGH86}.
More recent measurements have increased our knowledge about these underdense regions, now known as cosmic voids or voids in the cosmic web.
Sloan Digital Sky Survey (SDSS) measurements show that cosmic voids dominate the volume of our Universe, constituting $60$\% of the volume in \citet{Pan2012voids}'s analysis.
Voids vary by about an order of magnitude in size, spanning scales from $10$ to $200$~{\hMpc}, especially as found in various SDSS-related analyses, with higher numbers of smaller voids, and quantitative population estimates sensitive to the methods of defining and detecting voids \citep{HoyleVOg02voids, HoyleVog04voids, Pan2012voids,NadHot2013, SutterLWWWP14, PisaniLSW14stack, Pisani15voidcounts, PisaniSW15vpec, Mao2017}.
Depending on the chosen catalogue and void detection algorithm, a modest fraction of galaxies can be considered as being located in voids.
For example, \citet{Pan2012voids} estimate that about $7$\% of galaxies are located in voids.
Since the fraction of volume occupied by voids may be as high as 80\% \citep[for theoretical estimates, see e.g.][]{Colberg08voidcomparison,Cautun15voidfrac80pc}, they are suspected to play a key role in relation to dark energy \citep[e.g.][]{BCKRW16}.

\postrefereechanges{In the cosmological galaxy formation} context, we denote the gravitational potential hill corresponding to a cosmic void on megaparsec scales as an \enquote*{elaphrocentre} in order to emphasise its gravitational role in opposition to that of a barycentre\footnote{From ancient Greek: \href{https://en.wiktionary.org/wiki/\%E1\%BC\%90\%CE\%BB\%CE\%B1\%CF\%86\%CF\%81\%CF\%8C\%CF\%82\#Ancient_Greek}{$\varepsilon\lambda\alpha\phi\rho o \varsigma$} (light) and \href{https://en.wiktionary.org/wiki/\%CE\%B2\%CE\%B1\%CF\%81\%CF\%8D\%CF\%82\#Ancient_Greek}{$\beta\alpha\rho\upsilon\varsigma$} (heavy), respectively.}.
\postrefereechanges{The impact of voids} on galaxy formation remains largely unknown.
Whereas galaxies in clusters and the walls of the cosmic web typically undergo a gravitationally very active evolution with many mergers, \postrefereechanges{void galaxies tend to have few} mergers.
Galaxy mergers usually lead to bursts of star formation, making galaxies briefly much brighter.
\postrefereechanges{It is typical in} modelling the formation of a galaxy in a gravitational barycentre -- a knot of the cosmic web -- to approximate the surrounding universe via the Newtonian iron-spheres theorem.
Relativistically, the conditions of Birkhoff's theorem, for spherical symmetry and an asymptotically 4-Ricci flat universe \citep{BirkhoffLanger23}, are not satisfied for a flat Friedmann--Lema{\^i}tre--Robertson--Walker (FLRW) model, but nevertheless provide heuristic motivation for applying the iron-spheres theorem.
However, the situation in a void is different: the density contrast is much weaker, and there is a mass deficit below the mean density, rather than a mass excess above the mean density.
Approximating a strong overdensity as being embedded in a surrounding empty universe that is Newtonian is likely to be less \postrefereechanges{inaccurate} than approximating a modest (in terms of linear mean density) underdensity in the same way, since gravity is attractive.
What is \postrefereechanges{effectively} antigravity in voids -- in comparison to the surroundings -- is unlikely to be well modelled by a Newtonian approximation.
Indeed, relativistically, to reach turnaround, an overdensity has to pass through a strongly positive spatial curvature phase \citep{RO19flatness,Ostrowski19,VignBuch19}, after which \postrefereechanges{it virialises} at an overdensity of a few hundred times the mean density \citep[e.g.][]{LaceyCole93MN}.
Since a void tends to have negative spatial curvature (for a flow-orthogonal spacetime foliation), an overdensity inside a void will have difficulty forming.
If it forms nevertheless, the negative spatial curvature environment will tend to weaken \postrefereechanges{matter infall, weakening} the star formation rate.
\postrefereechanges{A void can also be thought of as approximated by a spatially compact domain in a relativistic Milne model -- empty} of matter and spatially hyperbolic (often called ``open'') -- in which structure forms more slowly than in the idealised background FLRW model.

\postrefereechanges{A pseudo-Newtonian way} of thinking about this is that compared to a background FLRW model, a galaxy in the emptiest parts of a void -- the \enquote*{elaphrocentre} -- feels a weak antigravitational environmental force around it, since the void is underdense compared to idealised average regions of space.
Another toy model way of thinking about the elaphrocentric effect is as follows.
Let the core of a dark matter halo be modelled as forming via linear theory followed by the standard pseudo-Newtonian spherical collapse approximation in a background FLRW model, and then by slow uniform mass infall induced by the elaphrocentric environment.
This would appear as a history of mass infall that is spread out in time and gradual, rather than fast and sudden.
If slow gradual infall is interpreted as the late collapse of the outer parts of the halo, then this corresponds to collapse at late epochs, when the FLRW critical density has dropped, implying a greater virial radius for a fixed total halo mass.

\postrefereechanges{Based on these heuristic arguments we assume that} if a dark matter halo forms at or near an elaphrocentre, then the merger rate of small haloes into that halo and the overall infall rate of dark matter and gas into the halo should be weaker than the usual infall rates towards a halo of similar mass at a barycentre.
This effect should tend to create lower mass dark matter haloes at elaphrocentres compared to barycentres, which is generally the case: the most massive haloes form at the knots of the cosmic web.
For a high-mass dark matter halo at the present epoch of a fixed mass, this elaphocentric effect should tend to increase the probability of the halo having grown in mass after its initial collapse by weak infall and a weak merger rate over a long period, rather than by an initial short burst of mass accumulation.
\postrefereechanges{These effects} on the host haloes could lead, for a fixed mass \postrefereechangesB{and a fixed efficiency factor for converting available baryons into stars,} to galaxies closer to the elaphrocentre preferentially forming with weak star formation rates over long time \postrefereechanges{scales}.

\postrefereechangesB{While \citet{HoffmanSilkWyse92} proposed that voids play a major role in LSBG formation, here we briefly present a broader overview of LSBG formation scenarios.}
\postrefereechanges{LSBGs were discovered more than three decades ago \citep{Sandage84,Bothun87}, leading to the question of their formation mechanisms \citep*{Schombert2011LSBG,
  Schombert2013LSBG,
  SchombertMcGaugh2014a,
  SchombertMcGaugh2014b}.
Interest in LSBGs was recently reignited by \cite{vanDokkum15}, who found a high abundance of large LSBGs in the Coma cluster, that they called ultra diffuse galaxies (UDGs), to distinguish them from traditional LSBGs.
It is not yet clear whether LSBGs and UDGs share the same formation scenario, especially since this is dependent on their definitions.
A common scenario is that UDGs}  form in \postrefereechanges{high-spin haloes}.
\citet{Rong2017} find from simulations that UDGs form naturally in high-spin haloes within the $\Lambda$CDM model; \citet{JiHoon15} has started investigating this observationally.
\citet{Chan2018} show that they can reproduce the observed quantities of red UDGs by imposing quenching, without assuming high spin haloes.
\citet{DiCintio2017} showed that UDGs can be produced in isolated dwarf galaxy haloes with stellar feedback and episodes of gas outflow.
\citet{Jiang2019} extended this work by investigating field UDGs.
Compared to other galaxies, LSBGs tend to be more \postrefereechanges{spatially isolated, i.e. they tend to be somewhat} elaphrocentric.
\citet{Rosenbaum09} quantified this on a 2--5 Mpc scale, finding that LSBGs are located in regions with lower galaxy number densities than those in which high surface brightness galaxies are located.
\postrefereechanges{LSBGs} typically have low \postrefereechangesB{\ion{H}{I}} surface densities, below around $5 M_{\odot}\mathrm{pc}^{-2}$, yielding weak star formation, with star formation rates that are approximately constant with cosmological time, rather than the exponentially declining star formation rates typically associated with high surface brightness galaxies \postrefereechanges{\citep[][\SSS{}7.1, 7.2]{DiPaoloSalucci20}}.

Here, \postrefereechanges{we} focus on the degree to which the elaphrocentric location \postrefereechanges{may contribute to low surface brightness of void galaxies} for a given host halo mass, via (i) a total-matter infall rate closer to being constant rather than being exponentially declining, and (ii) an enlarged disk size of a galaxy due to high spin \citep{Rong2017} and/or an increase in the typical virial radius.
\postrefereechangesB{(iii)} \postrefereechanges{We} estimate the magnitude of elaphrocentric acceleration as a basis for more \postrefereechanges{detailed} studies.

To study this hypothesis, we present a highly reproducible \postrefereechanges{\citep[][]{Akhlaghi2020maneage}} galaxy formation simulation and analysis pipeline \citep*{PeperRoukemaBolejko2019}.
\postrefereechanges{The packages} in the pipeline are free-licensed packages, and should only require a POSIX-compatible operating system with sufficient memory and disk space for reproducing the full calculations, tables and figures, generating values that are statistically equivalent to those published here.
\postrefereechanges{We present} the software packages and the pipeline in \SSS\ref{s-method-pipeline}.

In \SSS\ref{s-method-elaphro} we \postrefereechanges{propose a Voronoi-cell based definition of the \enquote*{elaphrocentre} and discuss alternative definitions of the \enquote*{centre} of a void from the literature.}
We present two different parameters characterising void membership and propose a criterion for use in global population comparisons between void and non-void galaxies in \SSS\ref{e-method-void-membership}.
\postrefereechanges{We describe} how we study the dependence of the infall of matter into galaxies (\SSS\ref{s-method-infall}) and the dependence of galaxy sizes (\SSS\ref{s-method-size}) on these parameters and on the global void membership criterion.
The elaphrocentres themselves are the places that are the most difficult to study via particle distributions, so in \SSS\ref{s-method-elaphro-acc} we describe how we investigate accelerations near the elaphrocentres in preparation for future studies of elaphrocentric effects that might help form LSBGs.

We present our results in \SSS\ref{s-results}, \postrefereechanges{discussion} in \SSS\ref{s-discuss} and conclude in \SSS\ref{s-conclu}.
\postrefereechanges{The} reproducibility package for this paper is available at {\projectzenodohref} and in live\footnote{\projectgitrepository} and archived\footnote{\projectgitrepositoryarchived} {\sc git} repositories.
The commit hash of the version of the source package used to produce this paper is \projectversion.
\postrefereechanges{The package was configured, compiled and run on a {\machinebyteorder} {\machinearchitecture} architecture.}

\section{Method}

\subsection{Software pipeline} \label{s-method-pipeline}

We provide a highly reproducible software pipeline for generating a realisation of galaxies with merger-history-tree based galaxy disk formation histories (and star formation histories, though we do not analyse these in this work) starting from early universe initial conditions.
\postrefereechanges{This approach not only combines existing community tools, but can also help in improving those existing tools by embedding them in a controlled software environment.
Our pipeline is intended to be modular, so that the well established cosmological software tools currently chosen, can, in principle,} be replaced in a modular way, provided that the user manages the input and output formats correctly.
\postrefereechanges{Our results are intended to be statistically reproducible.
  Parallelisation in several steps of the computational pipeline currently prevents byte-for-byte reproducibility.}

Dark matter haloes are identified in an $N$-body simulation, a merger history tree is created, semi-analytical recipes are used to generate galaxy disks with mass infall histories, and voids in the dark matter distribution are detected in the $N$-body simulation.
In the following we give a brief description of the codes and the key parameters used in these models.
URLs and SHA512 checksums for the upstream versions of software used to produce this paper are listed in the reproducibility package.
\postrefereechanges{(An SHA512 checksum is a 512-bit integer computed from the bytes in a file using the SHA512 algorithm, aiming to provide a data integrity check on the file contents that is sensitive to small changes in the file.)}

\postrefereechanges{The reproducibility} structure is based on the Maneage template that aims for a high level of reproducibility \citep{Akhlaghi2020maneage}.
We follow \citet{RougierHinsen2017Repdefn} for the definitions of the \enquote*{reproducibility} of a research paper -- in which independent authors attempt to use the same input data and the same source code and analysis pipeline to obtain the paper's claimed results -- versus the paper's \enquote*{replicability} -- in which independent authors attempt to use different but similar input data and/or a different but equivalent analysis to obtain the claimed results.
Using these definitions, we believe that it should be straightforward for the reader to verify the reproducibility of our results.
We expect that our results will also be replicable.
Version identities of the software packages in the text below include {\sc git} commit hashes.
Modifications that we have made to upstream versions of codes are included as patch files in the reproducibility source package ({\projectzenodohref}).

\subsubsection{Initial conditions} \label{s-meth-initial-cond}
We use {\mpgraficname}-{\mpgraficversion} \citep{PrunetPichon08mpgrafic} to generate a set of standard initial conditions for a flat-space $N$-body simulation with the standard 3-torus topology (often called \enquote*{periodic boundary conditions}).
The {\mpgraficname} package is a well-tested, parallelised package that generates peculiar velocity offsets against an FLRW background model for a standard cosmological power spectrum with Gaussian random fluctuations using the Zel'dovich approximation \citep{ZA,ZA70b}.
Checks are made that the amplitude of the numerically generated power spectrum matches that of the input power spectrum.
We generate a simulation with $N={\Ncrootvalue}^3$ particles.
The comoving fundamental domain size, often called the \enquote*{box size}, is $L_{\mathrm{box}}={\Lboxvalue}$~Mpc/$h$.
These parameters give a dark matter particle mass of ${\sageparamPartMassvalue} \times 10^{10} ~ M_{\odot}$, which is a reasonable mass resolution for modest RAM and CPU resources.
\postrefereechanges{Together with the minimum number of particles per halo (\SSS\ref{s-meth-halo-det}), this sets a minimum halo mass and indirectly a minimum galaxy mass in the simulation.
We do not expect to detect dwarf galaxies in the results presented here.}
We use the $\Lambda$CDM model as a proxy model that fits many observations.
The FLRW cosmological parameter settings include the current values of the matter density parameter $\Ommzero={\OmegaMvalue}$, the dark energy parameter $\OmLamzero={\OmegaLvalue}$ and the Hubble--Lema\^{\i}tre constant $H_0={\Hubblevalue}$~km/s/Mpc.

\subsubsection{Simulations} \label{s-meth-simulations}
For our $N$-body simulation, we chose \ramsesscalavname-{\ramsesscalavversion}, a fork of the widely used adaptive mesh code {\sc ramses}-3.0 \citep{Teyssier02}.
The {\ramsesscalavname} fork has modifications to comply with the MPI~3.0 recommended standards for inclusion of the MPI header file\footnote{\url{https://mpi-forum.org/docs/mpi-3.0/mpi30-report.pdf}} and optional extensions related to scalar averaging \citep{Roukema17silvir}.
The adaptive mesh structure of {\sc ramses} is designed to allow fast calculations that can resolve detailed gravitational behaviour in high density regions.
The maximum {\sc ramses} resolution, which effectively corresponds to a softening length, is set at {\ramsesparamLevelmaxname} $= \ramsesparamLevelmaxvalue$.
We produce snapshots starting at $t_{\mathrm{i}}={\TimeStartvalue}$~Myr with an equally spaced time step of $\Delta t={\TimeStepvalue}$~Myr, and convert to scale factor values using {\cosmdistname}-{\cosmdistversion}.
Newtonian gravitational potentials against the FLRW background are calculated in {\sc ramses} using the Poisson equation and a cloud-in-cell algorithm to estimate the matter density.
These potentials are used to decide how to accelerate and shift particles.
We output these values for later calculation of accelerations in relation to the elaphrocentre (\SSS\ref{s-method-elaphro}).

\subsubsection{Halo detection} \label{s-meth-halo-det}

For detecting dark matter haloes, we use \rockstarname-{\rockstarversion} \citep{Behroozi13rockstar}, which uses a $7$-dimensional friends-of-friends (FoF) algorithm.
In contrast to other FoF halo finders, {\rockstarname} not only uses particles' spatial locations and peculiar velocity information to decide on physically meaningful dark matter haloes, but also uses temporal information to identify groups of particles that are persistent in time, rather than only using the instantaneous characteristics of a virialised object.
\citet{Knebe2011} found that {\rockstarname} performs excellently in recovering haloes from an $N$-body simulation.
We run {\rockstarname} using a linking length of $\rockstarparamFOFLINKINGLENGTHvalue$ and a minimum of $\rockstarparamMINHALOOUTPUTSIZEvalue$ particles per halo.
We set the virial radius criterion for {\rockstarname} detection to {\tt \rockstarparamMASSDEFINITIONvalue}, i.e., $200$ times the critical density.

\subsubsection{Merger history trees}
We construct halo merger trees from the simulations \citep[][and references thereof]{1993ASPC...51..298R,Roukema1993PhD,RPQR97,1993ApJ...418L...1R}.
In this paper, instead of using the original Fortran77 routines from 1992, we use a more modern package, \ctreesname-{\ctreesversion} \citep{Behroozi13MHT}, which was designed to perform on outputs from {\rockstarname}.
In order to use these merger history trees for simulating galaxy evolution using {\sagename}, which was developed for following up simulations such as the Millenium simulation, we need to convert the trees to the {\sc LHaloTree} format.
We do the conversion with {\convertctreesname}-{\convertctreesversion}.

\subsubsection{\postrefereechanges{Galaxy formation and matter infall}} \label{s-meth-galaxies}

To form galaxies within our dark matter haloes, we use semi-analytical galaxy formation recipes \citep{1993ASPC...51..298R,KWG93,RPQR97,KauffCDW99}.
Again, rather than using the original code from 1992, we use {\sagename} \citep{Croton16SAGE}.
For an introductory review, see \citet{Baugh06review}.
We use the built-in functions of {\sagename} for estimating the \enquote*{size} and the infall rate history of each galaxy at the final output time.
All galaxies are assumed to form as disk galaxies in {\sagename}, with a disk radius of
\begin{equation}
  \rdiskscale = \frac{\lambda}{\sqrt{2}} R_{\mathrm{vir}},
\label{sage-disk-radius}
\end{equation}
where $R_{\mathrm{vir}}$ is the virial radius (set in {\rockstarname} at {\tt \rockstarparamMASSDEFINITIONvalue}) and $\lambda$ is the dimensionless spin parameter computed using {\rockstarname}'s halo properties.
The introduction of the parameter $\lambda$ is generally attributed to \citet[][eqs~(35), (37)]{Peebles69spin}.
The parameter is now widely used \citep[e.g.,][eq.~(10)]{MoShudeWhite98spinparam}, and is often normalised by a $\sqrt{2}$ factor \citep[][eq.~(5)]{Bullock01spin}, which is the convention chosen in {\sagename}, and adopted in this work:
\begin{equation}
  \lambda := 2^{-1/2}\,J\,|E|^{1/2}\,G^{-1}\,M^{-5/2} \,,
  \label{e-defn-spin}
\end{equation}
where $J$ is the total angular momentum \postrefereechangesB{(calculated as the vector sum of individual particles' angular momenta by functions including {\sc add\_ang\_mom} in {\sc properties.c} in {\rockstarname})}, $E$ is the total energy in the non-relativistic sense \postrefereechangesB{(calculated as the sum of individual particles' Newtonian kinetic and potential energies by functions including {\sc estimate\_total\_energy} in {\sc properties.c} in {\rockstarname})}, $G$ is the Newtonian gravitational constant, and $M$ is the halo mass.

The infalling mass is defined at a given time step by
\begin{equation}
  \Delta M_{\mathrm{infall}}(t_i) := f_{\mathrm{reion}} \, f_{\mathrm{b}} \, M_{\mathrm{vir}} - M_{\mathrm{tot}},
\end{equation}
where $f_{\mathrm{reion}}$ is the reionisation factor, which estimates the effect of ionisation of the intergalactic medium from early stars;
the baryon fraction $f_{\mathrm{b}} = \sageparamBaryonFracvalue$ determines what fraction of total matter is baryonic (assumed to be the same for any dark matter halo/galaxy pair); $M_{\mathrm{vir}}$ is the virial (total matter) mass of the halo;
and $M_{\mathrm{tot}}$ is the sum of all reservoirs of baryonic matter from the previous timestep (except at the initial timestep, when it is zero).
Cases where $M_{\mathrm{infall}} < 0$ are interpreted to mean that baryonic mass is ejected from the galaxy.
We extended {\sagename} in order to estimate infall rate, star formation rate (SFR) and outflow rate histories.
The history of any of these parameters for a given galaxy at a given time, traced backwards in cosmological time, is assumed to be the sum of the histories of all the separate pre-merger progenitors of the galaxy, appropriately matched by cosmological time.
At each merger event, this physically corresponds to the components (dark matter, hot gas, cold gas, stars) of the progenitors being conserved in the merger.
The summed star formation rate traced back for a given galaxy is what was originally used together with evolutionary stellar population synthesis to calculate galaxy spectral energy distributions and absolute magnitudes \citep{1993ASPC...51..298R,RPQR97}; we do not carry out evolutionary stellar population synthesis in this work.
To evaluate these sums, we identify all galaxies present at the present epoch, $a(t)=1$, and trace their progenitors' history back in time along the merger tree.
For example, if a progenitor is itself the result of a merger at an earlier timestep, then its own history is the sum of its own progenitors.
This procedure is continued recursively back in time in the merger tree for a given present-epoch galaxy.
As in the original implementation \citep{1993ASPC...51..298R}, other effects from mergers than conservation of mass, such as merger-induced starbursts, are also assumed in {\sagename}, but with a power law dependence on the satellite mass rather than direct proportionality \citep[][eq.~(27)]{Croton16SAGE}.

\subsubsection{Voids} \label{s-method-voids}
We identify the void environment of galaxies using the {\revolvername} watershed void finder based on {\sc zobov} \citep{Neyrinck2008Zobov,Nadathur19BOSSRevolver}, which provides a nearly parameter-free void finder that does not require assumptions about void shapes.
In contrast to other works, we use the full dark matter (DM) particle distribution as tracers.
\citet{NadathurHotch15voidsII} showed that the voids identified using the galaxy distribution as tracers differ from those traced using a randomly subsampled particle distribution.
The authors recommend using the galaxy distribution as tracers in order to match observations.
However, since our priority here is gravitational effects, we use the DM particle distribution rather than the galaxy distribution.
It is likely that we will detect voids that are smaller and more numerous than those observed in the galaxy distribution \citep[e.g.][]{Mao2017}, since the full DM particle distribution will show positive fluctuations in the density field that may be too weak to form DM haloes and galaxies, but will be detected by the watershed void finder and interpreted as boundaries of voids.

We introduce several small changes into {\revolvername}.
In addition to \citet{NadathurHotch15voidsI}'s definition of the circumcentre, we calculate the position of the elaphrocentre, as defined below in \SSS\ref{s-method-elaphro}.
We add a routine to read in simulation data in {\sc Gadget-2} format \citep{Springel05GADGET2}, the default output format that we chose for the {\sc ramses} $N$-body simulation.
We output lists of particle identities of the DM particles that constitute each void.
This information is needed to decide the extent to which a galaxy's host halo is located in a void.

\postrefereechangesB{We adopt the {\revolvername} definition of the effective radius of a void, $R_{\mathrm{eff}}$, which is not directly related to the choice of a definition of the void centre.
  The effective radius $R_{\mathrm{eff}}$ is defined by {\revolvername} as the radius of a hypothetical sphere that has the same volume as the total volume of all the Voronoi cells that constitute the void, i.e. $R_{\mathrm{eff}} := \frac{3}{4 \pi} \left( \sum_i V_i \right)^{1/3}$ where $V_i$ are the volumes of the Voronoi cells that determine the void.}

\subsection{Elaphrocentre and other definitions of void centres} \label{s-method-elaphro}
To investigate if a galaxy's position in a void -- its elaphrocentric location -- has a significant effect on \postrefereechanges{the formation and evolution of the galaxy}, we first need to clarify earlier terminology regarding void centres from the literature, and we need to define the elaphrocentre.

\citet[][\SSS{}2.3]{NadathurHotch15voidsI} define the \enquote{\em{circumcentre}} for a given void using the Voronoi cell with the lowest density and the three lowest density adjacent Voronoi cells.
The intersection of these four Voronoi cells determines the circumcentre.
By construction, the circumcentre is the centre of the largest sphere that can be inscribed in the tetrahedron determined by the particles in these four (neighbouring) Voronoi cells, and the centre of the largest empty sphere that can be inscribed in the void.
\citet[][\SSS{}2.3 (ii)]{Nadathur2017VoidCentre} rename this the \enquote{\em{void centre}} and show that it correlates strongly with the local maxima of the gravitational potential with respect to the background FLRW model.

We define the {\em elaphrocentre} similarly.
In a void identified by the watershed algorithm, we identify the particle at which the potential is highest, \postrefereechanges{using the potentials estimated by {\sc ramses} (\SSS\ref{s-meth-simulations}).}
  We then identify the three adjacent Voronoi cells \postrefereechanges{whose particles have the highest potentials.}
  \postrefereechanges{Together with the cell of the highest potential particle, we} again form a tetrahedron between the four particles that respectively define the four cells.
The centre of this tetrahedron is the elaphrocentre.
While we expect a strong spatial correlation between elaphrocentres and circumcentres, they will differ in general, in particular for small, highly non-spherical voids.
By definition, elaphrocentres are appropriate for studying elaphrocentric effects on galaxy formation.
A group of test particles at an elaphrocentre will, in the Newtonian sense, be accelerated away from the elaphrocentre, and disperse rather than cluster together.
Thus, the elaphrocentre would seem to be a good environment to form a large, diffuse galaxy, provided that the mass that forms the future galaxy is low compared to the mass deficit determining the gravitational properties of the void as a whole.

A third centre commonly defined in void studies is the \enquote*{macrocentre} or \enquote*{volume-weighted barycentre}.
This is defined (\citealt{Sutter15VIDE}, \SSS{}3, eq.~(4); \citealt{NadathurHotch15voidsI}, \SSS{}2.3, eq.~(2)) as the volume-weighted mean of the position vectors $\vec{x}_i$ of all DM particles identified as being in the void, i.e., $ \vec{c}_{\mathrm{vwb}} := \left({\sum_i {{V_i}\vec{x}_i}}\right)/{\sum_i V_i}$, where $V_i$ is the Voronoi cell volume associated with the $i$-th particle.
Since the position is not mass-weighted, it is unrelated to the normal Newtonian definition of a barycentre for a particle distribution, $ \vec{c}_{\mathrm{m}} := \left({\sum_i {{m_i}\vec{x}_i}}\right)/{\sum_i m_i}$, where $m_i$ is the mass of the $i$-th particle.
In the continuous limit to arbitrarily high particle resolution (assuming a continuous fluid), the volume-weighted barycentre approaches $\vec{c}_{\mathrm{vwb}} = {\int_{\cal D} \vec{x} \,\diffd V}/{\int_{\cal D} \diffd V}$ over a spatial domain ${\cal D}$.
Clearly, in the continuous limit, $\vec{c}_{\mathrm{vwb}}$ is the geometrical centroid of the domain $\cal D$, which in geometry is often termed the \enquote*{barycentre}.
In general, this corresponds to the astronomical barycentre only if the density distribution in the domain $\cal D$ is uniform.
In other words, the volume-weighted barycentre contains no information about the density distribution within $\cal D$ apart from numerical noise.

By definition, apart from discretisation and numerical effects, $\vec{c}_{\mathrm{vwb}}$ only determines the geometrical mean position (the centroid) of the overall shape of the void, as defined by the outermost particles of the void.
Adding a few particles with big Voronoi cells adjacent to a single side of a void would significantly modify the global shape (union of Voronoi cells) of the void.
This would shift the volume-weighted barycentre -- the centroid -- significantly.
Thus, \citet{NadathurHotch15voidsI} are correct that $\vec{c}_{\mathrm{vwb}}$ depends on the presence rather than the absence of tracers.
However, the fundamental problem with using $\vec{c}_{\mathrm{vwb}}$ in the context of cosmological voids is that it indicates a void centre that (apart from numerical effects) has no dependence on the density variations in the interior of the polyhedron (union of all Voronoi cells) that defines the void; only the void boundary affects $\vec{c}_{\mathrm{vwb}}$.
Since there is no reason for the centroid of the polyhedron bounding a void to have any tight relation with the position of minimal density if the void is even mildly asymmetrical, it is unsurprising that \citet{NadathurHotch15voidsI} found the density at $\vec{c}_{\mathrm{vwb}}$ to be higher than at the circumcentre.
Thus, in the context of cosmological voids, the physical relevance of $\vec{c}_{\mathrm{vwb}}$ is unclear, and if used, we recommend that it be described by the term \enquote*{geometrical centroid} (or \enquote*{boundary centroid}) rather than \enquote*{macrocentre} or \enquote*{volume-weighted barycentre}.

\subsection{Analysis} \label{s-method-analysis}

To study our hypothesis that the elaphrocentric location of galaxies in voids plays a significant role in their evolution, we analyse the simulated haloes, galaxies and voids produced by our pipeline in relation to the voids' elaphrocentres as follows.

\subsubsection{Void membership and elaphrocentric distance} \label{e-method-void-membership}
Identifying which galaxies are located in a void is non-trivial.
For a given galaxy, we could find the void that gives the shortest elaphrocentric or circumcentric distance, and consider the galaxy to be a member of the void if the circumcentric or elaphrocentric distance is below a given fraction of the void effective radius.
A distance between two positions in this work is calculated using the shortest of the multiple 3-torus ($\mathbb{R}^3/\mathbb{Z}\times \mathbb{Z} \times \mathbb{Z} \equiv \mathbb{S}^1 \times \mathbb{S}^1 \times \mathbb{S}^1$) spatial geodesic comoving distances.
This is often described more loosely as \enquote*{the comoving distance with periodic boundary conditions}.

However, identifying galaxy membership in a void by the elaphrocentric or circumcentric distance would only be accurate for voids that are spherically symmetric.
Although voids tend to evolve to become more spherical, as shown analytically by \citet{Icke84} by reverting a simple toy model for collapsing density perturbations and numerically by \citet{Sheth04} in $N$-body simulations, a void will in general be non-spherical.
Moreover, the elaphrocentre will not, in general, coincide exactly with the circumcentre.
Thus, a more accurate way of deciding on void membership should, in principle, be possible by using knowledge of the particle positions.

The void membership criterion proposed here, as with the \mbox{$>50$\%} merging identity criterion initially published in 1993 for merger history trees (\citealt{1993ApJ...418L...1R}, \SSS3; \citealt{RPQR97}, \SSS2.2.1), is a simple proposal that we expect to be improved upon later.
Our voids are detected as a union of Voronoi cells -- each containing a DM particle -- by {\revolvername}.
Thus, for any given void, we have a list of particles that approximately define the void.
Since we know from {\rockstarname} which particles are members of a halo in which a given galaxy forms, we can check which of these halo particles are present in the list of void member particles for any given void.
We restrict the list of void particles to those that are within $\relaphro \leq {\revolverparamiRadFracElaphrovalue}\, R_{\mathrm{eff}}$ from the elaphrocentre, where $R_{\mathrm{eff}}$ is the void effective radius calculated by {\revolvername} \postrefereechangesB{(the radius of the hypothetical sphere having the same volume as the void's Voronoi cells; see \SSS\ref{s-method-voids}).}
The $\relaphro \leq {\revolverparamiRadFracElaphrovalue}\, R_{\mathrm{eff}}$ restriction should remove some of the sharpest regions adjacent to the knots of the cosmic web and exclude the outermost regions of voids of high ellipticity.
In this sense, it will counteract the space-filling nature inherent to any watershed voidfinder to some degree.

Our void membership criterion is that we require that a fraction $\haloinvoidfrac$ strictly greater than $\haloinvoidminfrac = {\HaloIsInVoidMinFractionvalue}$ of the particles in a halo $\cal H$ be members of a void $\cal V$ for a galaxy in that halo to be considered a member of the void.
As in the case of the \mbox{$>50\%$} merging identity criterion, which prevents a halo from having multiple descendants, this void membership criterion is strong enough to prevent a galaxy in a halo that lies on a wall, filament or knot from being allocated to more than one void.
This criterion could be strengthened to force a selection of galaxies that are placed further in the interior of the voids, at the cost of reducing the total number of galaxies recognised as being members of voids.
A galaxy that does not satisfy this criterion is considered to be a non-void galaxy.

\subsubsection{Infall dependence on environment} \label{s-method-infall}
We wish to see if infall rates -- of dark and baryonic matter in general -- are affected by the host halo's location in a void.
The infall history should affect the star formation rate, which requires baryonic matter to first collapse into the centre of its host dark matter halo's potential well.
We consider the infall rate traced backwards in time for any halo at the final output time.
This infall rate is the sum of the mass accumulation histories of the component haloes of the final halo's merger tree.
We can write this backtraced history, over predecessor haloes destined to merge together, as the mass evolution assigned to the final halo, $M(t)$, so that $\diffd M/\diffd t \approx \Delta M/\Delta t\, (t)$ is the infall rate -- of small haloes and diffuse matter together.

The hypothesis that the elaphrocentre of a void (corresponding to a spatially compact part of a hyperbolic, super-Friedmannian expanding region) would weaken the infall rate can now be formalised.
For a given mass $M$, the average (mean) infall rate is, by definition, independent of location.
To distinguish the archetypal case of a typical disk galaxy, with an initial burst of star formation followed by an exponential decay, from that of an archetypal LSBG, with an approximately flat star formation rate, we attempt to fit $\diffd M/\diffd t$ by an exponential, of the form
\begin{equation}
  \frac{\diffd M}{\diffd t} (t) = {\InfallAmplitude} \exp({-t/\InfallDecayRate})\,,
  \label{e-infall-model}
\end{equation}
where ${\InfallAmplitude}$ is the infall amplitude and ${\InfallDecayRate}$ is a decay time scale.

Typical disk galaxies should have low time scales ${\InfallDecayRate}$, while \postrefereechanges{we hypothesise that void galaxies} should \postrefereechanges{on average have higher} ${\InfallDecayRate}$, corresponding to approximately flat infall rates.

\postrefereechanges{The exponential form of the fit for the matter infall rate is a heuristic choice inspired by the traditionally used declining exponential for modelling the star formation rate \citep{Bruzual1983}.}
In reality, merger histories are more complex than these simplified extremes, so we expect a fair fraction of automated fits to fail, especially since we apply this to all timesteps defined in \SSS\ref{s-method-pipeline}, starting from the first time step in which a galaxy is modelled to form in a dark matter halo.
\postrefereechanges{Nevertheless, this automated fit procedure should be able to distinguish whether the infall of matter is closer to a brief, quickly weakening series of early events or rather a more steady infall extended over a long period.
  We expect that a flat infall rate would correspond to a roughly constant star formation rate over time.
  The early, brief, burst scenario of the infall of matter would allow a high star formation rate immediately after the infall, whereas a constant rate of matter infall should yield an approximately constant star formation rate.
  Long cooling times would modify this relation, especially for galaxies in high-mass haloes in voids \citep[][Einstein--de~Sitter case]{HoffmanSilkWyse92}.}

We first compare ${\InfallAmplitude}$ and ${\InfallDecayRate}$ for galaxies depending on their classification as void or non-void galaxies.
The infall histories are calculated using our modification of {\sagename}.
The prediction for creating LSBGs is that void galaxies should tend to have low amplitude ${\InfallAmplitude}$ and a high (slow) decay rate ${\InfallDecayRate}$, and vice versa for non-void galaxies.

\begin{figure}
  \includegraphics[width=0.9\columnwidth]{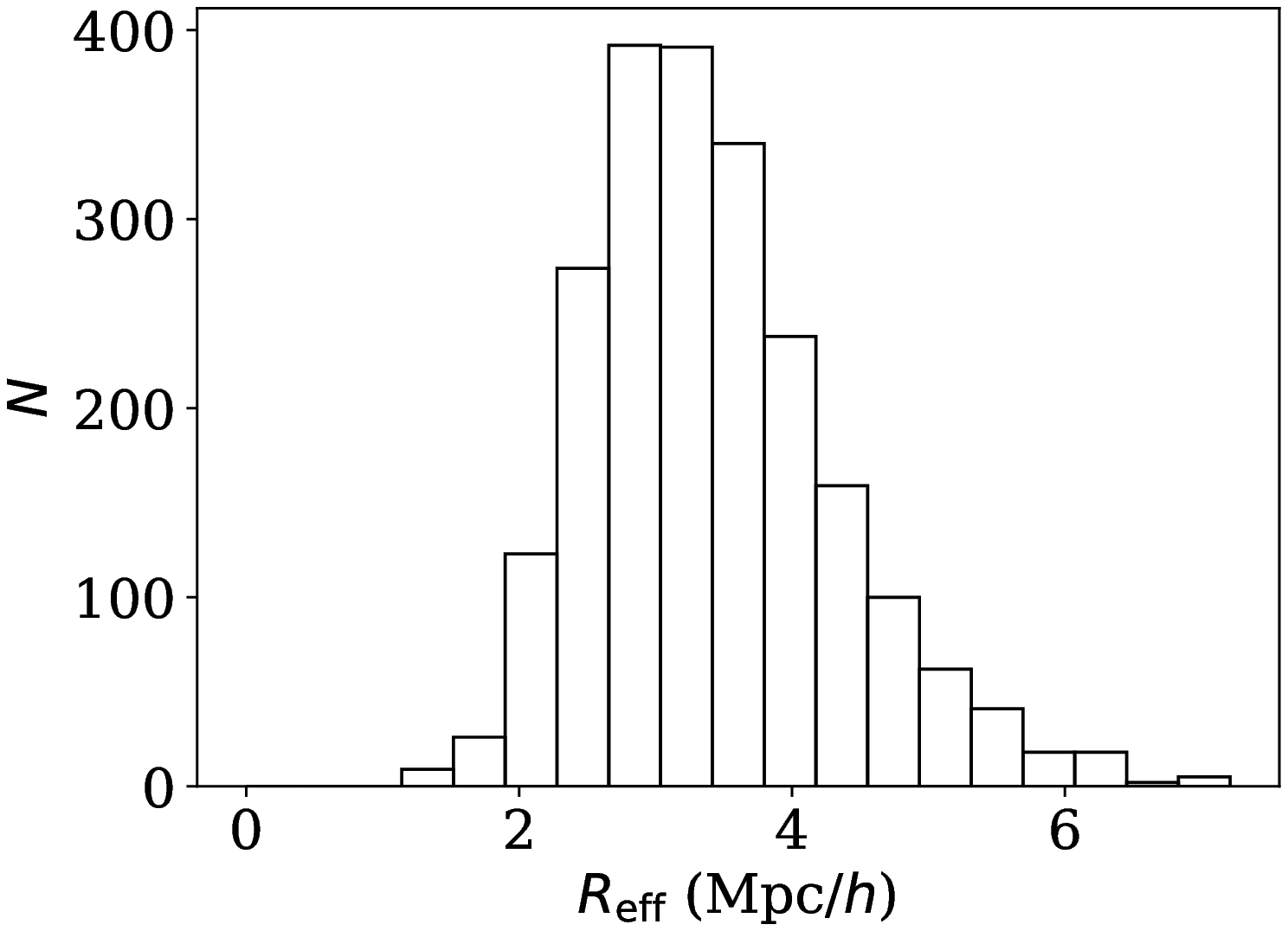}
  \caption{Histogram of void effective radii $R_{\mathrm{eff}}$ \postrefereechangesB{(\protect\SSS\ref{s-method-voids})}, using the full dark matter particle distribution, which leads to voids much smaller than typically observed or found in simulations when traced by galaxies.
\label{void_population}}
\end{figure}

\begin{figure}
\includegraphics[width=0.9\columnwidth]{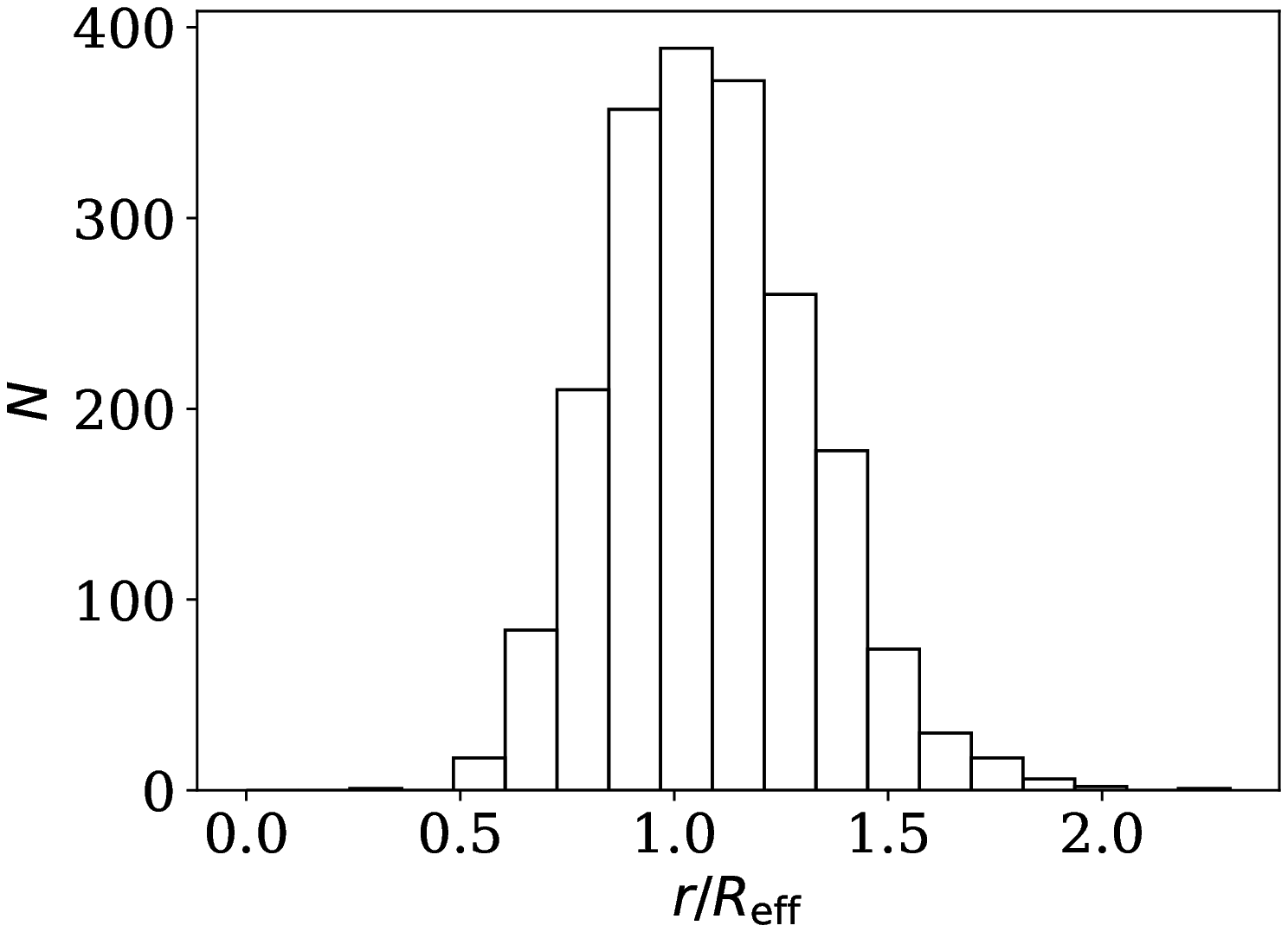}
\caption{Histogram of {\CentreAdjectivevalue} distance of galaxies identified as being located in voids.
  The distribution at $r/R_{\mathrm{eff}} < 1$ is typical of observational and simulated void profiles, with most galaxies located near the effective radius.
  The distribution at $r/R_{\mathrm{eff}} \ge 1$ can be interpreted as showing galaxies in the outer parts of voids that are generally quite asymmetrical.
  \postrefereechangesB{A perfectly symmetrical void would have all its member galaxies at $r/R_{\mathrm{eff}} < 1$.}
\label{f-results-galaxy-hist}}
\end{figure}

\subsubsection{Galaxy size dependence on environment} \label{s-method-size}

The other parameter that may indicate an elaphrocentric contribution to a disk galaxy becoming an LSBG is the disk scale length $\rdiskscale$ (for a density profile $\rho \propto \exp(-r/\rdiskscale)$).
This can be converted to a disk half-mass radius using the relation $r_{1/2} = \nu \rdiskscale$, where $\nu$ solves $(\nu+1) \exp(-\nu) = 1/2$ \citep[e.g.][$\nu=1.678$]{Kravtsov_2013}.
As stated above (\SSS\ref{s-meth-galaxies}; Eq.~\eqref{sage-disk-radius}), {\sagename} calculates $\rdiskscale$.
Since voids are underdensities, dark matter haloes forming in voids will tend to collapse somewhat later than in overdensities \citep[e.g.][App.~A]{LaceyCole93MN}.
In an expanding FLRW universe, the critical density $\rho_{\mathrm{crit}}$ decreases, so for a fixed virialisation overdensity threshold and fixed mass, $R_{\mathrm{vir}}$ increases with time.
Thus, modelling galaxy disk scale lengths as being proportional to the halo virial radius (Eq.~\eqref{sage-disk-radius}), it would be reasonable to expect void galaxies to have greater $\rdiskscale$ than non-void galaxies, for a fixed value of the spin parameter $\lambda$.
Elaphrocentric galaxies will typically undergo a more isolated evolution than barycentric galaxies, with fewer merger events.
\citet{DOnghia08} found, based on $N$-body simulations, that the spin parameter of haloes in equilibrium is not influenced by merger events.
If the role of the spin parameter is indeed weak, then void galaxies should be marginally larger than non-void galaxies.
We consider both $\rdiskscale$ directly, and $R_{\mathrm{vir}}$ and $\lambda$ individually.

\subsubsection{Elaphro-acceleration} \label{s-method-elaphro-acc}

As a complement to the direct analyses of simulated galaxies via {\sagename}, we also investigate accelerations near the elaphrocentres, as preparation for future studies of elaphrocentric effects that might \postrefereechanges{have an effect on galaxy formation in voids and might} help form \postrefereechanges{giant} LSBGs.
We estimate the acceleration (compared to the FLRW reference model) of test particles directed away from the elaphrocentre, in the direction of the boundaries of the void.
This acceleration can be thought of as counteracting the self-gravity of an overdensity that is destined to collapse into a dark matter halo and allow the formation of a galaxy within the halo.
This acceleration is the effective antigravity that we have assumed could enlarge the size of a galaxy and decrease its infall rate, provided that the simulation has a detectable galaxy near the elaphrocentre.

Here, we describe how we estimate this \enquote*{elaphro-acceleration} without requiring the presence of a halo or galaxy.
\postrefereechanges{We will compare our results with the Newtownian estimate for the gravitational pull towards the centre of a halo of a Malin-1--like galaxy \citep{Bothun87}.}
In this simplified model, we assume \postrefereechanges{a high-mass test halo} at the elaphrocentre \postrefereechanges{of a void identified in the simulation, without modifying the underlying DM distribution}.
In the real Universe and in simulations, it is rather unlikely for a galaxy to form exactly at an elaphrocentre.
Nevertheless, we feel that this calculation will be a useful guide, since the elaphro-acceleration should be maximal in amplitude at the elaphrocentre.
\postrefereechanges{For our canonical high-mass test halo we adopt parameters that are motivated by observations of Malin 1 \citep{Seigar2008,Junais2020}.
  For an order of magnitude estimate, we adopt $M\tagTest = 10^{12} \mathrm{M}_\odot$ for the mass of the halo and $\rtest=\ElaphrocentreHaloOriginalRadiusMpcvalue$~{Mpc} for the region from which dark matter originated.}

\postrefereechanges{For any given void}, we interpolate the potential $\phi$ linearly, and calculate the acceleration as the gradient of the potential, \postrefereechanges{$\dot{v} \propto -\nabla \phi$}.
We use the full dark matter particle distribution, since gravitationally this should be more accurate than that of collapsed haloes (or galaxies) alone.
We use the gravitational potential estimates calculated by {\sc ramses} (\SSS\ref{s-meth-simulations}).
Since {\sc ramses} only provides potential estimates at particle positions, the resolution limit implied by using these is determined by the particle number density.
By definition, the number density is very low inside a void, so there are very few particle positions available for interpolating the potential.
\postrefereechanges{We sample the elaphro-acceleration} at six positions which lie on the \postrefereechanges{sphere} with radius $\rtest$, i.e. at $((\pm \rtest,0,0),(0,\pm \rtest,0),(0,0,\pm \rtest)$, where the elaphrocentre is the origin of the coordinate system.
For a given void, we calculate the radial (signed) and tangential (amplitude) components of the velocity vector $\vec{v}$ for each of the six positions, and find the mean values $\dot{v}_{\parallel}$ and $\dot{v}_\perp$, respectively.
As stated in \SSS\ref{s-method-voids}, some of the six positions may fall outside of a void when the void is too small; we ignore the void in such cases.

\subsection{Reproducibility versus cosmic variance} \label{s-method-reprod-cosvar}

We present results below (\SSS\ref{s-results}) with a preference for reproducibility over cosmic variance.
Our pipeline, using the Maneage template \citep{Akhlaghi2020maneage}, includes a step for verification, in the sense of verifying that when the reader recalculates our complete pipeline, s/he should obtain statistically equivalent results to our original results, within some tolerances.
Estimation of these tolerances effectively requires an approximate estimate of cosmic variance in the parameters of interest.
In principle, we could use these repeated full runs of the pipeline to obtain mean or median estimates of our main parameters of interest, rather than presenting the values from a single simulation, and the random uncertainties derived from that simulation.
However, that would reduce the reproducibility of our results, in the sense that readers wishing to run our pipeline would also have to perform multiple full runs.

Thus, here we favour reproducibility over cosmic variance, \postrefereechanges{using the latter only for reproducibility purposes, not \postrefereechangesB{for obtaining results} of the paper.}
\postrefereechangesB{We first run a fixed version of our full package $10$ times, for different random seeds and with physical randomisation induced by parallel computation.
  We obtain a full set of the results presented in this paper from each run.
  For any given parameter, we calculate the standard deviation $\sigmacosvar$ of the values of the parameter, where the subscript \enquote*{CV} refers to \enquote*{cosmic variance}.
  The particular version of the source package used for these verification runs has the string {\CosmicVarCommitID} as its {\sc git} commit identity (commit hash).}
\postrefereechanges{We use these repeat runs only for verification in the reproducibility sense, not for results.
A given parameter in a fresh realisation that aims to reproduce our results can then be verified for consistency by requiring that it agree with our published value to within $\NVerifySigmas \sqrt{2} $ times the stated random error $\sigmarandom$, where the $\sqrt{2}$ factor represents the assumption that both have independent Gaussian errors drawn from the same distribution.
For parameters with high $\sigmacosvar$ (i.e., those known to fail this verification), we require agreement between the fresh value and the mean from the ensemble of runs} within $\NVerifySigmas$ times $\sigmacosvar$, and we state $\sigmacosvar$ as an additional uncertainty, using the notation \enquote*{$\pmcosvar$}.
We use this formalised verification procedure for the parameters that we judge to be the more physically relevant.

The simulation presented below was chosen randomly.

\section{Results} \label{s-results}

\subsection{Simulation pipeline} \label{s-pipeline}

\begin{figure}
  \includegraphics[width=0.9\columnwidth]{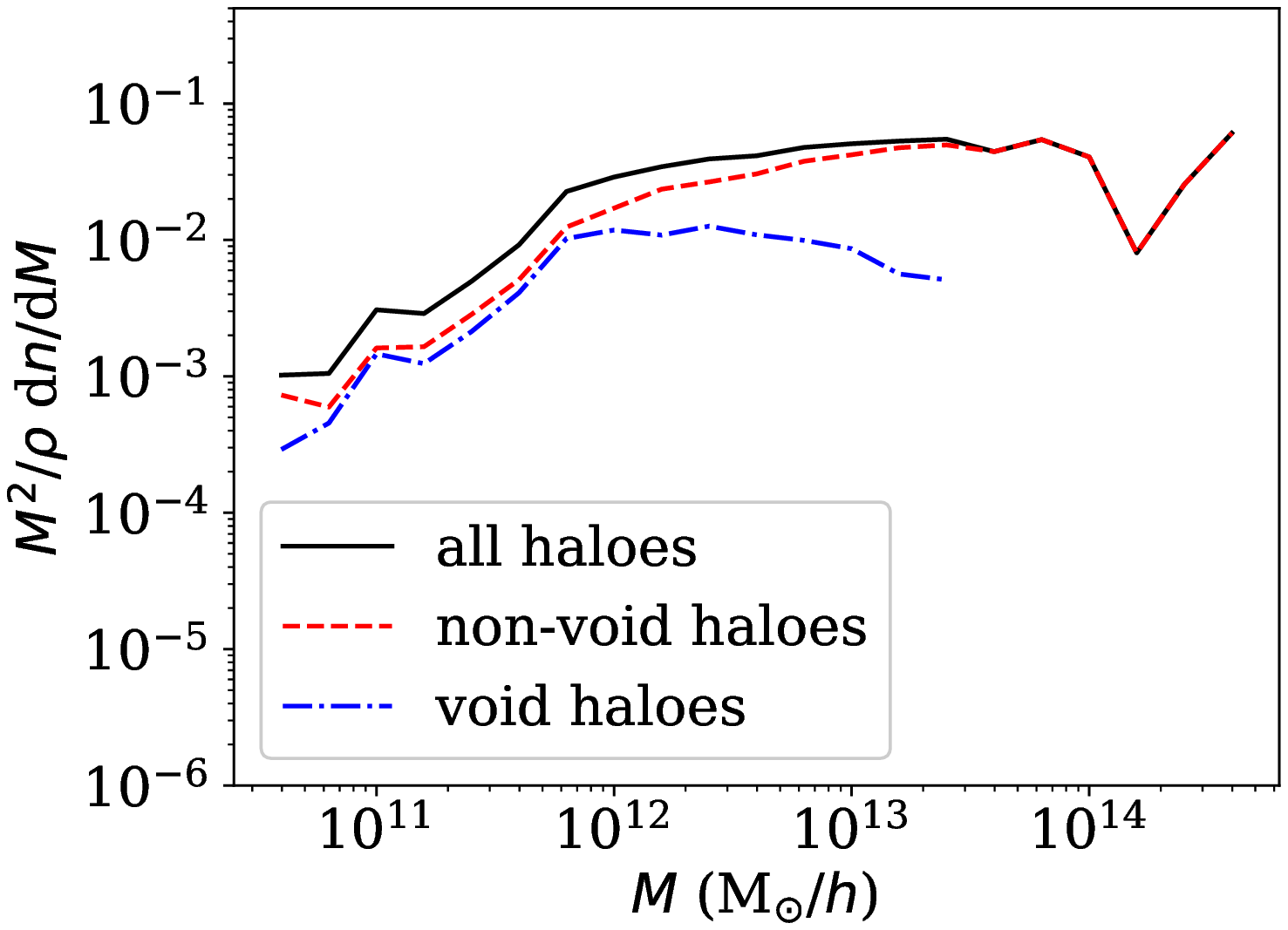}
  \caption{\protect\postrefereechangesB{Differential halo number counts $M^2/\rho \,\diffd n/\diffd M$ versus halo mass $M$ for the haloes that host galaxies in the simulation, where $n$ is the number density of haloes in a given mass interval and $\rho$ is the mean mass density of the simulation.
    The solid, black (top) curve represents all haloes; the dashed, red (middle) curve is for haloes that are not in a void; and the dash-dotted, blue (bottom) curve is for haloes that are identified as being located in a void.
    The gradual decline in the numbers of haloes towards the lower mass scales and the sharp cut at the lowest mass scale are artefacts of the limited resolution of our simulation.}
    \label{f-diff_halo_mass_fct}}
\end{figure}

In the final time step of our $N={\Ncrootvalue}^3$-particle simulation, we detected {\NHaloesFinalTimevalue} haloes, with $\PlotsSizeiNGalsTotalvalue$ galaxies evolved along the merger history trees for these haloes.
Among these galaxies, $\PlotsSizeiNGalsInIntervalvalue$ have virial mass $M$ in the range ${\MassgapLowerLimitvalue}$--${\MassgapUpperLimitvalue} \mathrm{M}_\odot /h$, which we study with the aim of seeking a factor in the formation of \postrefereechanges{high-mass galaxies}.
Applying the $\haloinvoidfrac > {\HaloIsInVoidMinFractionvalue}\%$ definition in \SSS\ref{e-method-void-membership}, we identify ${\PlotsSizeiNVoidGalsTotalvalue}$ galaxies in voids, among which ${\PlotsSizeiNVoidGalsvalue}$ of these galaxies have a mass in the selected range.
This fraction of galaxies identified as void galaxies is larger than the 7\% estimated by \citet{Pan2012voids}.
A more detailed analysis of the galaxies that are identified to be in voids is given below.
We investigate key quantities dependent on the fraction of their host haloes particles in a void $\haloinvoidfrac$ and dependent on their relative distance to the voids centre $r/R_{\mathrm{eff}}$.
\postrefereechangesB{We use Theil--Sen robust linear fits \citep{Theil50,Sen68} on each key quantity to see if it has a statistically significant dependence on either of the two void location parameters.}
There are $\nvoidsiNvoidsvalue$ voids in the final time step.
The void size distribution is shown in Fig.~\ref{void_population}.
Since we detect voids physically, following the dark matter particular distribution, as described in \SSS\ref{s-method-voids}, it is unsurprising that the void population is dominated by voids a few Mpc/$h$ in size, with a correspondingly higher number density than that typically seen in void catalogues calculated using galaxies as tracers.
\postrefereechangesB{The distribution of void galaxy elaphrocentric distances is shown in Fig.~\ref{f-results-galaxy-hist}.}

\begin{figure}
  \iftoggle{removesomefits}{
    \includegraphics[width=0.9\columnwidth]{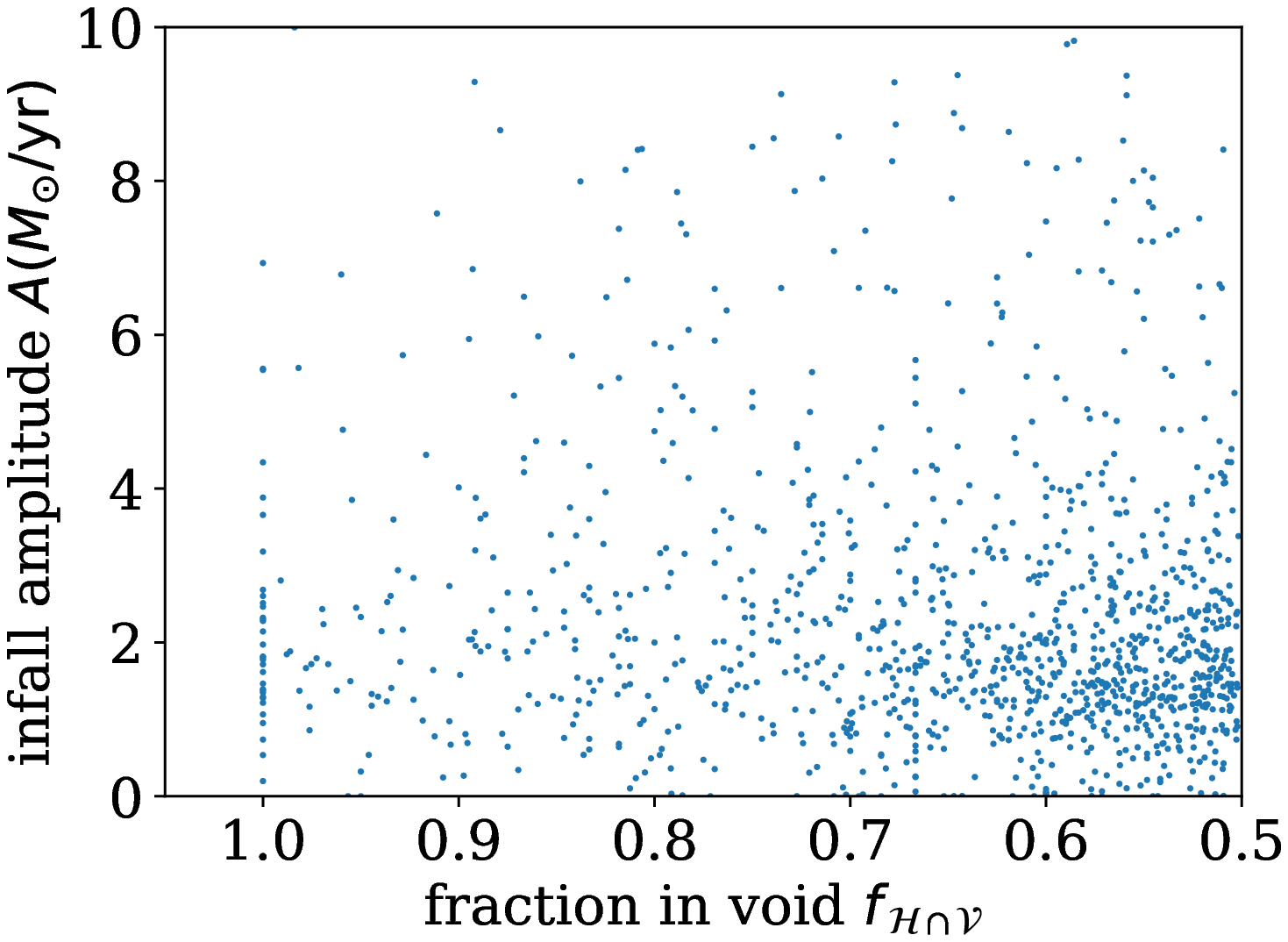}}{
    \includegraphics[width=0.9\columnwidth]{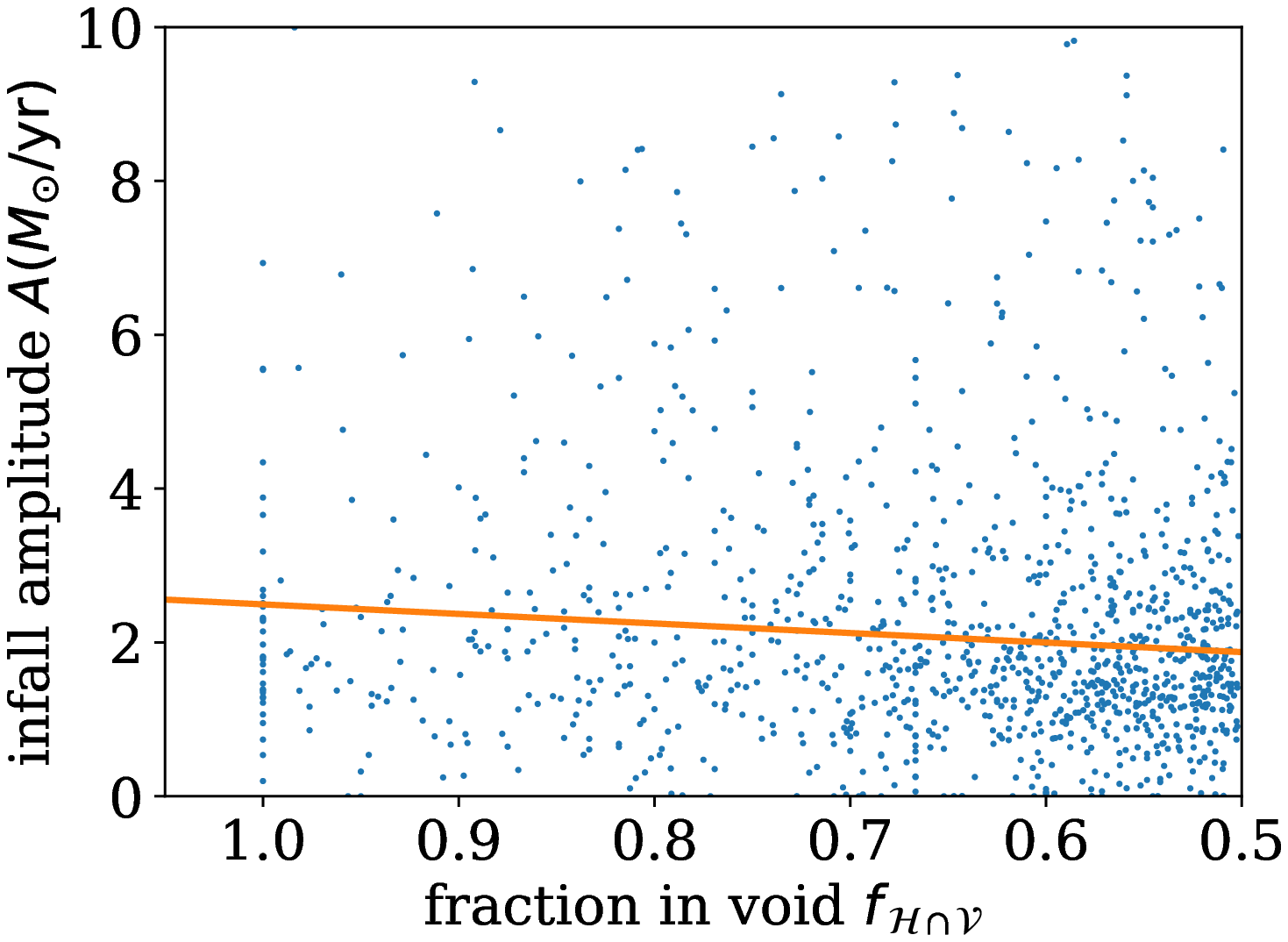}}
  \caption{Amplitude of infall rate ${\InfallAmplitude}$ versus fraction $\haloinvoidfrac$ of a galaxy's host halo composed of void particles.
    A Theil--Sen robust linear fit to the relation, ${\InfallAmplitude} = \left[(\InfallAmpFracZerovalue \pm \InfallAmpFracSigZerovalue) + (\InfallAmpFracSlopevalue \pm \InfallAmpFracSigSlopevalue \pmcosvar \InfallAmpFracSlopevalueCosmStdDev) \haloinvoidfrac\right]\,M_{\odot}/\mathrm{yr} $, is \postrefereechanges{shown}.
    Galaxies towards the left ($\haloinvoidfrac = 1$) are those best identified as being in voids.
    \protect\postrefereechangesB{Fits are made for this figure through to Fig.~\protect\ref{f-frac-rreff}.
      These} are robust best fits for studying statistical relations; they are not predictive models.
    \iftoggle{removesomefits}{\protect\postrefereechangesB{We display} the fits in \postrefereechangesB{almost} all these scatter plots, quantifying them and their uncertainties in the captions
      \protect\postrefereechangesB{(fits are not displayed in this figure, nor in Figs.~\protect\ref{f-infall-rreff-amp}, \protect\ref{f-galaxy-size-frac}, \protect\ref{f-acc-tan-rreff} and \protect\ref{f-frac-rreff}).}}{To avoid selection bias, we \protect\postrefereechangesB{display} the fits in all these scatter plots, quantifying them and their uncertainties in the captions.}
    See the text for discussion of which relations have significantly non-zero slopes.
    Plain text table for this figure through to Fig.~\ref{f-infall-rreff-tau}: \mbox{\href{\projectzenodofilesbase/voidgals_infall.dat}{\projectzenodoid/voidgals\_infall.dat}}.
    \label{f-infall-frac-amp}}
\end{figure}

\begin{figure}
  \includegraphics[width=0.9\columnwidth]{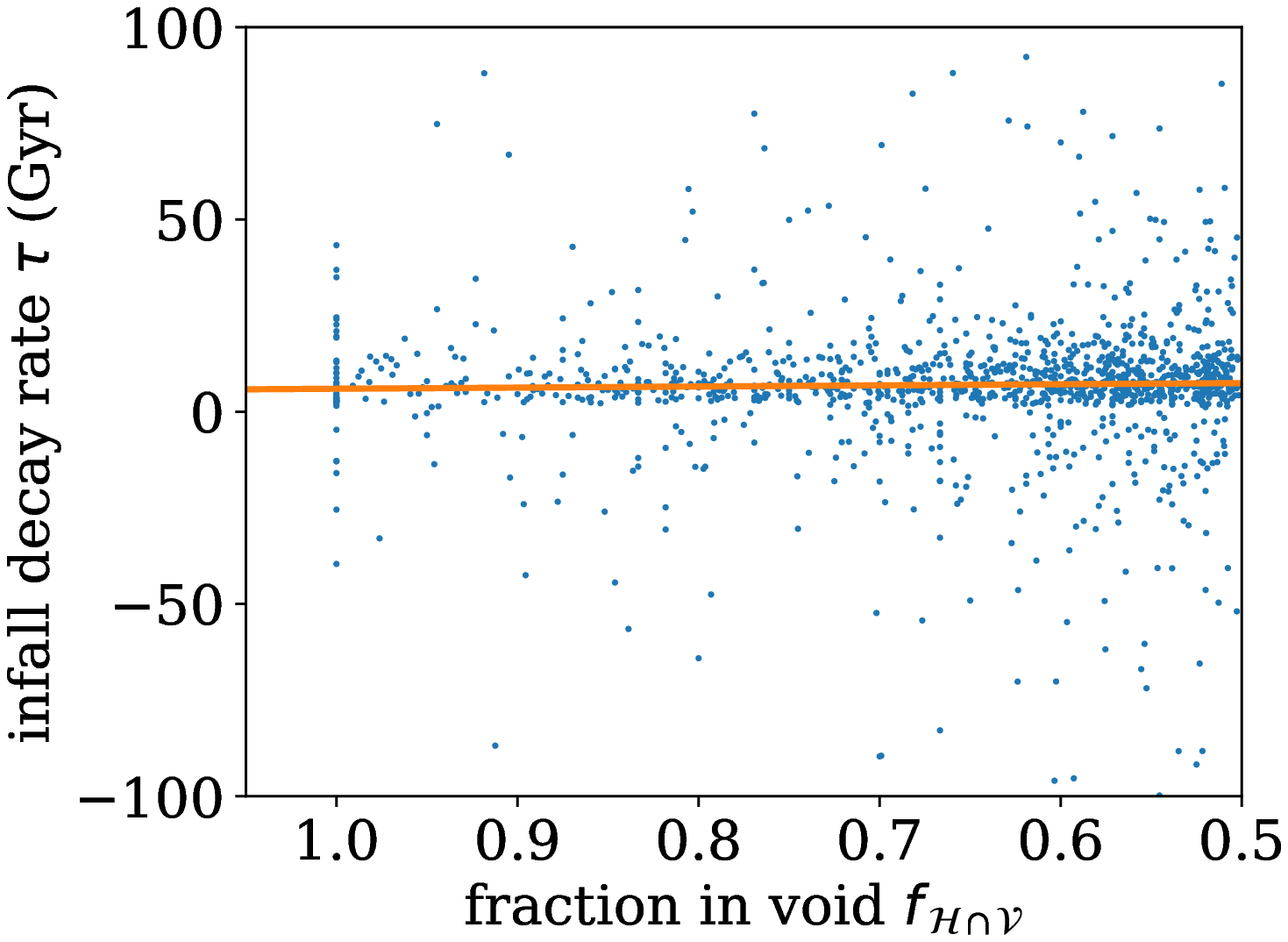}
  \caption{Infall decay rate ${\InfallDecayRate}$ versus fraction $\haloinvoidfrac$ of a galaxy's host halo composed of void particles, as in Fig.~\protect\ref{f-infall-frac-amp}, with a Theil--Sen robust linear fit ${\InfallDecayRate} = \left[(\InfallTauFracZerovalue \pm \InfallTauFracSigZerovalue) + (\InfallTauFracSlopevalue \pm \InfallTauFracSigSlopevalue \pmcosvar \InfallTauFracSlopevalueCosmStdDev) \haloinvoidfrac\right] \, \mathrm{Gyr}$, is shown.
    \label{f-infall-frac-tau}}
\end{figure}

\begin{figure}
  \iftoggle{removesomefits}{
  \includegraphics[width=0.9\columnwidth]{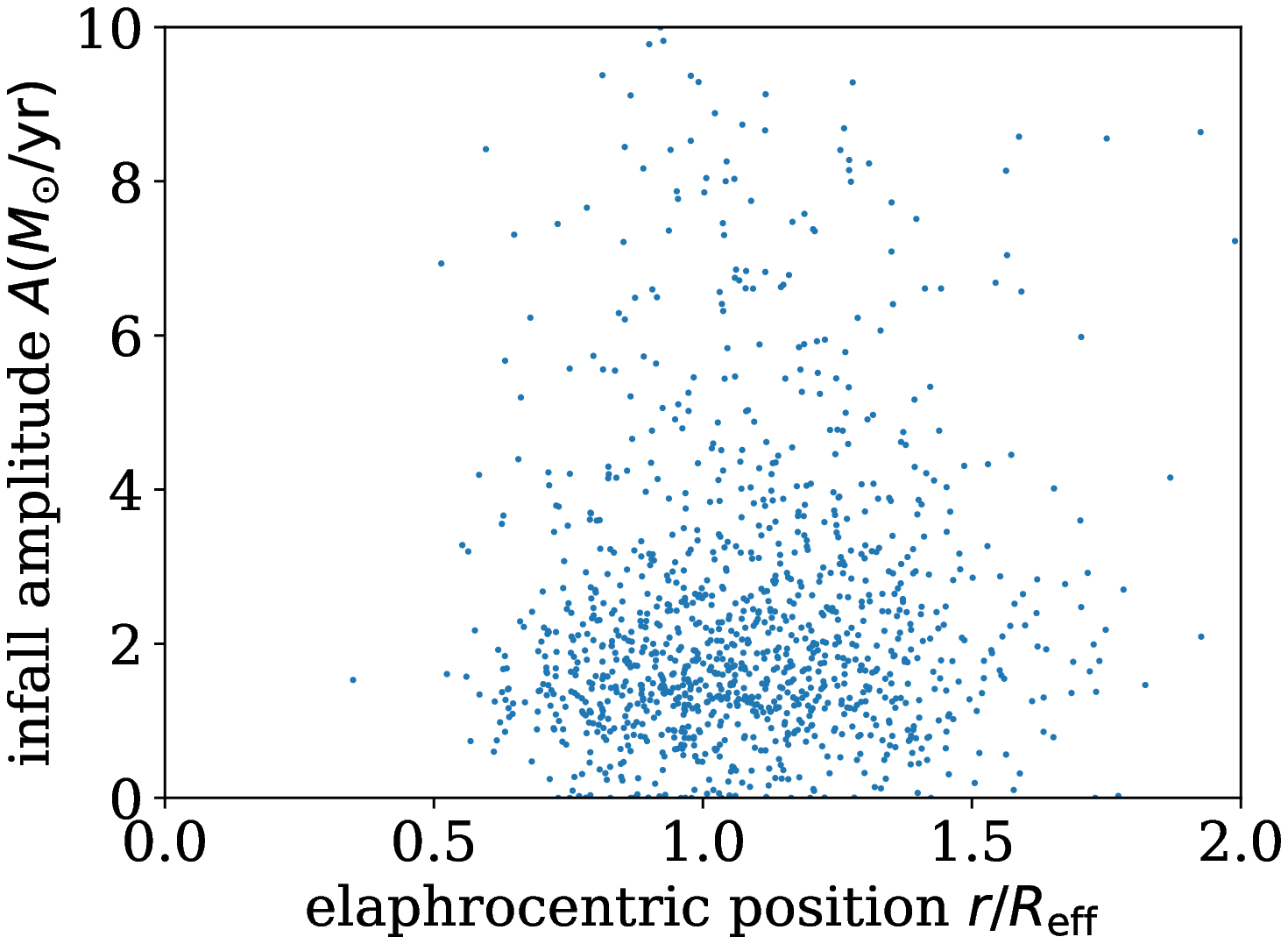}}{
  \includegraphics[width=0.9\columnwidth]{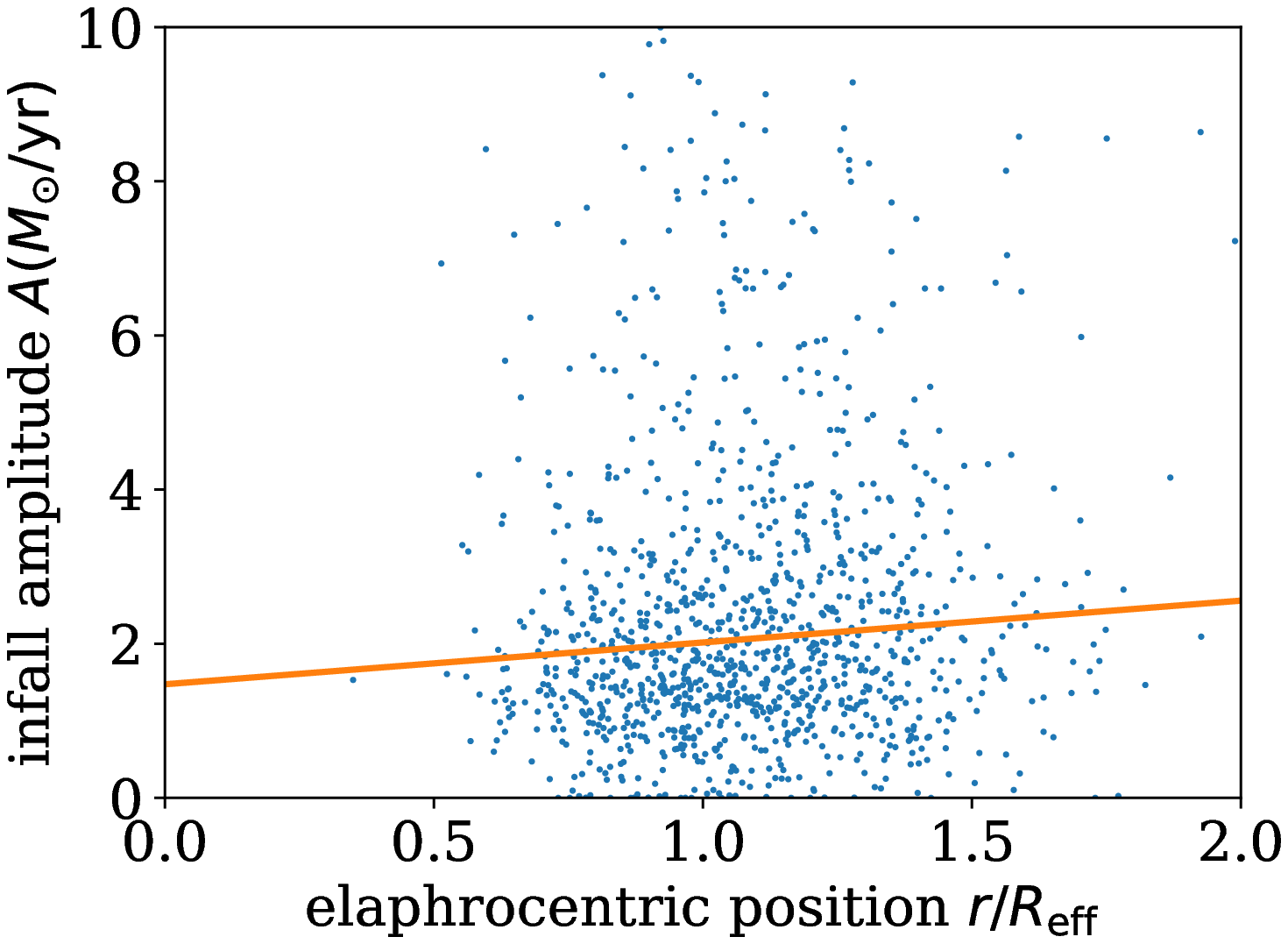}}
  \caption{Amplitude of infall rate ${\InfallAmplitude}$ versus elaphrocentric location $r/R_{\mathrm{eff}}$ of a galaxy's host halo.
    The fit is ${\InfallAmplitude} = \left[(\InfallAmpReldistZerovalue \pm \InfallAmpReldistSigZerovalue) + (\InfallAmpReldistSlopevalue \pm {\InfallAmpReldistSigSlopevalue} \postrefereechanges{\pmcosvar \InfallAmpReldistSlopevalueCosmStdDev}) \,r/R_{\mathrm{eff}}\right] \,M_{\odot}/\mathrm{yr}$.
    As in Fig.~\protect\ref{f-infall-frac-amp}, galaxies towards the left are those best identified as being in voids, but voidness is characterised in this plot by a lower elaphrocentric distance $r/R_{\mathrm{eff}}$, instead of by a higher void fraction $\haloinvoidfrac$.
    \label{f-infall-rreff-amp}}
\end{figure}

\begin{figure}
  \includegraphics[width=0.9\columnwidth]{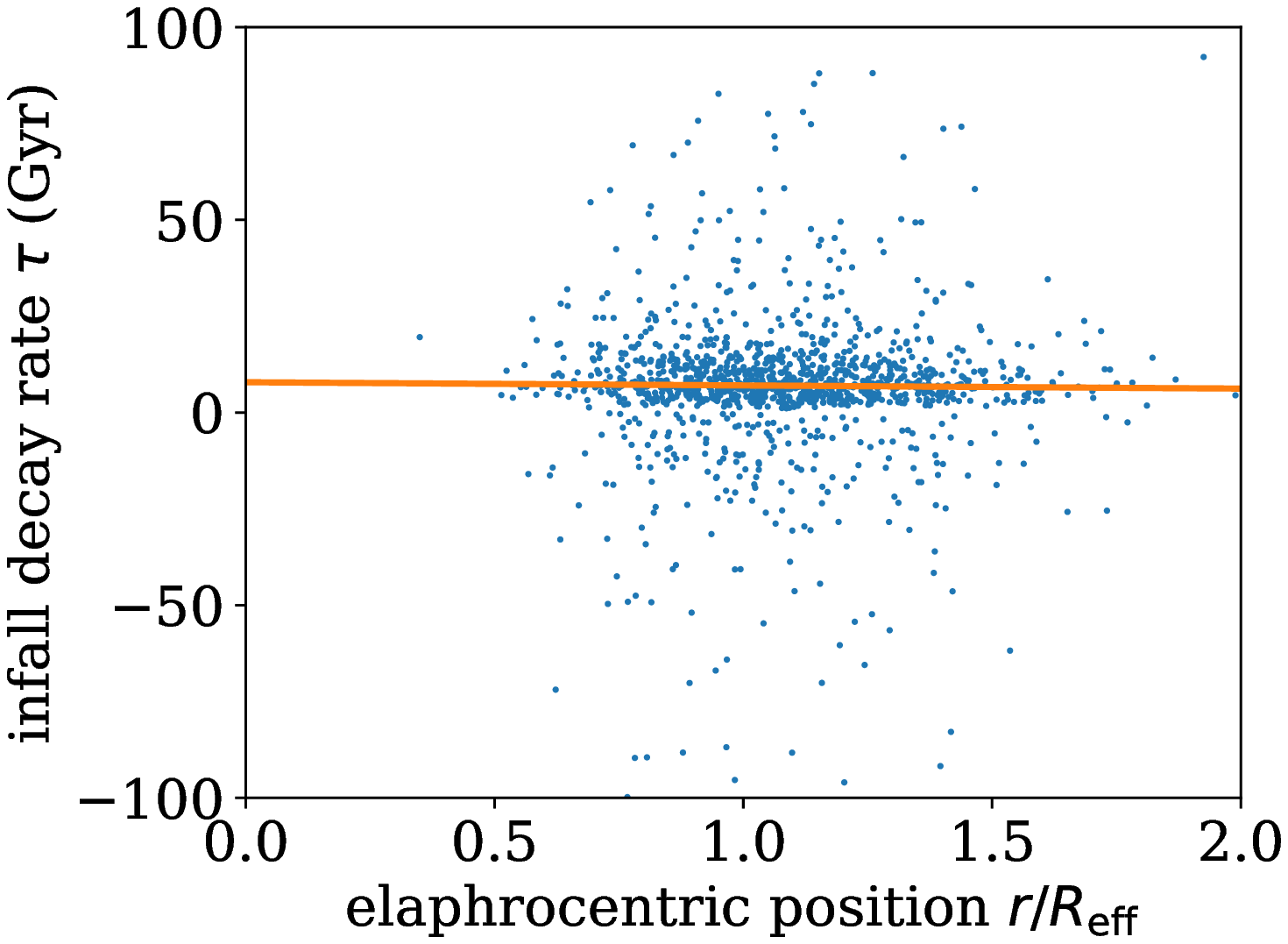}
  \caption{Infall decay rate ${\InfallDecayRate}$ versus versus elaphrocentric location $r/R_{\mathrm{eff}}$ of a galaxy's host halo, as in Fig.~\protect\ref{f-infall-rreff-amp}, with robust fit ${\InfallDecayRate} = \left[ (\InfallTauReldistZerovalue \pm \InfallTauReldistSigZerovalue) + (\InfallTauReldistSlopevalue \pm \InfallTauReldistSigSlopevalue) \,r/R_{\mathrm{eff}}\right] \,\mathrm{Gyr}$.
    \label{f-infall-rreff-tau}}
\end{figure}

\subsection{Infall rate}  \label{s-results-infall}
As described in \SSS\ref{s-method-infall}, we first compared infall rates for galaxies as separate void and non-void populations.
For each galaxy mass infall history $\diffd M/\diffd t\,(t)$, we first find a linear least-squares best fit to $\log_{10}(\diffd M/\diffd t)$ versus $t$ for time steps where $\diffd M/\diffd t\,(t) > 0$.
The optimal parameters of this fit are used to find a non-linear least-squares best fit of $\diffd M/\diffd t (t)$ to a decaying exponential (Eq.~\eqref{e-infall-model}), starting from the first time step with $\diffd M/\diffd t\,(t) > 0$ and no longer excluding time steps with $\diffd M/\diffd t = 0$.
As stated above, merger histories are complex, and many galaxies' infall histories are poorly fit by this procedure.
This applies both to void and non-void galaxies, and we do not attempt to analyse these cases.
The limitations of this simplified approach should similarly affect both populations and should not affect our comparison of the successfully fit subsets of the two populations.

\begin{table}
  \centering
  \caption{Exponential decaying fit parameters (medians and standard error in the median) for infall rates for void and non-void galaxies.
    A large standard error in the time scale, of the order of the age of the Universe, indicates that many fits represent nearly constant infall rates.
    \label{Infall-rate-table}}
  $\begin{array}{l c c}
    \hline
    & \mbox{in voids} & \mbox{not in voids} \\
    \hline
    \log_{10}(A) \,({M_{\odot}/\mathrm{yr}})
    & {{\InfallMedianValueAmplitudeVoidsvalue} \pm {\InfallRobustErrorAmplitudeVoidsvalue}}
    & {{\InfallMedianValueAmplitudeNonVoidsvalue} \pm {\InfallRobustErrorAmplitudeNonVoidsvalue}}\rule{0ex}{2.7ex}\\
    \InfallDecayRate \,(\mathrm{Gyr})
    & {\InfallMedianValueDecayVoidsvalue} \pm {\InfallRobustErrorDecayVoidsvalue}
    & {\InfallMedianValueDecayNonVoidsvalue} \pm {\InfallRobustErrorDecayNonVoidsvalue} \\
    \hline
  \end{array}$
\end{table}

We find $\mvoidgood={\VoidFitiGoodParametervalue}$ valid fits for the void galaxies, and $\mvoidbadtot={\VoidFitiBadParameterTotalvalue}$ fits that are rejected either as failed fits ($\mvoidbadfit = {\VoidFitiBadParameterFailedFitvalue}$) or as having an unreasonably high amplitude, ${\InfallAmplitude} > {\VoidFitiParameterThresholdInputvalue}$~$M_{\odot}/$yr, indicating a physically unrealistic fit ($\mvoidbadthresh = {\VoidFitiBadParameterThresholdvalue}$ cases).
For non-void galaxies we find $\mnonvoidgood={\NonVoidFitiGoodParametervalue}$ valid fits and $\mnonvoidbadtot={\NonVoidFitiBadParameterTotalvalue}$ fits rejected either as failed fits ($\mnonvoidbadfit = {\NonVoidFitiBadParameterFailedFitvalue}$) or as having unphysically high amplitudes ($\mnonvoidbadthresh = {\NonVoidFitiBadParameterThresholdvalue}$ cases).
For galaxies in host haloes with virial masses in the range ${\MassgapLowerLimitvalue}$--${\MassgapUpperLimitvalue} \mathrm{M}_\odot /h$, we find the medians listed in Table~\ref{Infall-rate-table}, where the uncertainties are standard errors in the median.
Throughout this work, uncertainties in the median are given as the standard error in the median, unless otherwise stated.
The median host halo mass for the void galaxies is $(\HaloMassVoidsMedvalue \pm \HaloMassVoidsStderrMedvalue) \times 10^{11} M_{\odot}$, lower than that of the non-void galaxies, $(\HaloMassNonvoidsMedvalue \pm \HaloMassNonvoidsStderrMedvalue) \times 10^{11} M_{\odot}$.
\postrefereechangesB{Figure~\ref{f-diff_halo_mass_fct} shows the differential mass function of the haloes in the form of the halo multiplicity function.
  We show the differential halo masses for all haloes, for haloes that are classified as being associated with a void and for haloes that not associated with any void.
  As stated in \SSS\ref{s-meth-initial-cond}, we cannot resolve dwarf galaxies; this is seen in the sharp cut at the lowest masses.
  The gradual decline in low-mass haloes and the sharp cut are both consistent with the use of the {\rockstarname} halo finder (\SSS\ref{s-meth-halo-det}).
  The void and non-void halo mass functions clearly differ in the higher mass ranges.
  The most massive haloes form in voids very rarely, while among the lowest mass haloes, the probabilities of forming in a void or not are of similar magnitude.}

\begin{table}
  \centering
  \caption{Median disk galaxy scale length, spin parameter and virial radius for void and non-void galaxies in the mass interval ${\MassgapLowerLimitvalue}$--${\MassgapUpperLimitvalue} \mathrm{M}_\odot /h$ and on all mass scales.
    \postrefereechanges{Two parameters are known to have high cosmic variance, as given in the table (see \SSS\protect\ref{s-method-reprod-cosvar}).}
    \label{Galaxy-size-table}}
  $\begin{array}{l c c}
    \hline
    &\mbox{in voids} & \mbox{not in voids} \\
    \hline
\multicolumn{3}{c}{\mbox{restricted mass interval}} \\
    \hline
    \rdiskscale (\mbox{kpc}/h)
    & {\PlotsSizeiRAvgVoidsvalue} \pm {\PlotsSizeiErrorRAvgVoidsvalue}
    & {\PlotsSizeiRAvgNonvoidsvalue} \pm {\PlotsSizeiErrorRAvgNonvoidsvalue}\rule{0ex}{2.7ex}\\
    \lambda
    & {\SpiniSpinAvgVoidsvalue} \pm {\SpiniErrorSpinAvgVoidsvalue}
    & {\SpiniSpinAvgNonvoidsvalue} \pm {\SpiniErrorSpinAvgNonvoidsvalue} \\
    R_{\mathrm{vir}} (\mbox{kpc}/h)
    & {\RviriRvirAvgVoidsvalue} \pm {\RviriErrorRvirAvgVoidsvalue}
    & {\RviriRvirAvgNonvoidsvalue} \pm {\RviriErrorRvirAvgNonvoidsvalue} \\
    \hline
    \multicolumn{3}{c}{\mbox{all mass scales}} \\
    \hline
    \rdiskscale (\mbox{kpc}/h)
    & {\PlotsSizeiRAvgVoidsTotalvalue} \pm {\PlotsSizeiErrorRAvgVoidsTotalvalue} \postrefereechanges{\pmcosvar {\PlotsSizeiRAvgVoidsTotalvalueCosmStdDev}}
    & {\PlotsSizeiRAvgNonvoidsTotalvalue} \pm {\PlotsSizeiErrorRAvgNonvoidsTotalvalue}\rule{0ex}{2.7ex}\\
    \lambda
    & {\SpiniSpinAvgVoidsTotalvalue} \pm {\SpiniErrorSpinAvgVoidsTotalvalue}
    & {\SpiniSpinAvgNonvoidsTotalvalue} \pm {\SpiniErrorSpinAvgNonvoidsTotalvalue} \\
    R_{\mathrm{vir}} (\mbox{kpc}/h)
    & {\RviriRvirAvgVoidsTotalvalue} \pm {\RviriErrorRvirAvgVoidsTotalvalue} \postrefereechanges{\pmcosvar {\RviriRvirAvgVoidsTotalvalueCosmStdDev}}
    & {\RviriRvirAvgNonvoidsTotalvalue} \pm {\RviriErrorRvirAvgNonvoidsTotalvalue} \\

\hline
  \end{array}$
\end{table}

We find (Table~\ref{Infall-rate-table}) no significant difference in either the amplitude ${\InfallAmplitude}$ or the time scale ${\InfallDecayRate}$ of infall between the void and non-void galaxy populations.
The dispersion in infall patterns within each population is as great as both ${\InfallAmplitude}$ and ${\InfallDecayRate}$ themselves.
We found that, to very high significance, void galaxies typically form later than non-void galaxies, as expected, since they form in underdensities.
We find median collapse epochs (in standard FLRW cosmological time) of $t^{\mathrm{f}}_{\mathrm{v}}=\FormationTimeVoidsMedvalue \pm \FormationTimeVoidsStderrMedvalue$~Gyr and $t^{\mathrm{f}}_{\mathrm{nv}}=\FormationTimeNonvoidsMedvalue \pm \FormationTimeVoidsStderrMedvalue $~Gyr for the void and non-void galaxies, respectively.
By the collapse epoch of a galaxy, we mean the first epoch at which the mass infall rate calculated by {\sagename} for the galaxy is non-zero.

We investigate the results for ${\InfallAmplitude}$ and ${\InfallDecayRate}$ more closely by checking if either ${\InfallAmplitude}$ or ${\InfallDecayRate}$ has a dependence on either the fraction of a host halo's particles that are identified as void particles, $\haloinvoidfrac$, or on the host halo's elaphrocentric distance $r/R_{\mathrm{eff}}$.
Figures~\ref{f-infall-frac-amp}--\ref{f-infall-rreff-tau} show the dependence of ${\InfallAmplitude}$ and ${\InfallDecayRate}$ on $\haloinvoidfrac$ and $r/R_{\mathrm{eff}}$ for all void galaxies.
The $\haloinvoidfrac$ axis is shown with $\haloinvoidfrac$ decreasing from left to right, so that the galaxies that are best qualified as void galaxies are shown towards the left in all four figures.
There is no visually obvious dependence of the infall parameters on $\haloinvoidfrac$.
However, Fig.~\ref{f-infall-frac-amp} does show a modestly significant non-zero slope, \postrefereechangesB{i.e., the median of the infall amplitudes ${\InfallAmplitude}$ is somewhat higher for galaxies better identified as void galaxies (having a higher value of $\haloinvoidfrac$).}
This would tend to oppose the hypothesis of a general tendency to form LSBGs in voids.
The other slopes of the best fit linear relations, using robust statistics as above, indicated numerically in the figure captions, are not significantly non-zero.

\subsection{Galaxy Sizes} \label{s-results-size}

While we do not detect significant elaphrocentric effects on infall rates, effects on galaxy sizes could play an important role in forming large diffuse galaxies.
As stated above (\SSS\ref{s-method-size}), to check the size of a galaxy at the final output time step, we use the disk scale length $\rdiskscale$ provided by {\sagename}.
The results for the galaxies divided into void and non-void populations are shown in Table~\ref{Galaxy-size-table}, where we list the disk exponential scale length $\rdiskscale$, the spin parameter $\lambda$ and the virial radius $R_{\mathrm{vir}}$.
Table~\ref{Galaxy-size-table} shows a significant difference for the overall scale length $\rdiskscale$.
Our results show that, as a population, void galaxies form significantly smaller disks, both for our selected mass interval and for the full sample.
Although this might seem to support \citet[][fig.~2, left]{vdWeygaert11VGS}'s finding that, for a given absolute magnitude, void galaxies tend to be smaller than the general galaxy population, we do not (yet) model stellar populations and estimate absolute magnitudes, so this qualitative agreement is promising but not conclusive.

A likely explanation for the smaller sizes of void galaxies is shown by the other rows in \postrefereechanges{Table~\ref{Galaxy-size-table}:} the void galaxy population has a much smaller median virial radius $R_{\mathrm{vir}}$ than the non-void population, but an insignificantly higher spin parameter $\lambda$.
The slightly greater spin parameter appears insufficient to compensate or override the lower $R_{\mathrm{vir}}$ of the void galaxies.
The values of the spin parameter are reasonable in relation to standard values in the literature.
Our non-void host halo values of $\lambda$ listed in Table~\ref{Galaxy-size-table} are consistent with the friends-of-friends halo estimate of \citet[][\SSS2.4, para.\/ 8]{ZjupaSpringel2017}, $\lambda_{\mathrm{B,FOF}} = 0.0414$, while our void host halo values of $\lambda$ are slightly higher.

\begin{figure}
  \iftoggle{removesomefits}{
    \includegraphics[width=0.9\columnwidth]{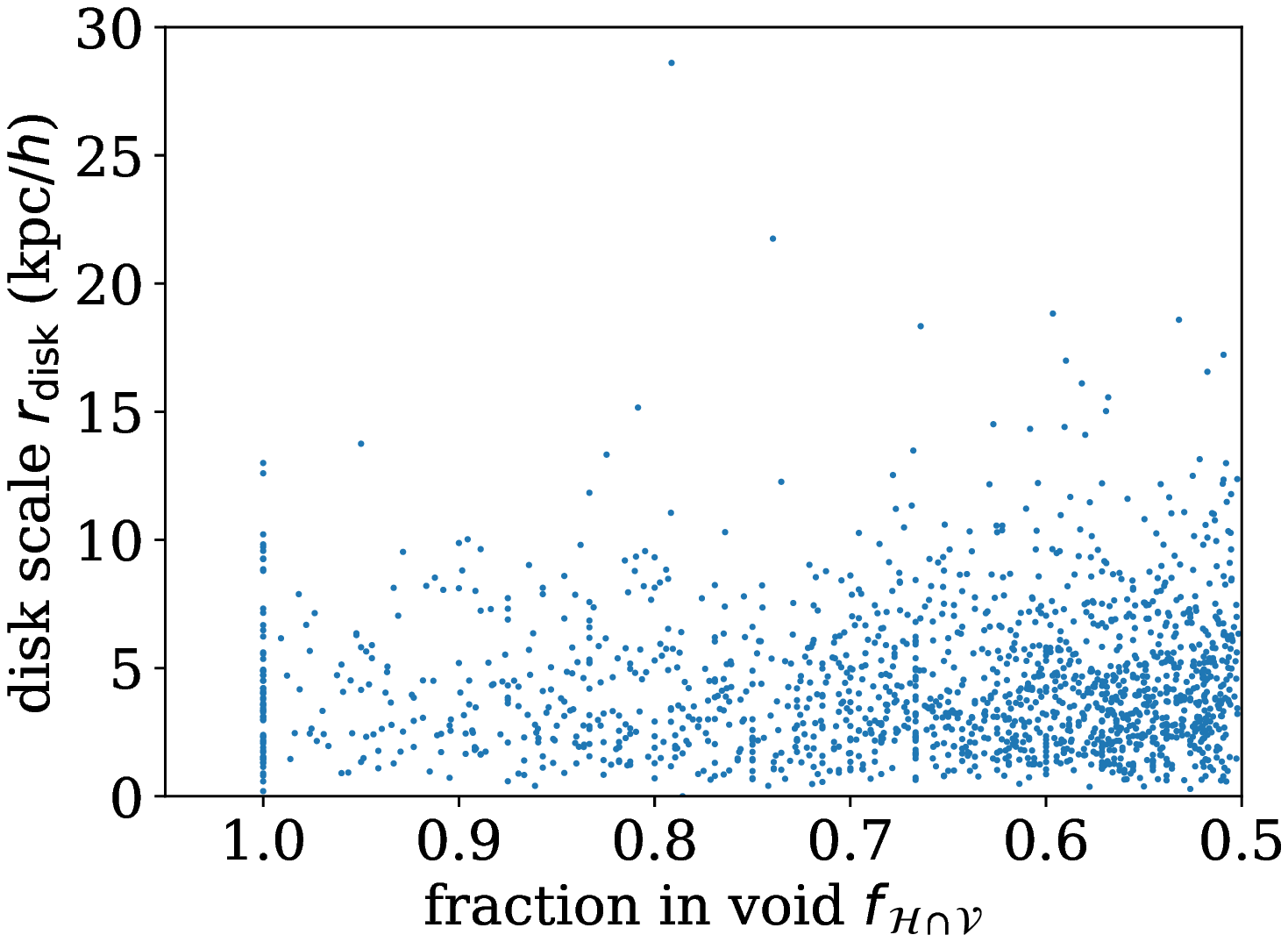}}{
    \includegraphics[width=0.9\columnwidth]{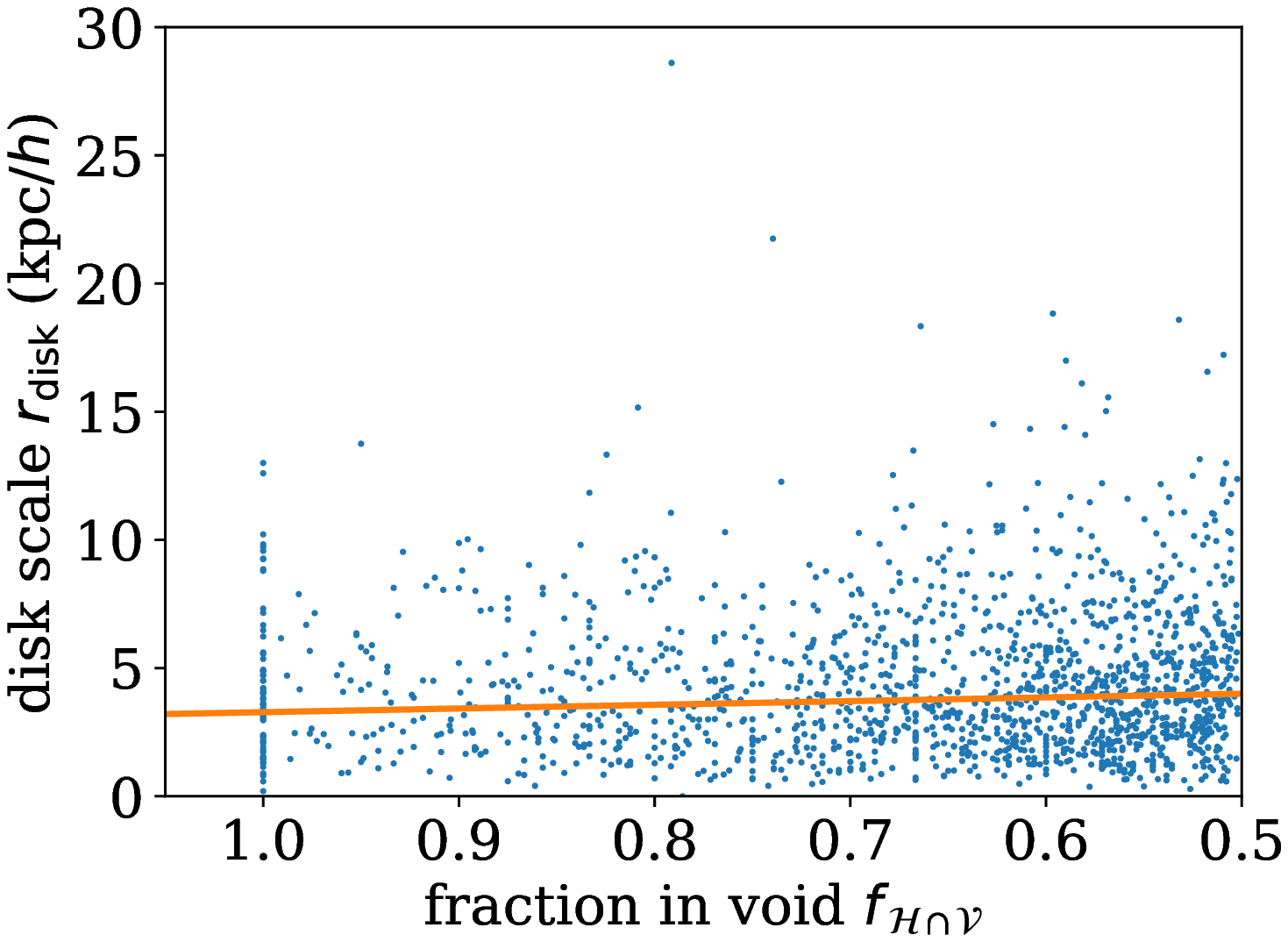}}
  \caption{Galaxy disk scale length $\rdiskscale$ versus fraction $\haloinvoidfrac$ of a galaxy's host halo composed of void particles.
    The fit is $\rdiskscale = \left[ (\GalsizeFracZerovalue \pm \GalsizeFracSigZerovalue) + (\GalsizeFracSlopevalue \pm \GalsizeFracSigSlopevalue \pmcosvar \GalsizeFracSlopevalueCosmStdDev) \haloinvoidfrac \right] \,\mathrm{kpc}/h$.
    Plain text table for this figure through to Fig.~\ref{f-galaxy-Rvir-rreff}: \mbox{\href{\projectzenodofilesbase/voidgals_infall.dat}{\projectzenodoid/voidgals\_infall.dat}}.
    \label{f-galaxy-size-frac}}
\end{figure}

\begin{figure}
  \includegraphics[width=0.9\columnwidth]{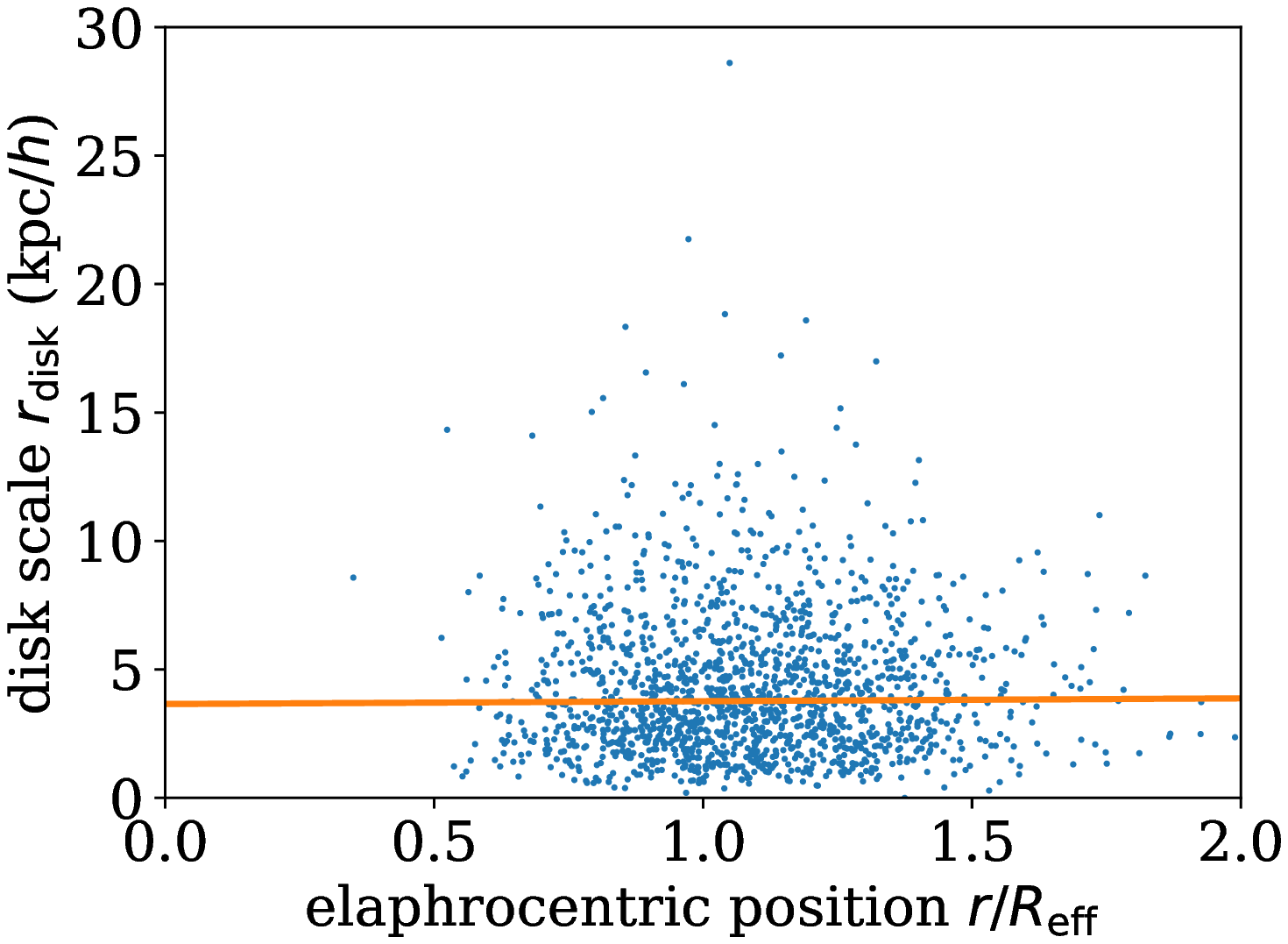}
  \caption{Disk scale length $\rdiskscale$ versus elaphrocentric location $r/R_{\mathrm{eff}}$, as in Fig.~\protect\ref{f-galaxy-size-frac}.
    The fit is $\rdiskscale = \left[ (\GalsizeReldistZerovalue \pm \GalsizeReldistSigZerovalue) + (\GalsizeReldistSlopevalue \pm \GalsizeReldistSigSlopevalue) \,r/R_{\mathrm{eff}} \right] \,\mathrm{kpc}/h$.
    \label{f-galaxy-size-rreff}}
\end{figure}

\begin{figure}
  \includegraphics[width=0.9\columnwidth]{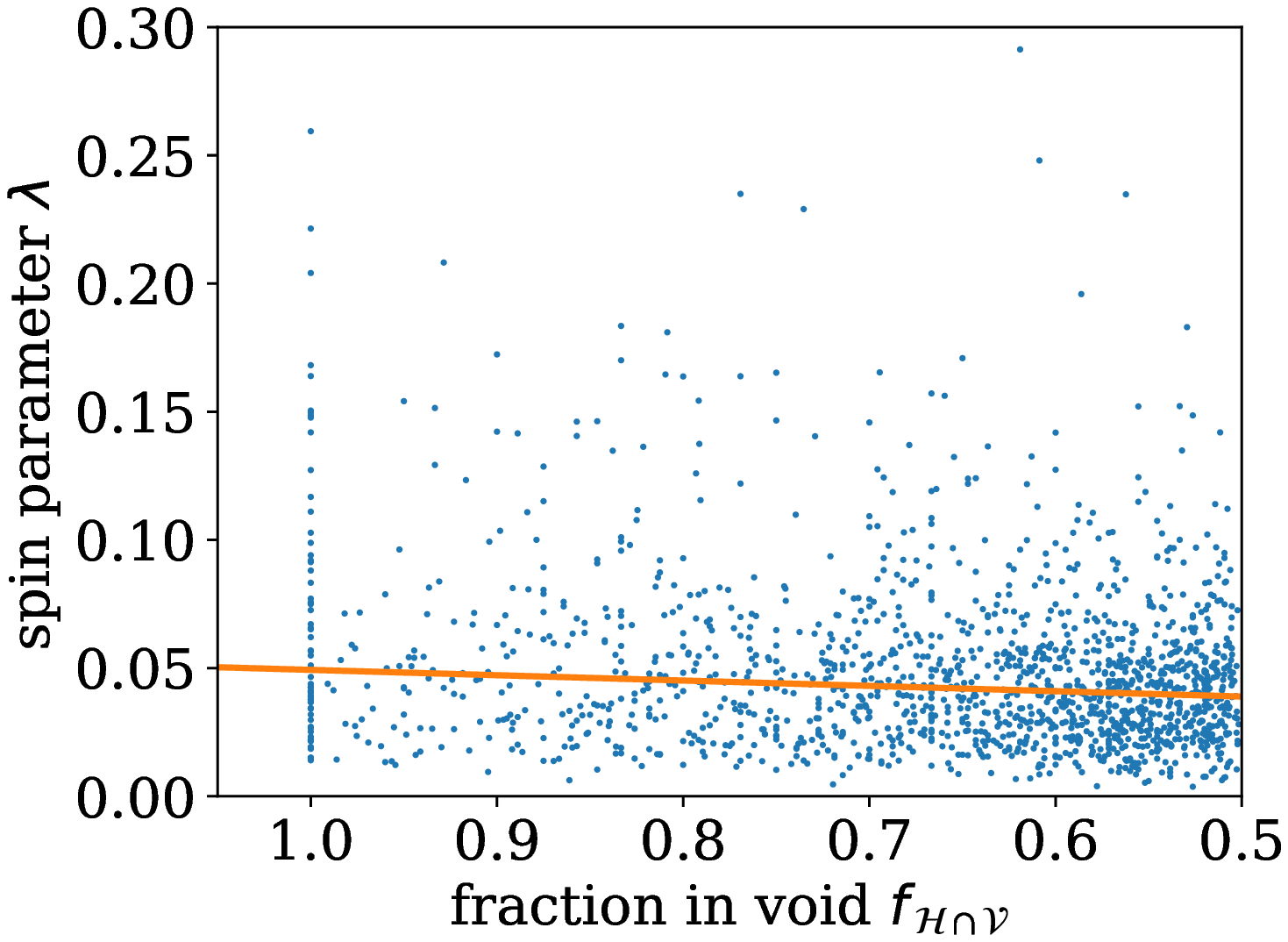}
  \caption{Dimensionless spin parameter $\lambda$ (Eq.~\protect\eqref{e-defn-spin}) versus $\haloinvoidfrac$.
    The fit is $\lambda = (\GalspinFracZerovalue \pm \GalspinFracSigZerovalue) + (\GalspinFracSlopevalue \pm \GalspinFracSigSlopevalue \pmcosvar \GalspinFracSlopevalueCosmStdDev) \haloinvoidfrac$.
    \label{f-galaxy-spin-frac}}
\end{figure}

\begin{figure}
  \includegraphics[width=0.9\columnwidth]{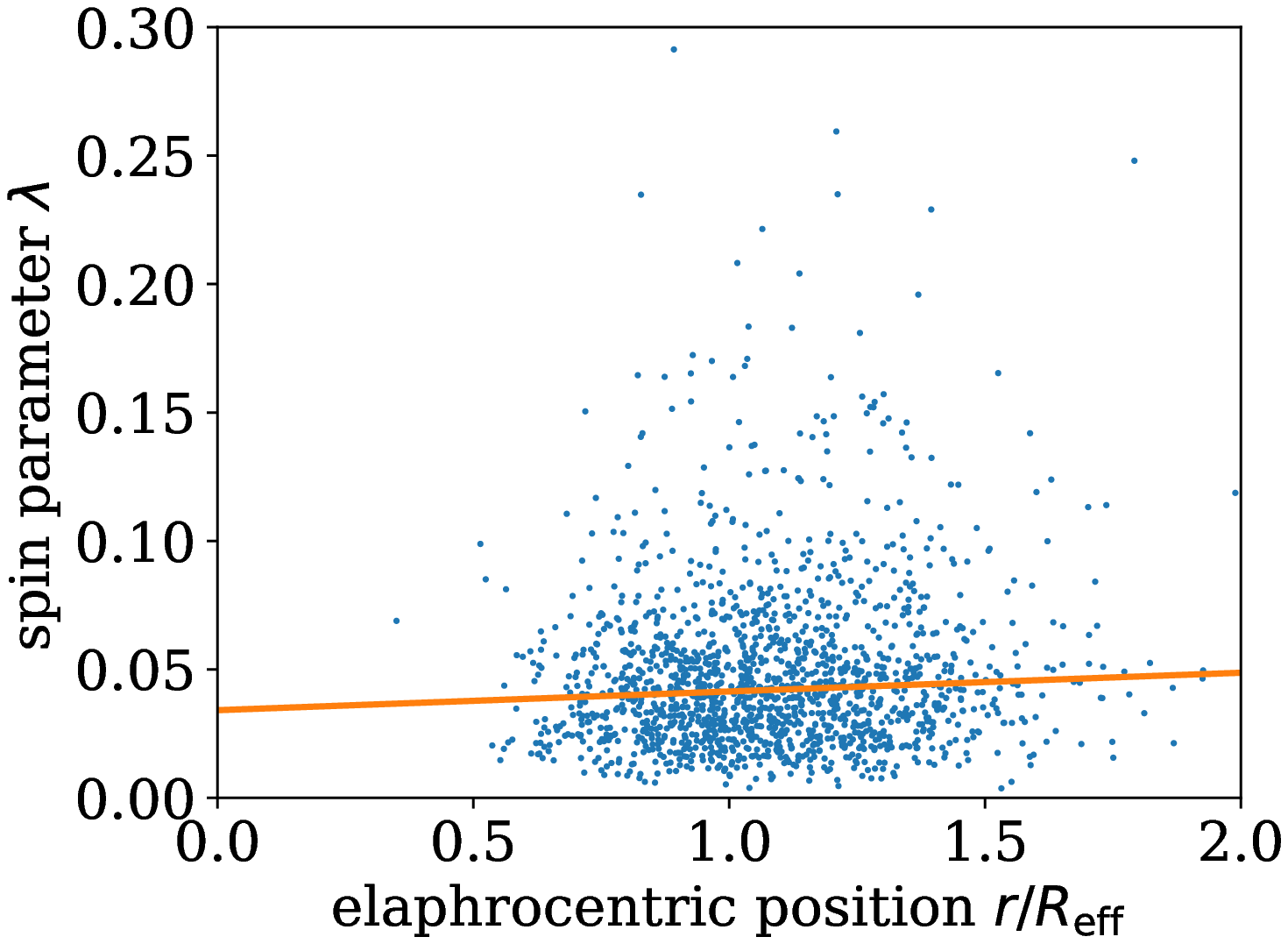}
  \caption{Dimensionless spin parameter $\lambda$ (Eq.~\protect\eqref{e-defn-spin}) versus elaphrocentric location $r/R_{\mathrm{eff}}$, as in Fig.~\protect\ref{f-galaxy-spin-frac}.
    The fit is $\lambda = (\ElaphcenGalspinReldistZerovalue \pm \ElaphcenGalspinReldistSigZerovalue) + (\ElaphcenGalspinReldistSlopevalue \pm \ElaphcenGalspinReldistSigSlopevalue \pmcosvar \ElaphcenGalspinReldistSlopevalueCosmStdDev) \,r/R_{\mathrm{eff}}$.
    \label{f-galaxy-spin-rreff}}
\end{figure}

\begin{figure}
  \includegraphics[width=0.9\columnwidth]{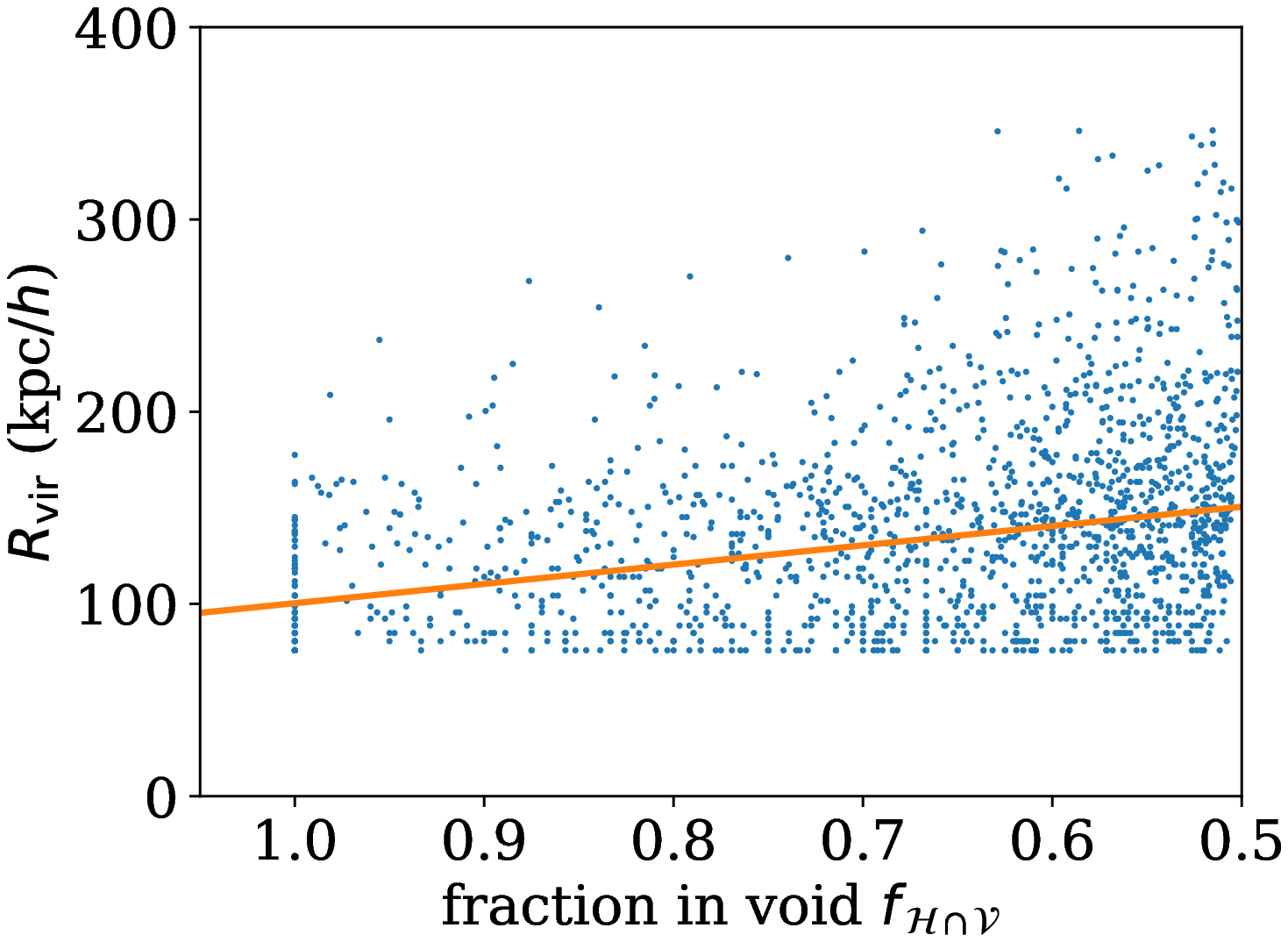}
  \caption{Virial radius $R_{\mathrm{vir}}$ (Eq.~\protect\eqref{e-defn-spin}) versus $\haloinvoidfrac$.
    The fit is $R_{\mathrm{vir}} = \left[ (\GalRvirFracZerovalue \pm \GalRvirFracSigZerovalue) + (\GalRvirFracSlopevalue \pm \GalRvirFracSigSlopevalue) \haloinvoidfrac \right] \,\mathrm{kpc}/h$.
    \postrefereechanges{The sharp lower limit in $R_{\mathrm{vir}}$ follows from the minimum detectable halo mass in the $N$-body simulation.}
    \label{f-galaxy-Rvir-frac}}
\end{figure}

\begin{figure}
  \includegraphics[width=0.9\columnwidth]{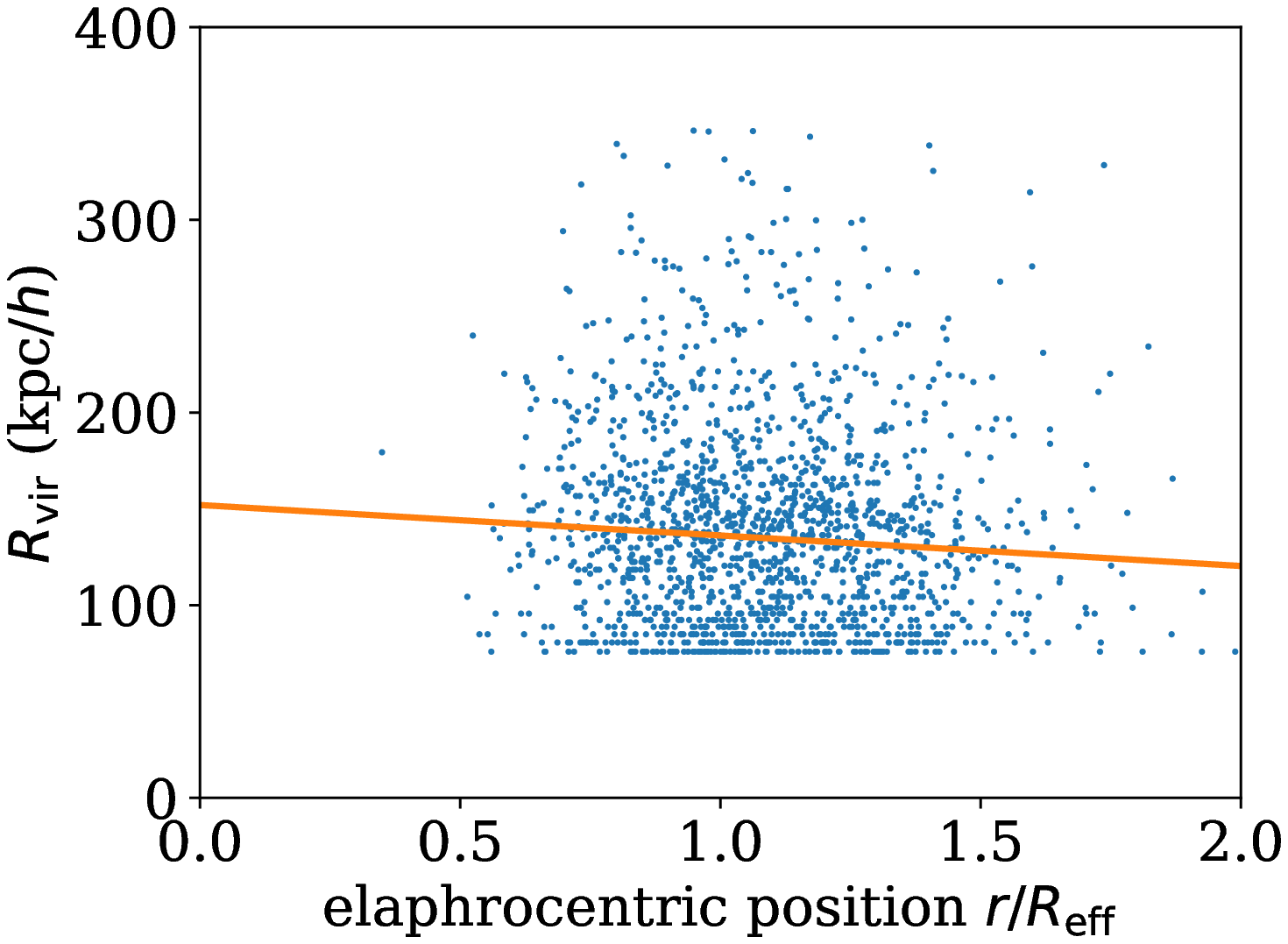}
  \caption{Virial radius $R_{\mathrm{vir}}$ versus elaphrocentric location $r/R_{\mathrm{eff}}$, as in Fig.~\protect\ref{f-galaxy-Rvir-frac}.
    The fit is $R_{\mathrm{vir}} = \left[ (\ElaphcenGalRvirReldistZerovalue \pm \ElaphcenGalRvirReldistSigZerovalue) + (\ElaphcenGalRvirReldistSlopevalue \pm \ElaphcenGalRvirReldistSigSlopevalue \pmcosvar \ElaphcenGalRvirReldistSlopevalueCosmStdDev) \,r/R_{\mathrm{eff}} \right] \,\mathrm{kpc}/h$.
    \label{f-galaxy-Rvir-rreff}}
\end{figure}

As stated above, the void galaxy host haloes are typically somewhat less massive than the non-void haloes and the collapse epochs of void galaxies are significantly later.
These two parameters should have opposite effects on the halo virial radii.
In Table~\ref{Galaxy-size-table}, we see that $R_{\mathrm{vir}}$ is significantly larger for non-void galaxies, showing that the higher mass of non-void galaxies plays the dominant role.

To see if a general trend of $\rdiskscale$ also exists as a function of a galaxy's void location, Figs~\ref{f-galaxy-size-frac} and \ref{f-galaxy-size-rreff} examine the dependence of $\rdiskscale$ on $\haloinvoidfrac$ and elaphrocentric distance for void galaxies.
The slope of the best fit in Fig.~\ref{f-galaxy-size-frac}, $\diffd \rdiskscale/\diffd \haloinvoidfrac = \GalsizeFracSlopevalue \pm \GalsizeFracSigSlopevalue \pmcosvar \GalsizeFracSlopevalueCosmStdDev$~kpc/$h$, is not significantly non-zero when we take into account cosmic variance (the distribution of $\diffd \rdiskscale/\diffd \haloinvoidfrac$ over repeated runs includes a strong tail of values that are not significantly non-zero).
The dependence on elaphrocentric position (Fig.~\ref{f-galaxy-size-rreff}) is not significant either.

In Figs~\ref{f-galaxy-spin-frac}--\ref{f-galaxy-Rvir-rreff}, we investigate whether the spin parameter $\lambda$ or the virial radius $R_{\mathrm{vir}}$ is more responsible for the modest reduction in the disk scale length of void galaxies as indicated in Table~\ref{Galaxy-size-table}.
\postrefereechanges{The sharp lower limit in Figs~\ref{f-galaxy-Rvir-frac} and \ref{f-galaxy-Rvir-rreff} is an artefact of the detection threshold of dark matter haloes in the $N$-body simulation.
The virial radius is calculated by {\sagename} from the halo mass.}
Out of the four figures (Figs~\ref{f-galaxy-spin-frac}--\ref{f-galaxy-Rvir-rreff}), two show highly significant slopes: Figs~\ref{f-galaxy-spin-frac} and \ref{f-galaxy-Rvir-frac}.
The slopes in Figs~\ref{f-galaxy-spin-rreff} and \ref{f-galaxy-Rvir-rreff} are not significantly different from zero.

The slopes in Figs~\ref{f-galaxy-spin-frac} and \ref{f-galaxy-Rvir-frac}, $\diffd \lambda/\diffd \haloinvoidfrac = \GalspinFracSlopevalue \pm \GalspinFracSigSlopevalue \pmcosvar \GalspinFracSlopevalueCosmStdDev$ and $\diffd R_{\mathrm{vir}}/\diffd \haloinvoidfrac = \GalRvirFracSlopevalue \pm \GalRvirFracSigSlopevalue$~kpc/$h$, respectively, are both very strong, but opposed.
Galaxies better identified as void galaxies have higher spins, but also lower virial radii and lower masses.
The overall effect, as shown in Fig.~\ref{f-galaxy-size-frac}, is that the two effects more or less cancel, in contrast to the full-population results shown in Table~\ref{Galaxy-size-table}.

Figure~\ref{f-galaxy-size-rreff} and its fit show that overall, the elaphrocentric distance $r/R_{\mathrm{eff}}$ has only a weak effect on galaxy disk scale lengths $\rdiskscale$.
The halo size and spin parameter both have weak, though apparently again opposite, dependences on $\rdiskscale$, with $\diffd \lambda/\diffd \left(r/R_{\mathrm{eff}}\right) = \GalspinReldistSlopevalue \pm \GalspinReldistSigSlopevalue \pmcosvar \ElaphcenGalspinReldistSlopevalueCosmStdDev$ in Fig.~\ref{f-galaxy-spin-rreff} and $\diffd R_{\mathrm{vir}}/\diffd \left(r/R_{\mathrm{eff}}\right) = \GalRvirReldistSlopevalue \pm \GalRvirReldistSigSlopevalue \pmcosvar \ElaphcenGalRvirReldistSlopevalueCosmStdDev$~kpc/$h$ in Fig.~\ref{f-galaxy-Rvir-rreff}.

In summary, the trends for $\rdiskscale$, $\lambda$, and $R_{\mathrm{vir}}$ found in Table~\ref{Galaxy-size-table} are similar, but strengthened, when void location of a galaxy is quantified by $\haloinvoidfrac$, and insignificant when void location is quantified by $r/R_{\mathrm{eff}}$.
The lack of a significant dependence of these parameters on the elaphrocentric distance, $r/R_{\mathrm{eff}}$, is somewhat surprising, since one might expect $\haloinvoidfrac$ and $r/R_{\mathrm{eff}}$ to be proxies for one another, equally valid for defining how high a galaxy is on the potential hill of a void.
We discuss this counterintuitive result further in \SSS\ref{s-discuss-sizes}.

\postrefereechanges{Since $R_{\mathrm{vir}}$ is obtained from $M_{\mathrm{vir}}$ by {\sagename} on the assumption of a detection threshold of $200$ times the critical density, Fig.~\ref{f-galaxy-Rvir-frac} can be qualitatively compared with observational estimates of the masses of void galaxies.
  Keeping in mind the fixed lower limit in mass resolution, the robust best fit relation can be used to describe the galaxies best located in a void ($\haloinvoidfrac = 1$) as having host haloes with $R_{\mathrm{vir}} \sim 100$~kpc/$h$, and those best located in walls ($\haloinvoidfrac = 0$) as having higher mass host haloes, with $R_{\mathrm{vir}} \sim 200$~kpc/$h$.
  Thus, the masses of galaxies' host haloes located in the walls should be typically eight times those of galaxies located in voids.
  \citet{Weistrop95} found that 12 H$\alpha$-emitting galaxies in the Bo{\"o}tes void were mostly quite luminous, and \citet{Szomoru96} found that a sample of 16 targetted Bo{\"o}tes void galaxies (along with 21 companion galaxies) appeared to be similar to corresponding late-type, gas-rich field galaxies and of similar masses.
  These Bo{\"o}tes void surveys would appear to be inconsistent with the mass difference found here.
  However, the more recent and bigger survey of 60 void galaxies in the Sloan Digital Sky Survey (SDSS) by \citet{Kreckel12} found that these have moderately low stellar masses, mostly around $10^{9}$--$10^{10} M_{\odot}$.
  The SDSS void galaxy survey would appear more likely to be consistent with our results.
  Future work, in which the remaining steps in galaxy formation and evolution modelling are added to the pipeline presented here, should enable quantitative comparisons to see if the Bo{\"o}tes and SDSS void galaxies \citep[see also][]{Pan2012voids,NadathurHotch15voidsI,Sutter15VIDE} are consistent with those modelled here.}

\begin{table}
  \centering
  \caption{Median radial $\dot{v}_{\parallel}$ and tangential $\dot{v}_{\perp}$ accelerations at $\rtest = \ElaphrocentreHaloOriginalRadiusMpcvalue$~Mpc/$h$ from the elaphrocentre.
    \label{t-elaphro-accel}}
  $\begin{array}{l c}
    \hline
    & \mbox{median acceleration}\\
    \hline
    \dot{v}_{\parallel} &
    \ElaphroAcciAccRadMedvalue \pm \ElaphroAcciAccRadStderrMedvalue \,\mathrm{km/s/Gyr}\rule{0ex}{2.7ex} \\
    \dot{v}_{\perp}  &
    \ElaphroAcciAccTanMedvalue \pm \ElaphroAcciAccTanStderrMedvalue \pmcosvar \ElaphroAcciAccTanMedvalueCosmStdDev \,\mathrm{km/s/Gyr} \\
    \hline
  \end{array}$
\end{table}

\begin{figure}
  \includegraphics[width=0.9\columnwidth]{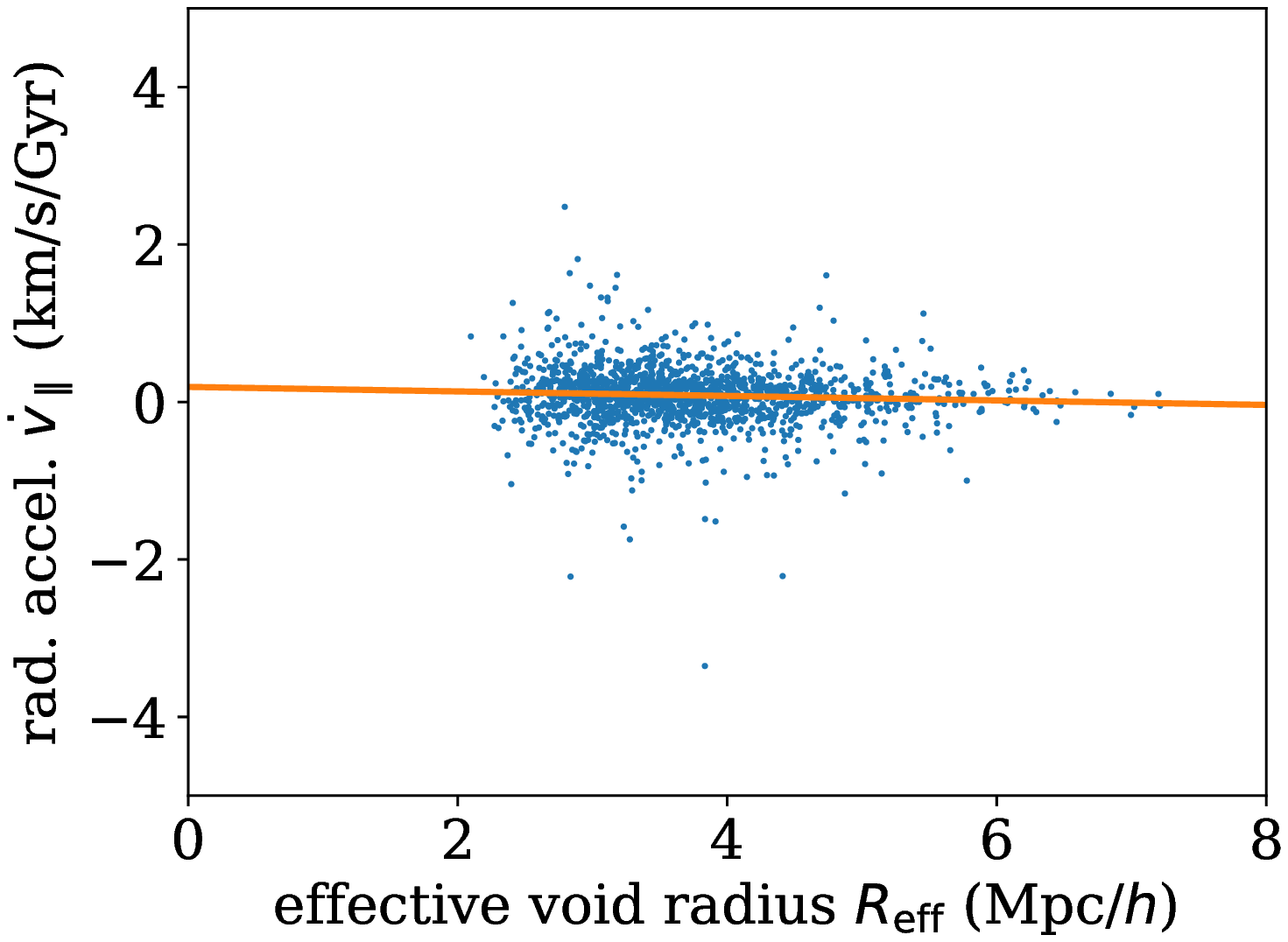}
  \caption{Dependence of the radial elaphro-acceleration $\dot{v}_{\parallel}$ as a function of effective void radius $R_{\mathrm{eff}}$, together with a robust linear fit.
    The fit is $\dot{v}_{\parallel} = [(\AccRadZerovalue \pm \AccRadSigZerovalue) + (\AccRadSlopevalue \pm \AccRadSigSlopevalue \pmcosvar \AccRadSlopevalueCosmStdDev)\,h/\mathrm{Mpc}\, R_{\mathrm{eff}}]$~km/s/Gyr.
\label{f-acc-rad-rreff}}
\end{figure}

\begin{figure}
  \iftoggle{removesomefits}{
    \includegraphics[width=0.9\columnwidth]{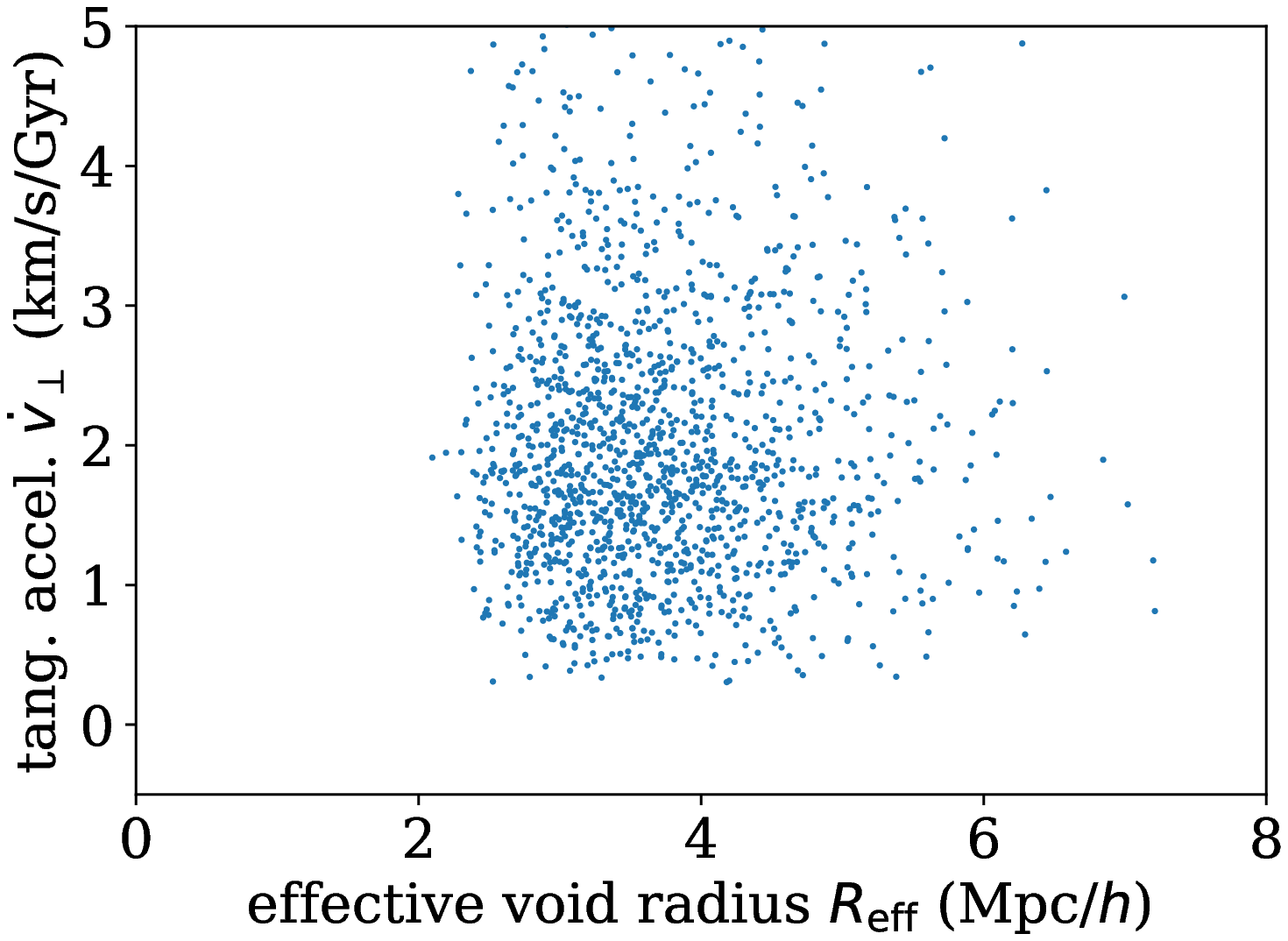}}{
    \includegraphics[width=0.9\columnwidth]{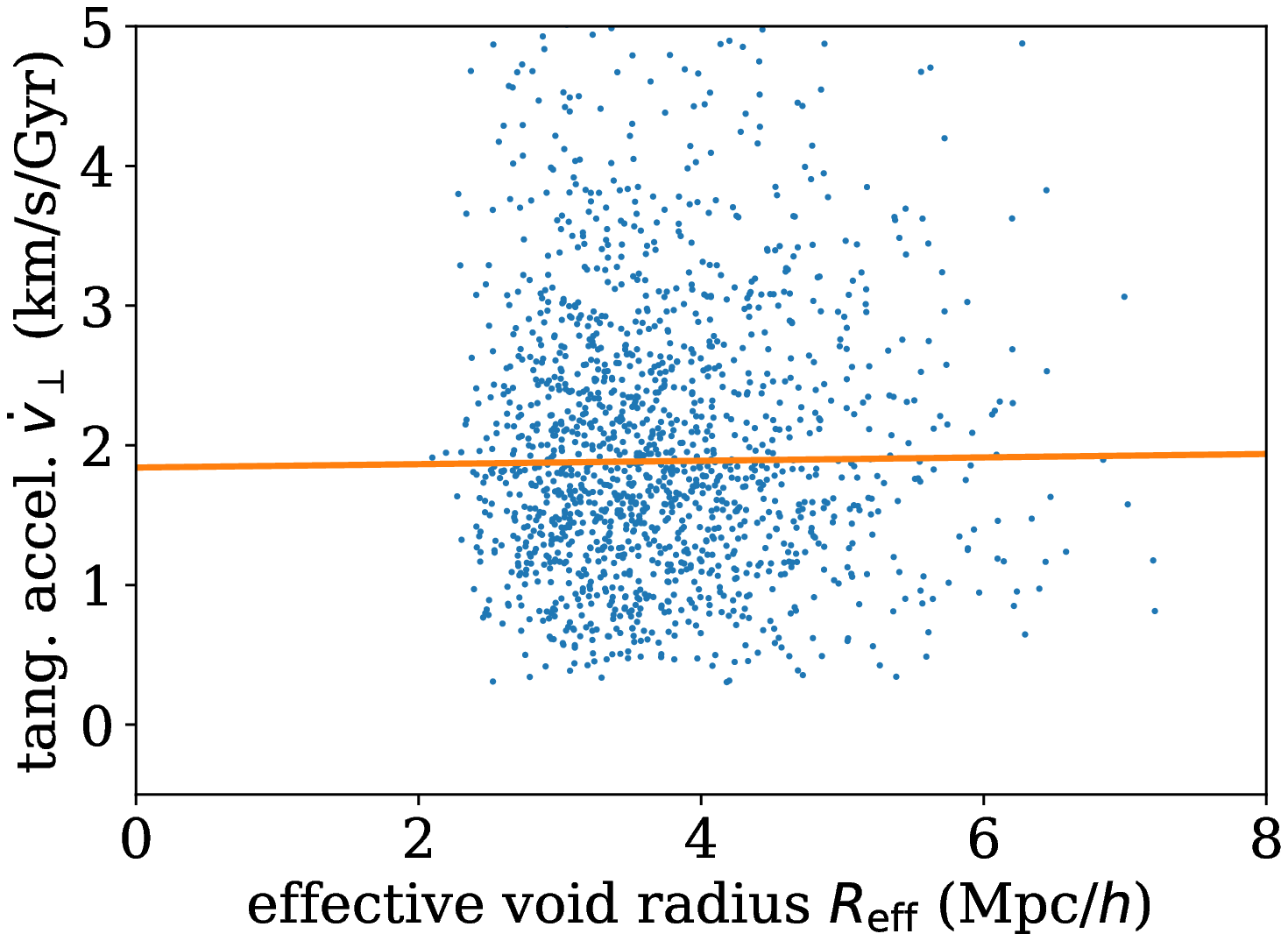}}
  \caption{Dependence of the tangential elaphro-acceleration $\dot{v}_{\parallel}$ as a function of $R_{\mathrm{eff}}$.
    The best robust linear fit is $\dot{v}_{\perp} = [(\AccTanZerovalue \pm \AccTanSigZerovalue) + (\AccTanSlopevalue \pm \AccTanSigSlopevalue \pmcosvar \AccTanSlopevalueCosmStdDev)\,h/\mathrm{Mpc}\,  R_{\mathrm{eff}}]$~km/s/Gyr.
\label{f-acc-tan-rreff}}
\end{figure}

\subsection{Elaphro-acceleration}  \label{s-results-elaphro}

Elaphro-accelerations were calculated as described in \SSS\ref{s-method-elaphro-acc}, at elaphrocentric distances of $\rtest = \ElaphrocentreHaloOriginalRadiusMpcvalue$~Mpc.
We found ${\ElaphroAcciCountedVoidsPotvalue}$ voids that allowed valid estimates.
The median radial acceleration for a test particle at these positions is given in Table~\ref{t-elaphro-accel}.
The amplitudes of these two values are not directly comparable, because $\dot{v}_{\parallel}$ is the median of signed values, while $\dot{v}_{\perp}$ is non-negative by construction.
The Newtonian estimate for the gravitational pull at $\rtest$ away from the centre of \postrefereechanges{the canonical high-mass test halo} of mass \postrefereechanges{$M\tagTest$}, a barycentre, is an inward-pointing acceleration, i.e. \postrefereechanges{$\dot{v}_{\parallel}^{\textrm{test-halo}} = \testAccel\, \mathrm{{km/s}/{Gyr}}$}.

The median estimate of $\dot{v}_{\parallel}$ given in Table~\ref{t-elaphro-accel} is an outward-pointing elaphro-acceleration to high statistical significance.
\postrefereechangesB{This is consistent with what could be expected of the elaphrocentre, defined as the location of the global maximum in the potential of a void, with mass typically moving away from the elaphrocentre.}
The amplitude is more than an order of magnitude weaker than that of the barycentric acceleration towards our canonical \postrefereechanges{high-mass} halo.
Figure~\ref{f-acc-rad-rreff} shows that the full spread of radial elaphro-accelerations is wide, including many negative values, and that dependence on the void effective radius $R_{\mathrm{eff}}$ is weak.
Together, these properties imply that, at least with the numerical techniques and simulation parameters adopted in this work, a systematic antigravitational effect at the elaphrocentre helping to oppose infall is likely to be modest.
This is consistent with our infall results above.

The median tangential acceleration $\dot{v}_{\perp}$ is given in Table~\ref{t-elaphro-accel} and the individual estimates and fit are shown in Fig.~\ref{f-acc-tan-rreff}.
These values are about an order of magnitude higher in amplitude than those of $\dot{v}_{\parallel}$, and similar to that for our canonical \postrefereechanges{high-mass} halo.
This supports the argument that rotational properties of the fluid flow such as shear and vorticity are likely to be significant in understanding voids.
Figures~\ref{f-acc-rad-rreff} and \ref{f-acc-tan-rreff} do not show significant dependence of $\dot{v}_{\parallel}$ and $\dot{v}_{\perp}$ on the size of a void.

\section{Discussion} \label{s-discuss}

\subsection{Infall rates} \label{s-discuss-infall}

We generally found a lack of statistically significant trends in the two infall parameters on $\haloinvoidfrac$ and $r/R_{\mathrm{eff}}$ \postrefereechanges{(Table~\ref{Infall-rate-table}, Figs~\ref{f-infall-frac-amp}--\ref{f-infall-rreff-tau})}, for those galaxies whose host haloes' infall rates could be fit with an exponential best fit.
\postrefereechanges{A moderately significant non-zero dependence is that} of ${\InfallAmplitude}$ on $\haloinvoidfrac$, \postrefereechanges{shown in Figure~\ref{f-infall-frac-amp}}, opposite to that expected: void galaxies have slightly higher amplitudes of their best fit infall rate histories.
The simplest interpretation is that in a halo destined to collapse with a given final mass, the infall history of matter into that halo is nearly independent of the environment.
\postrefereechanges{The modest amplitude of the median acceleration outwards from the elaphrocentre -- the {\em elaphro-acceleration} $\dot{v}_{\parallel}$ (Table~\ref{t-elaphro-accel}) -- is probably the main reason for this.}
This is something like a Newtonian numerical equivalent of Birkhoff-like \citep{BirkhoffLanger23} or \enquote*{finite infinity} \citep{Ellis84relcosmo,Wiltshire07clocks} arguments for modelling galaxies in isolation from their environment.
Our hypothesis that the void environment helps to form \postrefereechanges{giant} LSBGs by providing slow, weak infall is not supported by our numerical results.

\postrefereechanges{Although the parameters of our simplified infall fit are not affected by the position of a galaxy, we found a significant difference in the median galaxy collapse epoch (the first time step with a non-zero, non-negative infall rate) between void galaxies and non-void galaxies.}
\postrefereechanges{While we did not expect this} to play a major role in \postrefereechanges{galaxy} formation, it should.
The median collapse epochs (in standard FLRW cosmological time) that we found were $t^{\mathrm{f}}_{\mathrm{v}}=\FormationTimeVoidsMedvalue \pm \FormationTimeVoidsStderrMedvalue$~Gyr and $t^{\mathrm{f}}_{\mathrm{nv}}=\FormationTimeNonvoidsMedvalue \pm \FormationTimeVoidsStderrMedvalue $~Gyr for the void and non-void galaxies, respectively (\SSS\ref{s-results-infall}).
Thus, to very high significance, void galaxies typically form later than non-void galaxies.
We interpret this as a result of their formation in underdensities.
\postrefereechanges{Haloes that collapse later should, according to the standard spherical collapse model, have a lower matter density.
  Galaxies should thus form} with lower dark matter and baryonic matter densities (M$_{\odot}$/kpc$^3$), \postrefereechanges{which} may lead to lower surface densities (M$_{\odot}$/kpc$^2$) of galaxy disks and lower surface brightnesses (L$_{\odot}$/kpc$^2$) induced by star formation.

\postrefereechanges{This characteristic of void galaxies agrees} with \citet{Rong2017}, in the sense that these authors found that UDGs have a later formation time than typical dwarfs, and assuming that we associate UDGs as being located in voids.
\citet{Rong2017} found that UDGs at the current epoch have a median age of $7.1$~Gyr compared to typical dwarfs with a median age of \postrefereechanges{$9.6$~Gyr}.
For the current epoch estimated at $t_0 \sim 13.8$~Gyr, these correspond to median UDG and ordinary dwarf formation epochs of $6.7$~Gyr and \postrefereechanges{$4.2$~Gyr,} respectively.
Since (i) we consider high-mass host haloes and correspondingly high-mass galaxies, rather than the more typical UDGs and dwarfs, and (ii) we separate populations by their location in voids rather than by continuing through to stellar population synthesis, more precise correspondence with \citet{Rong2017}'s results would be unlikely with our current pipeline.
The qualitative agreement that void galaxies typically form later than non-void galaxies by about a Gigayear (our result) and that UDGs form about three Gigayears later than ordinary dwarf galaxies \citep{Rong2017} is a promising sign of progress towards a cohesive theory of LSBG formation.

\subsection{Galaxy sizes} \label{s-discuss-sizes}

\begin{figure}
  \iftoggle{removesomefits}{
    \includegraphics[width=0.9\columnwidth]{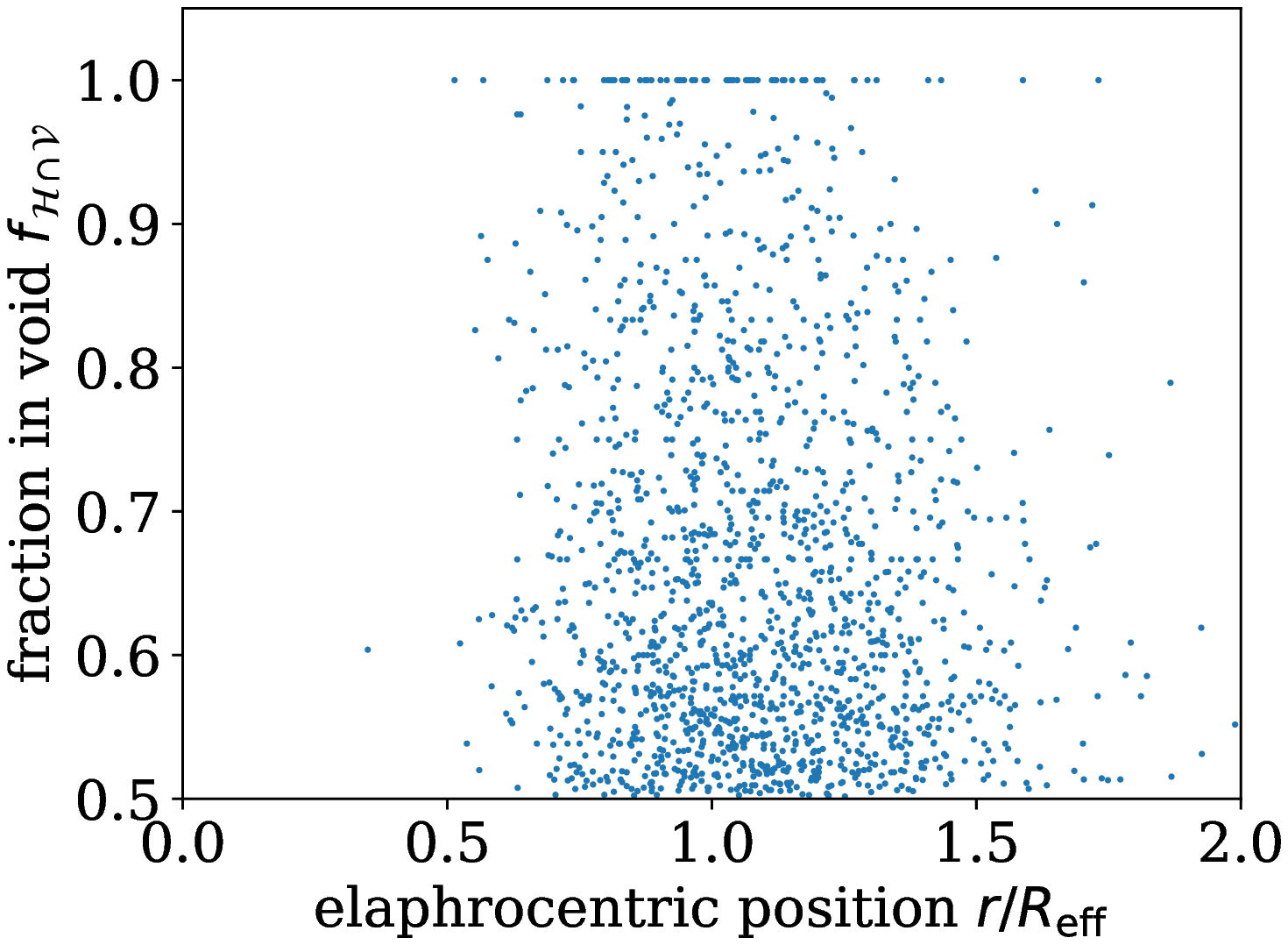}}{
    \includegraphics[width=0.9\columnwidth]{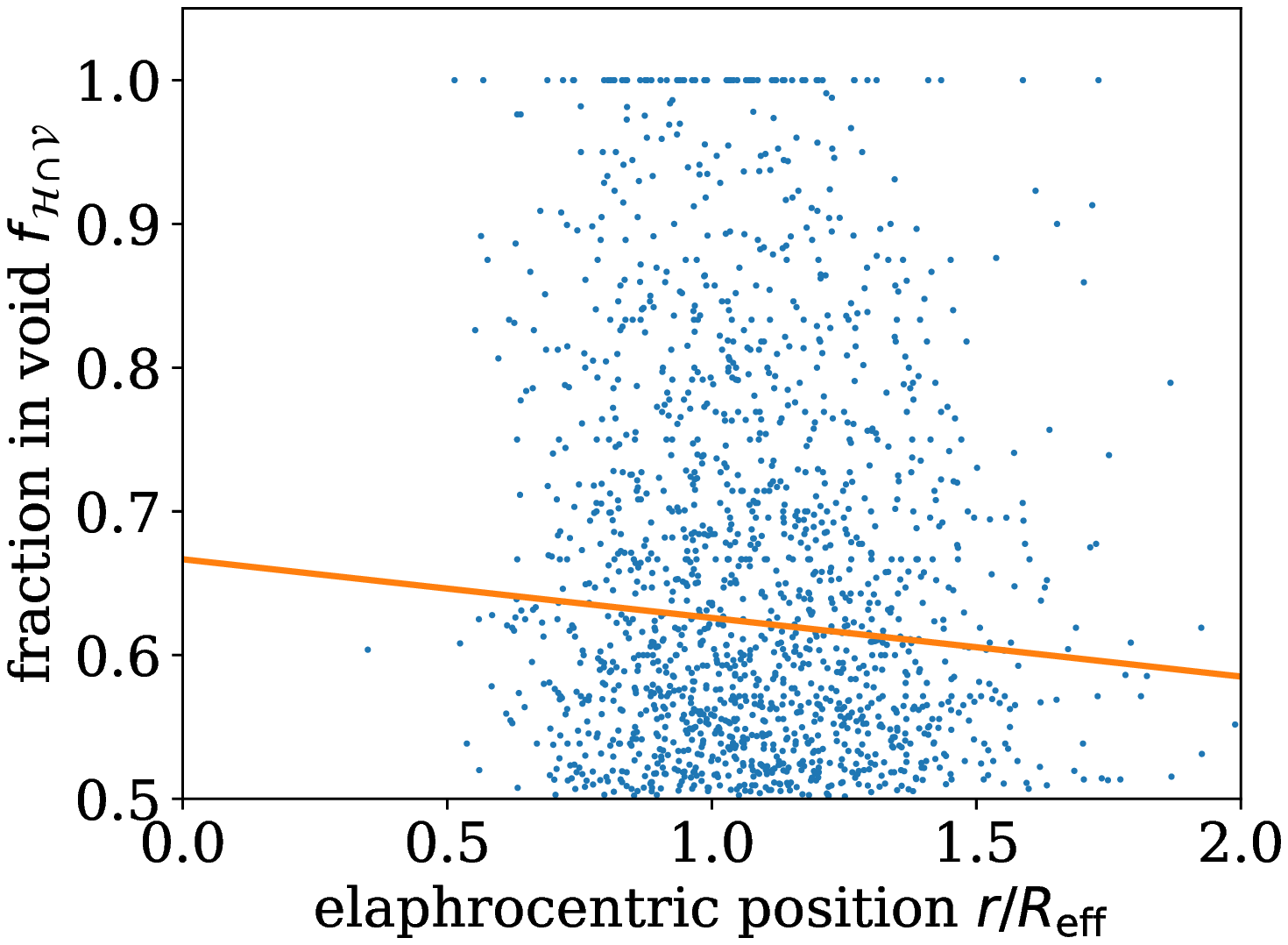}}
  \caption{Relation between fraction in void $\haloinvoidfrac$ and elaphrocentric position $r/R_{\mathrm{eff}}$.
    The best robust linear (Theil--Sen) fit is $\lambda = (\FracRreffZerovalue \pm \FracRreffSigZerovalue) + (\FracRreffSlopevalue \pm \FracRreffSigSlopevalue \pmcosvar \FracRreffSlopevalueCosmStdDev) \,r/R_{\mathrm{eff}}$.
    \postrefereechanges{Plain} text table for this figure: \mbox{\href{\projectzenodofilesbase/voidgals_infall.dat}{\projectzenodoid/voidgals\_infall.dat}}.
\label{f-frac-rreff}}
\end{figure}

\postrefereechanges{The most massive galaxies and} their host haloes (with the largest virial radii $R_{\mathrm{vir}}$) form in the tight knots of the cosmic web.
Although we found that galaxies in voids tend to be smaller when comparing the overall void to non-void populations (Table~\ref{Galaxy-size-table}), this was not detected in the dependence of the galaxy disk scale length $\rdiskscale$ on $\haloinvoidfrac$, nor on on $r/R_{\mathrm{eff}}$.
\postrefereechanges{In contrast,} we did find significant dependences of the two contributing parameters to $\rdiskscale$, the spin parameter $\lambda$ and the virial radius $R_{\mathrm{vir}}$, on $\haloinvoidfrac$, but insignificant dependence on $r/R_{\mathrm{eff}}$.
\postrefereechanges{While the dependence on $R_{\mathrm{vir}}$ is clearly} the dominant effect, we find that the spin parameter considered alone, which tends to form large galaxy disks, is higher for void galaxies.
This is seen most significantly in Fig.~\ref{f-galaxy-spin-frac}, via the dependence of $\lambda$ on $\haloinvoidfrac$.
The higher spin parameter effect could be interpreted as the result of either fewer merger events weakening the spin parameter, or of gravitational effects inside the void.
A likely candidate for the latter is the tangential acceleration $\dot{v}_{\perp}$ (Table~\ref{t-elaphro-accel}).
This is typically of the same order of magnitude as the gravitational pull of the source region of \postrefereechanges{the high-mass test} halo \postrefereechanges{that is, in its properties, inspired by Malin 1}.
However, in this work we have focussed on overall population properties and reproducibility of the pipeline.
Continuation through to disk surface densities, for comparison with \citet[][\SSS{}7.1, 7.2]{DiPaoloSalucci20}, and to stellar population evolution, remains a task for future work.
Moreover, the rare high-mass galaxies that are well identified as void galaxies may require high numbers of simulations, if realised randomly, since, by definition, they are rare.
Alternatively, a small number of big simulations may provide qualitative clues, as in the Malin 1 analogue found in the IllustrisTNG simulation by \citet{ZhuXuGaspari18Malin1}.

It may seem somewhat surprising that these significant dependences on $\haloinvoidfrac$ do not translate into significant dependences on $r/R_{\mathrm{eff}}$.
The explanation most likely lies in the fact that {\revolvername} traces voids using Voronoi tessellation and the watershed algorithm, and voids are in general aspherical.
Galaxies can lie in fairly empty parts of a void, with high $\haloinvoidfrac$, while lying, for example, in the far ends of a prolate void, where $r/R_{\mathrm{eff}} > 1$.
It would be useful to quantify the relation between $\haloinvoidfrac$ and $r/R_{\mathrm{eff}}$.

Figure~\ref{f-frac-rreff} shows visually that there is no obvious relation between $\haloinvoidfrac$ and $r/R_{\mathrm{eff}}$.
As indicated in the caption, the best robust fit indicates no statistically significant non-zero linear slope relating the two parameters.
The fact that most of the void galaxies lie at $r/R_{\mathrm{eff}} \gtapprox 1$ is consistent with the explanation suggested above.
This can also be thought of as follows.
Voids are defined by the absence of particles.
Galaxies are generally not found in the interior of voids, because then the void shapes would be defined differently, shifting those galaxies' host haloes from void status to near-boundary status in the redefined voids.
A halo located at approximately $r/R_{\mathrm{eff}}  \gtapprox 1$ in a highly aspherical void is not necessarily located in a locally low density region, so it is not constrained to contain a high number of particles identified as void particles.

Thus, $\haloinvoidfrac$ and $r/R_{\mathrm{eff}}$ appear to be statistically independent parameters.
The significant trend of $\rdiskscale$ on $\haloinvoidfrac$, and the fact that $\haloinvoidfrac$ has a more local physical meaning than $r/R_{\mathrm{eff}}$, suggest that $\haloinvoidfrac$ is the more physically useful parameter to choose.
Numerical and observational studies that measure the local number density around a given density will tend to correspond to the use of $\haloinvoidfrac$ as a parameter for characterising the void nature of a galaxy.

\iftoggle{CompareCentres}{
\begin{table}
  \centering
  \caption{\protect\postrefereechanges{Robust best fit parameters for $\rdiskscale$, $R_{\mathrm{vir}}$ and $\lambda$ with respect to the void-centric distance for the three different centre definitions discussed in \SSS\ref{s-method-elaphro}.}
    \label{t-centres-comparison}}
  \postrefereechanges{$\begin{array}{l c}
    \hline
    \multicolumn{2}{c}{\mbox{elaphrocentre}} \\
    \hline
    \rdiskscale &
    (\ElaphcenGalsizeReldistZerovalue \pm \ElaphcenGalsizeReldistSigZerovalue) + (\ElaphcenGalsizeReldistSlopevalue \pm \ElaphcenGalsizeReldistSigSlopevalue) \,r/R_{\mathrm{eff}} \rule{0ex}{2.7ex} \\ R_{\mathrm{vir}} &
    (\ElaphcenGalRvirReldistZerovalue \pm \ElaphcenGalRvirReldistSigZerovalue) + (\ElaphcenGalRvirReldistSlopevalue \pm \ElaphcenGalRvirReldistSigSlopevalue) \,r/R_{\mathrm{eff}} \\
    \lambda &
    (\ElaphcenGalspinReldistZerovalue \pm \ElaphcenGalspinReldistSigZerovalue) + (\ElaphcenGalspinReldistSlopevalue \pm \ElaphcenGalspinReldistSigSlopevalue) \,r/R_{\mathrm{eff}} \\
    \hline
    \multicolumn{2}{c}{\mbox{circumcentre}} \\
    \hline
    \rdiskscale &
    (\CirccenGalsizeReldistZerovalue \pm \CirccenGalsizeReldistSigZerovalue) + (\CirccenGalsizeReldistSlopevalue \pm \CirccenGalsizeReldistSigSlopevalue) \,r/R_{\mathrm{eff}} \rule{0ex}{2.7ex} \\ R_{\mathrm{vir}} &
    (\CirccenGalRvirReldistZerovalue \pm \CirccenGalRvirReldistSigZerovalue) + (\CirccenGalRvirReldistSlopevalue \pm \CirccenGalRvirReldistSigSlopevalue) \,r/R_{\mathrm{eff}} \\
    \lambda &
    (\CirccenGalspinReldistZerovalue \pm \CirccenGalspinReldistSigZerovalue) + (\CirccenGalspinReldistSlopevalue \pm \CirccenGalspinReldistSigSlopevalue) \,r/R_{\mathrm{eff}} \\
    \hline
    \multicolumn{2}{c}{\mbox{geometrical centroid}} \\
    \hline
    \rdiskscale &
    (\GeomcenGalsizeReldistZerovalue \pm \GeomcenGalsizeReldistSigZerovalue) + (\GeomcenGalsizeReldistSlopevalue \pm \GeomcenGalsizeReldistSigSlopevalue) \,r/R_{\mathrm{eff}} \rule{0ex}{2.7ex} \\ R_{\mathrm{vir}} &
    (\GeomcenGalRvirReldistZerovalue \pm \GeomcenGalRvirReldistSigZerovalue) + (\GeomcenGalRvirReldistSlopevalue \pm \GeomcenGalRvirReldistSigSlopevalue) \,r/R_{\mathrm{eff}} \\
    \lambda &
    (\GeomcenGalspinReldistZerovalue \pm \GeomcenGalspinReldistSigZerovalue) + (\GeomcenGalspinReldistSlopevalue \pm \GeomcenGalspinReldistSigSlopevalue) \,r/R_{\mathrm{eff}} \\
    \hline
  \end{array}$}
\end{table}
\postrefereechanges{To see if earlier definitions of void centres, as discussed in \SSS\ref{s-method-elaphro}, could have more significant effects on galaxy sizes, we repeated our calculations for distances from the circumcentre and from the geometrical centroid instead of the elaphrocentre.
Table~\ref{t-centres-comparison} shows the robust best fits to the dependence of $\rdiskscale$, $R_{\mathrm{vir}}$ and $\lambda$ on void-centric distance for the three definitions.
The full scatter plots (not shown) are visually indistinguishable from Figs~\ref{f-galaxy-size-rreff}, \ref{f-galaxy-spin-rreff} and \ref{f-galaxy-Rvir-rreff}.
The differences between the three cases are statistically negligible.
Given that the geometrical centroid (macrocentre or volume-weighted barycentre) only encodes information about the void's periphery, with no information from its interior, it may seem surprising that this gives similar results to the other two centres.
However, this is probably explained by the near-total absence of galaxies located within the central half-radius of the void; the detected galaxies' radial distances from the centre only vary mildly with different definitions of the centre.}
}{
}

\subsection{Elaphro-acceleration} \label{s-discuss-elaphro}

We found a significantly non-zero positive median acceleration towards the edges of a void.
However, this median outwards acceleration, $\dot{v}_{\parallel}$ (Table~\ref{t-elaphro-accel}), is of the order of only a few percent of the inward gravitational pull at $\rtest = \ElaphrocentreHaloOriginalRadiusMpcvalue$~Mpc that the source mass excess for our canonical \postrefereechanges{high-mass} host halo would create, $\dot{v}_{\parallel}\tagTest = \testAccel\, \mathrm{{km/s}/{Gyr}}$.
Moreover, Fig.~\ref{f-acc-rad-rreff} shows a wide scatter between outward and inward accelerations from the elaphrocentre.
Thus, while a modest average effect in opposing infall could be expected for galaxies that are close to the elaphrocentre of a void, the effect would be sensitive to the wide distribution in values and to relations between the elaphrocentric acceleration and other dynamical parameters.

Future work in placing a test halo near an elaphrocentre, with the assumption that the test halo has no dynamical effect on the underlying void properties, may use these results as a guide to judging the likely strength of the effect.
For example, the probability that a test halo of a given mass placed at the elaphrocentre of a random void has its infall rate weakened sufficiently to make it a candidate LSBG could be estimated.
This could be compared with the infall rate behaviour from haloes from the $N$-body realisation itself, as we presented in \SSS\ref{s-results-infall}, at elaphrocentric positions far from the elaphrocentre.

The median tangential acceleration $\dot{v}_{\perp}$ is much higher than the radial acceleration, and might be used to study the higher spin parameter $\lambda$ found for void galaxies when identified by $\haloinvoidfrac$, as shown in Fig.~\ref{f-galaxy-spin-frac}.
Since this is of the same order of magnitude as our canonical radial acceleration, $\dot{v}_{\parallel}\tagTest$, it is likely that $\dot{v}_{\perp}$ could play an important role for galaxies forming in voids.

\subsection{Future extensions}

An obvious further development, not yet included in the present work, would be to analyse the star formation rate histories and to extend the pipeline with evolutionary stellar population synthesis methods.
This would allow us to identify LSBGs in our galaxy population in a way more closely comparable to observational results, while continuing to benefit from the reproducibility and modularity of the pipeline presented in this work.
\postrefereechanges{The inclusion of metallicity evolution, in particular that of O/H, would allow comparison with the populations of extremely metal-poor gas-rich, dwarf galaxies that seem to characterise a difference between void and non-void galaxy formation \citep{Pustilnik2020SALT}.}

Another extension would be to extend or replace the gravitational simulation.
Using a relativistic simulation, rather than a standard (Newtonian) simulation, would provide a major theoretical improvement towards more realistic results.
The scalar averaging extensions provided by {\sc inhomog} through the {\sc ramses}/{\ramsesscalavname} front end to check background-independent dynamical properties \citep{Roukema17silvir}, using the relativistic Zel'dovich approximation \citet*{BuchRZA1,BuchRZA2}, could easily be added.
Other options could include using either {\sc gevolution} \citep{AdamekDDK16code} or the fully relativistic {\sc Einstein Toolkit} \citep*{BentivegnaBruni15,Macpherson17}.
Hydrodynamical simulations would also be useful for comparison.
Given the aims of this project in providing a reproducible pipeline with modular, free-licensed components, it should, in principle, be straightforward to replace any of the pipeline steps or to start the pipeline at an intermediate step, such as analysing pre-calculated $N$-body simulation outputs.
The present form of the pipeline assumes {\sc gadget2} format for the $N$-body simulation output snapshots.

Alternatives in the statistical analysis of infall rate histories would also be useful to explore.
Here, we chose to fit the infall rate history with decaying exponentials, which include nearly constant rates as a special case with very long time scales $\InfallDecayRate$, but the reality of the mass infall rate history, and the corresponding star formation rate history, is in generally much more complex, depending especially on merger events.
A more general quantitative way of characterising the global population of mass infall or star formation rate histories would bring this pipeline closer to physical reality.

\section{Conclusion} \label{s-conclu}

\postrefereechanges{We} have presented a complete, {\em ab initio}, reproducible galaxy formation pipeline starting from a standard post--recombination-epoch spectrum of initial perturbations, aiming to identify key factors \postrefereechanges{in void galaxy formation that might contribute to the formation of giant} low surface brightness galaxies in voids (\SSS\ref{s-method-pipeline}).
We introduced the term {\em elaphrocentre} to clarify its opposite physical nature to the barycentre and we clarified the confusing use of the latter term in void studies (\SSS\ref{s-method-elaphro}).

\postrefereechanges{We} did not find statistically significant numerical evidence that the elaphrocentre, or the void location of a galaxy more generally, plays a key role in forming major populations of large diffuse galaxies -- LSBGs -- via the parameters that we considered as the most likely to play a strong role -- ${\InfallAmplitude}$ and ${\InfallDecayRate}$ \postrefereechanges{(Figs~\ref{f-infall-frac-amp}--\ref{f-infall-rreff-tau})}.
\postrefereechanges{Since gravity is attractive, there is an asymmetry between the sharp nature of barycentres (potential wells) and the wide spread of elaphrocentres (potential hills), which could explain the lack of a strong effect.}

\postrefereechanges{We found that} the fractional elaphrocentric distance of a void galaxy $r/R_{\mathrm{eff}}$ is, statistically, a less useful independent variable than $\haloinvoidfrac$, the fraction of a galaxy's host halo particles that are identified as being in a single void.
\postrefereechanges{This is important for observational} studies of void galaxies.
The characterisation of galaxies as void galaxies by $\haloinvoidfrac$, which should roughly correspond to a low local dark matter density, or by relative elaphrocentric radius $r/R_{\mathrm{eff}}$, which would require identification of voids in a catalogue, will in general \postrefereechanges{give uncorrelated results; the two parameters show no significant linear correlation (Fig.~\ref{f-frac-rreff})}.

\postrefereechanges{A serendipitous result is that void galaxies were found to be significantly smaller in virial radius (host halo mass) than non-void galaxies (Table~\ref{Galaxy-size-table}, Figs~\ref{f-galaxy-size-frac}--\ref{f-galaxy-Rvir-rreff}).
This complicates the question of giant LSBG formation, because the disk scale length $\rdiskscale$, as calculated in Eq.~\eqref{e-defn-spin}, is dominated by the virial radius.
We did find that galaxies better identified in voids have a higher spin parameter.}
This finding of a higher spin parameter for high $\haloinvoidfrac$ is \postrefereechangesB{qualitatively} consistent with \citet{Rong2017}'s result that a higher spin is a key feature of UDGs \postrefereechanges{and thus indirectly supports the hypothesis of void location constituting a significant factor in LSBG formation.}
\postrefereechangesB{Higher resolution simulations, extending to lower mass galaxies, would be needed to see if the higher spin of UDGs is quantitatively explained as a consequence of void location.}

We also found \postrefereechanges{that the median galaxy collapse epoch differs} to very high statistical significance between void and non-void populations ($t^{\mathrm{f}}_{\mathrm{v}}=\FormationTimeVoidsMedvalue \pm \FormationTimeVoidsStderrMedvalue$~Gyr and $t^{\mathrm{f}}_{\mathrm{nv}}=\FormationTimeNonvoidsMedvalue \pm \FormationTimeVoidsStderrMedvalue $~Gyr for void and non-void galaxies, respectively; \SSS\ref{s-results-infall}).
\postrefereechanges{For a standard spherical collapse model, the later collapse of void galaxies should lead to these galaxies being less dense, quite likely resulting in lower surface densities and star formation rates.}

\postrefereechanges{In summary, despite not finding} direct numerical evidence for LSBG formation in our overall populations, the higher spin parameter $\lambda$ for the overall population of void galaxies, especially when characterised by $\haloinvoidfrac$, and the later formation epoch of void galaxies, are \postrefereechangesB{qualitatively} consistent with \postrefereechanges{\citet{Rong2017}'s findings for UDGs,} \postrefereechangesB{assuming that the extension to lower masses remains valid.}
\postrefereechanges{Together with these two key features that contribute to the formation of diffuse galaxies, the smaller size of void galaxies suggests that, in contrary to our hypothesis of giant LSBG formation in voids, the role of voids is to preferentially form diffuse, somewhat smaller galaxies.}
Moreover, we hope that by providing our complete software pipeline\footnote{DOI-stamped record: {\projectzenodohref}; live {\sc git} repository: {\projectgitrepository}} using the Maneage template that aims at full reproducibility \citep{RougierHinsen2017Repdefn,Akhlaghi2020maneage}, rather than only giving the names and URLs of cosmological software packages, our work will encourage the community to avoid unnecessary effort spent in guessing the precise details of the computational software used in this and other extragalactic research.

\section*{Data Availability Statement}
As stated in the introduction, the full reproducibility package for this paper, including all input data (parameters), is available at {\projectzenodohref} and in live\footnote{\projectgitrepository} and archived\footnote{\projectgitrepositoryarchived} {\sc git} repositories.
The numerical output data for our main results is available as a DOI-identified record at \mbox{\href{\projectzenodofilesbase/voidgals_infall.dat}{\projectzenodoid/voidgals\_infall.dat}}.
The specific version of the source package used to produce this paper can be identified by its {\sc git} commit hash \projectversion.

\section*{Acknowledgments}
The authors wish to thank Krzysztof Bolejko, Mariana Jaber, Matteo Cinus, Mohammad Akhlaghi, Ra{\'u}l Infante--Sainz, M.\/ Ahsan Nazer, \postrefereechanges{Seshadri Nadathur, Nicolas Peschken,} \postrefereechangesB{and an anonymous referee} for many very useful suggestions.
Work on this paper has been supported by the ``A next-generation worldwide quantum sensor network with optical atomic clocks'' project of the TEAM IV programme of the Foundation for Polish Science co-financed by the European Union under the European Regional Development Fund.
This paper has been supported by Polish MNiSW grant DIR/WK/2018/12.
This paper has been supported by Pozna\'n Supercomputing and Networking Center (PSNC) computational grant 314.
Part of this work was supported by Universitas Copernicana Thoruniensis in Futuro under NCBR grant POWR.03.05.00-00-Z302/17.

{}We gratefully acknowledge the use of the following free-software programs and libraries in this research project: Boost 1.73.0, Bzip2 1.0.6, convertctrees 0.0-522dac5, cosmdist 0.3.8.2, ctrees 1.01-e49cbf0, cURL 7.71.1, Dash 0.5.10.2, Discoteq flock 0.2.3, Eigen 3.3.7, Expat 2.2.9, fftw2 2.1.5-4.2, FFTW 3.3.8 \citep{fftw}, File 5.39, Fontconfig 2.13.1, FreeType 2.10.2, Git 2.28.0, GNU Autoconf 2.69.200-babc, GNU Automake 1.16.2, GNU AWK 5.1.0, GNU Bash 5.0.18, GNU Binutils 2.35, GNU Compiler Collection (GCC) 10.2.0, GNU Coreutils 8.32, GNU Diffutils 3.7, GNU Findutils 4.7.0, GNU gettext 0.21, GNU gperf 3.1, GNU Grep 3.4, GNU Gzip 1.10, GNU Integer Set Library 0.18, GNU libiconv 1.16, GNU Libtool 2.4.6, GNU libunistring 0.9.10, GNU M4 1.4.18-patched, GNU Make 4.3, GNU Multiple Precision Arithmetic Library 6.2.0, GNU Multiple Precision Complex library, GNU Multiple Precision Floating-Point Reliably 4.0.2, GNU Nano 5.2, GNU NCURSES 6.2, GNU Patch 2.7.6, GNU Readline 8.0, GNU Scientific Library 2.6, GNU Sed 4.8, GNU Tar 1.32, GNU Texinfo 6.7, GNU Wget 1.20.3, GNU Which 2.21, GPL Ghostscript 9.52, HDF5 library 1.10.5, ImageMagick 7.0.8-67, Less 563, Libbsd 0.10.0, Libffi 3.2.1, libICE 1.0.10, Libidn 1.36, Libjpeg v9b, Libpaper 1.1.28, Libpng 1.6.37, libpthread-stubs (Xorg) 0.4, libSM 1.2.3, Libtiff 4.0.10, libXau (Xorg) 1.0.9, libxcb (Xorg) 1.14, libXdmcp (Xorg) 1.1.3, libXext 1.3.4, Libxml2 2.9.9, libXt 1.2.0, LibYAML 0.2.5, Lzip 1.22-rc2, Metastore (forked) 1.1.2-23-fa9170b, mpgrafic 0.3.19-4b78328, OpenBLAS 0.3.10, Open MPI 4.0.4, OpenSSL 1.1.1a, PatchELF 0.10, Perl 5.32.0, pkg-config 0.29.2, Python 3.8.5, ramses-scalav 0.0-482f90f, revolver 0.0-3b15335, rockstar 0.99.9-RC3+-6d16969, sage 0.0-2be3027, Unzip 6.0, util-Linux 2.35, util-macros (Xorg) 1.19.2, X11 library 1.6.9, XCB-proto (Xorg) 1.14, xorgproto 2020.1, xtrans (Xorg) 1.4.0, XZ Utils 5.2.5, Zip 3.0 and Zlib 1.2.11. 
Part of this project used the following {\sc python} modules: Astropy 4.0 \citep{astropy2013,astropy2018},  BeautifulSoup 4.7.1,  Cycler 0.10.0, Cython 0.29.21 \citep{cython2011},  h5py 2.10.0,  HTML5lib 1.0.1,  Kiwisolver 1.0.1, Matplotlib 3.3.0 \citep{matplotlib2007}, mpi4py 3.0.3 \citep{mpi4py2011}, Numpy 1.19.1 \citep{numpy2011}, pkgconfig 1.5.1,  pybind11 2.5.0,  pyFFTW 0.12.0,  PyParsing 2.3.1,  python-dateutil 2.8.0,  PyYAML 5.1, Scipy 1.5.2 \citep{scipy2007,scipy2011},  Setuptools 41.6.0,  Setuptools-scm 3.3.3,  Six 1.12.0,  SoupSieve 1.8 and  Webencodings 0.5.1. 
This paper was prepared using the following \LaTeX{} and related typesetting software packages: alegreya 54512 (revision), biber 2.16, biblatex 3.16, bitset 1.3, caption 56771 (revision), courier 35058 (revision), csquotes 5.2l, datetime 2.60, ec 1.0, enumitem 3.9, environ 0.3, etoolbox 2.5k, fancyhdr 4.0.1, fmtcount 3.07, fontaxes 1.0e, fontspec 2.7i, footmisc 5.5b, fp 2.1d, kastrup 15878 (revision), lastpage 1.2m, latexpand 1.6, letltxmacro 1.6, listings 1.8d, logreq 1.0, mnras 3.1, mweights 53520 (revision), newtx 1.642, pdfescape 1.15, pdftexcmds 0.33, pgf 3.1.8b, pgfplots 1.17, preprint 2011, setspace 6.7a, tcolorbox 4.42, tex 3.141592653, texgyre 2.501, times 35058 (revision), titlesec 2.13, trimspaces 1.1, txfonts 15878 (revision), ulem 53365 (revision), xcolor 2.12, xkeyval 2.8, xstring 1.83 and xstring 1.83. 
{}

\end{document}